\documentclass[12pt]{article}
\usepackage{indentfirst} 
\usepackage{tikz}
\usetikzlibrary{decorations.pathmorphing,patterns} 
\usepackage{amsfonts,amssymb,amsmath}
\usepackage{bm}          
\usepackage{cite}
\usepackage{url}
\usepackage{graphicx}
\usepackage{color}
\usepackage{rotating}  
\usepackage{dsfont}    
\usepackage[colorlinks=true,%
            citecolor=black,%
            linkcolor=black,%
            urlcolor=blue]{hyperref}

\numberwithin{equation}{section}  
\begin{document}

\bibliographystyle{plain}

\def\today{April 24, 2018 \\[1mm] Revised July 03, 2018} 

\title{The three-state Potts antiferromagnet on plane quadrangulations} 

\author{
  {\small Jian--Ping Lv${}^{1,2}$,  
          Youjin Deng${}^{3,4}$, 
          Jesper Lykke Jacobsen${}^{5,6,7}$, and 
          Jes\'us Salas${}^{8,9}$} \\[4mm]
  {\small\it ${}^1$Department of Physics, Anhui Normal University, 
             Wuhu 241000, China}\\[-2mm]
  {\small\it ${}^2$Anhui Province Key Laboratory of Optoelectric Materials 
             Science and Technology,}\\[-2mm] 
  {\small\it Wuhu 241000, China}\\[-2mm]
  {\small\tt phys.lv@gmail.com}\\[1mm]
  {\small\it ${}^3$ Hefei National Laboratory for Physical Sciences at 
             Microscale and} \\[-2mm]
  {\small\it Department of Modern Physics, 
             University of Science and Technology of China,}\\[-2mm]
  {\small\it Hefei, Anhui 230026, China} \\[-2mm]
  {\small\it ${}^4$CAS Center for Excellence and Synergetic Innovation Center
             in Quantum Information}\\[-2mm]
  {\small\it and Quantum Physics, University of Science and Technology of 
             China,}\\[-2mm] 
  {\small\it Hefei, Anhui 230026, China}\\[-2mm]
  {\small\tt yjdeng@ustc.edu.cn}               \\[1mm]
  {\small\it ${}^5$Laboratoire de Physique Th\'eorique, D\'epartement de 
             Physique de l'ENS,}\\[-2mm] 
  {\small\it \'Ecole Normale Sup\'erieure, Sorbonne Universit\'e, CNRS, 
             PSL Research University,}\\[-2mm]
  {\small\it 75005 Paris, France} \\[-2mm]
  {\small\it ${}^6$Sorbonne Universit\'e, \'Ecole Normale Sup\'erieure, 
             CNRS,}\\[-2mm]
  {\small\it Laboratoire de Physique Th\'eorique (LPT ENS), 
             75005 Paris, France}\\[-2mm] 
  {\small\it ${}^7$Institut de Physique Th\'eorique, CEA Saclay, 91191 
             Gif-sur-Yvette, France}\\[-2mm] 
  {\small\tt jesper.jacobsen@ens.fr}\\[1mm]
  {\small\it ${}^8$Grupo de Modelizaci\'on, Simulaci\'on Num\'erica
                   y Matem\'atica Industrial,}  \\[-2mm]
  {\small\it Universidad Carlos III de Madrid, 
             Avda.\/ de la Universidad 30, 28911 Legan\'es, Spain} \\[-2mm]
  {\small\it ${}^9$Grupo de Teor\'{\i}as de Campos y F\'{\i}sica
             Estad\'{\i}stica,}\\[-2mm]
  {\small\it Instituto Gregorio Mill\'an, Universidad Carlos III de
             Madrid,}\\[-2mm]
  {\small\it Unidad Asociada al Instituto de Estructura de la Materia, CSIC,
             Madrid, Spain}\\[-2mm]
  {\small\tt jsalas@math.uc3m.es}  \\[1mm]
  {\protect\makebox[5in]{\quad}}  
  \\
}

\maketitle

\thispagestyle{empty}   

\begin{abstract}
We study the antiferromagnetic 3-state Potts model on general (periodic) plane 
quadrangulations $\Gamma$. Any quadrangulation can be built 
from a dual pair $(G,G^*)$. Based on the duality properties of $G$, we 
propose a new criterion to predict the phase diagram of this model. 
If $\Gamma$ is of self-dual type (i.e., if $G$ is isomorphic to
its dual $G^*$), the model has a zero-temperature critical point 
with central charge $c=1$, and it is disordered at all positive temperatures.
If $\Gamma$ is of non-self-dual type (i.e., if $G$ is not 
isomorphic to $G^*$), three ordered phases coexist at low temperature, and
the model is disordered at high temperature. In addition, there is a 
finite-temperature critical point (separating these two phases) which 
belongs to the universality class of the ferromagnetic 3-state Potts model 
with central charge $c=4/5$. We have checked these 
conjectures by studying four (resp.\/ seven) quadrangulations of self-dual 
(resp.\/ non-self-dual) type, and using three complementary high-precision
techniques: Monte-Carlo simulations, transfer matrices, and critical 
polynomials. In all cases, we find agreement with the conjecture.
We have also found that the Wang--Swendsen--Koteck\'y Monte Carlo algorithm 
does not have (resp.\/ does have) critical slowing down at the corresponding
critical point on quadrangulations of self-dual (resp.\/ non-self-dual) type. 
\end{abstract}


\clearpage

%
%
%
%
\newcommand{\be}{\begin{equation}}
\newcommand{\ee}{\end{equation}}
\newcommand{\<}{\langle}
\renewcommand{\>}{\rangle}
\newcommand{\widebar}{\overline}
\def\spose#1{\hbox to 0pt{#1\hss}}
\def\ltapprox{\mathrel{\spose{\lower 3pt\hbox{$\mathchar"218$}}
 \raise 2.0pt\hbox{$\mathchar"13C$}}}
\def\gtapprox{\mathrel{\spose{\lower 3pt\hbox{$\mathchar"218$}}
 \raise 2.0pt\hbox{$\mathchar"13E$}}}
\def\textprime{${}^\prime$}
\def\proof{\par\medskip\noindent{\sc Proof.\ }}
\def\qed{\hbox{\hskip 6pt\vrule width6pt height7pt depth1pt \hskip1pt}\bigskip}
\def\proofof#1{\bigskip\noindent{\sc Proof of #1.\ }}
\def\half{\frac{1}{2}}
\def\third{\frac{1}{3}}
\def\twothird{\frac{2}{3}}
\def\smfrac#1#2{\textstyle \frac{#1}{#2}}
\def\smhalf{\smfrac{1}{2} }

\newcommand{\restrict}{\upharpoonright}
\newcommand{\drop}{\setminus}
\renewcommand{\emptyset}{\varnothing}
\newcommand{\real}{\mathop{\rm Re}\nolimits}
\renewcommand{\Re}{\mathop{\rm Re}\nolimits}
\newcommand{\imag}{\mathop{\rm Im}\nolimits}
\renewcommand{\Im}{\mathop{\rm Im}\nolimits}

\newcommand{\eval}[1]{\left\langle {#1} \right\rangle}
\newcommand{\leval}[1]{\langle {#1} \rangle}
\newcommand{\reval}[1]{\overline{#1}}
\newcommand{\bigast}[1]{\underset{#1}{\textrm{{\huge $\ast$}}}}
\newcommand{\deenne}[2]{\frac{\partial^#2}{\partial #1 ^#2}}
\newcommand{\atopp}[2]{\genfrac{}{}{0pt}{}{#1}{#2}}     
\newcommand{\tinyfrac}[2] {\genfrac{}{}{}{1}{#1}{#2} }
\newcommand{\Lfrac}[2] {\genfrac{}{}{}{0}{#1}{#2} }
\newcommand{\dx}[1] {\mathrm{d}{#1}}

\newenvironment{sarray}{
          \textfont0=\scriptfont0
          \scriptfont0=\scriptscriptfont0
          \textfont1=\scriptfont1
          \scriptfont1=\scriptscriptfont1
          \textfont2=\scriptfont2
          \scriptfont2=\scriptscriptfont2
          \textfont3=\scriptfont3
          \scriptfont3=\scriptscriptfont3
        \renewcommand{\arraystretch}{0.7}
        \begin{array}{l}}{\end{array}}

\newenvironment{scarray}{
          \textfont0=\scriptfont0
          \scriptfont0=\scriptscriptfont0
          \textfont1=\scriptfont1
          \scriptfont1=\scriptscriptfont1
          \textfont2=\scriptfont2
          \scriptfont2=\scriptscriptfont2
          \textfont3=\scriptfont3
          \scriptfont3=\scriptscriptfont3
        \renewcommand{\arraystretch}{0.7}
        \begin{array}{c}}{\end{array}}

%
%
\newcommand{\C}{{\mathbb C}}
\newcommand{\D}{{\mathbb D}}
\newcommand{\Z}{{\mathbb Z}}
\newcommand{\N}{{\mathbb N}}
\newcommand{\R}{{\mathbb R}}
\newcommand{\Q}{{\mathbb Q}}

%
%
\newcommand{\TT}{{\mathsf T}}
\newcommand{\HH}{{\mathsf H}}
\newcommand{\VV}{{\mathsf V}}
\newcommand{\JJ}{{\mathsf J}}
\newcommand{\PP}{{\mathsf P}}
\newcommand{\DD}{{\mathsf D}}
\newcommand{\QQ}{{\mathsf Q}}
\newcommand{\RR}{{\mathsf R}}

%
%
\newcommand{\bsigma}{{\boldsymbol{\sigma}}}
\newcommand{\vecbsigma}{{\vec{\boldsymbol{\sigma}}}}
\newcommand{\bpi}{{\boldsymbol{\pi}}}
\newcommand{\vecbpi}{{\vec{\boldsymbol{\pi}}}}
\newcommand{\btau}{{\boldsymbol{\tau}}}
\newcommand{\bphi}{{\boldsymbol{\phi}}}
\newcommand{\bvarphi}{{\boldsymbol{\varphi}}}
\newcommand{\bGamma}{{\boldsymbol{\Gamma}}}

%
%
\newcommand{\psibar}{ {\bar{\psi}} }
\newcommand{\varphibar}{ {\bar{\varphi}} }

%
%
\newcommand{\ba}{ {\bf a} }
\newcommand{\bb}{ {\bf b} }
\newcommand{\bc}{ {\bf c} }
\newcommand{\bp}{ {\bf p} }
\newcommand{\br}{ {\bf r} }
\newcommand{\bs}{ {\bf s} }
\newcommand{\bt}{ {\bf t} }
\newcommand{\bu}{ {\bf u} }
\newcommand{\bv}{ {\bf v} }
\newcommand{\bw}{ {\bf w} }
\newcommand{\bx}{ {\bf x} }
\newcommand{\by}{ {\bf y} }
\newcommand{\bz}{ {\bf z} }
\newcommand{\bone}{ {\mathbf 1} }

%
%
\newcommand{\scra}{{\mathcal{A}}}
\newcommand{\scrb}{{\mathcal{B}}}
\newcommand{\scrc}{{\mathcal{C}}}
\newcommand{\scrd}{{\mathcal{D}}}
\newcommand{\scre}{{\mathcal{E}}}
\newcommand{\scrf}{{\mathcal{F}}}
\newcommand{\scrg}{{\mathcal{G}}}
\newcommand{\scrh}{{\mathcal{H}}}
\newcommand{\scri}{{\mathcal{I}}}
\newcommand{\scrj}{{\mathcal{J}}}
\newcommand{\scrk}{{\mathcal{K}}}
\newcommand{\scrl}{{\mathcal{L}}}
\newcommand{\scrm}{{\mathcal{M}}}
\newcommand{\scrn}{{\mathcal{N}}}
\newcommand{\scro}{{\mathcal{O}}}
\newcommand{\scrp}{{\mathcal{P}}}
\newcommand{\scrq}{{\mathcal{Q}}}
\newcommand{\scrr}{{\mathcal{R}}}
\newcommand{\scrs}{{\mathcal{S}}}
\newcommand{\scrt}{{\mathcal{T}}}
\newcommand{\scru}{{\mathcal{U}}}
\newcommand{\scrv}{{\mathcal{V}}}
\newcommand{\scrw}{{\mathcal{W}}}
\newcommand{\scrx}{{\mathcal{X}}}
\newcommand{\scry}{{\mathcal{Y}}}
\newcommand{\scrz}{{\mathcal{Z}}}

%
%
\newtheorem{theorem}{Theorem}[section]
\newtheorem{definition}[theorem]{Definition}
\newtheorem{proposition}[theorem]{Proposition}
\newtheorem{lemma}[theorem]{Lemma}
\newtheorem{corollary}[theorem]{Corollary}
\newtheorem{conjecture}[theorem]{Conjecture}
\newtheorem{result}[theorem]{Result}
\newtheorem{question}[theorem]{Question}

%
%
\newcommand{\stirlingsubset}[2]{\genfrac{\{}{\}}{0pt}{}{#1}{#2}}
\newcommand{\stirlingcycle}[2]{\genfrac{[}{]}{0pt}{}{#1}{#2}}
\newcommand{\associatedstirlingsubset}[2]{\left\{\!\!%
            \stirlingsubset{#1}{#2} \!\! \right\}}
\newcommand{\associatedstirlingsubsetBis}[2]{\big\{\!\!%
            \stirlingsubset{#1}{#2} \!\! \big\}}
\newcommand{\assocstirlingsubset}[3]{{\genfrac{\{}{\}}{0pt}{}{#1}{#2}}_{\!%
            \ge #3}}
\newcommand{\assocstirlingcycle}[3]{{\genfrac{[}{]}{0pt}{}{#1}{#2}}_{\ge #3}}
\newcommand{\associatedstirlingcycle}[2]{\left[\!\!%
            \stirlingcycle{#1}{#2} \!\! \right]}
\newcommand{\associatedstirlingcycleBis}[2]{\big[\!\!%
            \stirlingcycle{#1}{#2} \!\! \big]}
\newcommand{\euler}[2]{\genfrac{\langle}{\rangle}{0pt}{}{#1}{#2}}
\newcommand{\eulergen}[3]{{\genfrac{\langle}{\rangle}{0pt}{}{#1}{#2}}_{\! #3}}
\newcommand{\eulersecond}[2]{\left\langle\!\! \euler{#1}{#2} \!\!\right\rangle}
\newcommand{\eulersecondBis}[2]{\big\langle\!\! \euler{#1}{#2} \!\!\big\rangle}
\newcommand{\eulersecondgen}[3]%
      {{\left\langle\!\! \euler{#1}{#2} \!\!\right\rangle}_{\! #3}}
\newcommand{\binomvert}[2]{\genfrac{\vert}{\vert}{0pt}{}{#1}{#2}}
\newcommand{\nueuler}[3]{{\genfrac{\langle}{\rangle}{0pt}{}{#1}{#2}}^{\! #3}}

\newcommand{\doi}[1]{\href{http://dx.doi.org/#1}{\texttt{doi:#1}}}
\newcommand{\arxiv}[1]{\href{http://arxiv.org/abs/#1}{\texttt{arXiv:#1}}}
\newcommand{\seqnum}[1]{\href{http://oeis.org/#1}{#1}}

%
%
\section{Introduction} \label{sec.intro}

The $q$-state Potts model \cite{Potts_52,Wu_82,Wu_84}
plays an important role in the theory of critical phenomena,
especially in two dimensions \cite{Baxter_book,Nienhuis_84,DiFrancesco_97},
and has applications to various condensed-matter systems \cite{Wu_82,Wu_84}.
Ferromagnetic Potts models are by now fairly well understood,
thanks to universality; and much is known about their phase diagrams 
\cite{Wu_82,Wu_84}, critical exponents, and their relation to
conformal field theories (CFT) \cite{Nienhuis_84,DiFrancesco_97}.

On the other hand, the behaviour of antiferromagnetic (AF) Potts models
depends strongly on the microscopic lattice structure. One striking 
example is the AF 3-state Potts model: 
\begin{itemize}
\item On the honeycomb lattice, it is disordered at all temperatures $T\ge 0$ 
\cite{MC_hc}. 
\item On the square lattice, it has a critical point at $T=0$ (i.e., 
at $T=0$, the model has algebraic decay of correlations), and it is
disordered at any positive temperature $T>0$ 
\cite{Nijs_82,Kolafa_84,Burton_Henley_97,Salas_98}. 
\item On the diced lattice, it displays an ordinary finite-temperature 
critical point \cite{Kotecky-Salas-Sokal}. 
\item On the triangular lattice, it shows a finite-temperature weak 
first-order phase transition \cite{Adler_95}. 
\end{itemize}
Therefore, many basic questions about the phase diagram of this AF model 
must be investigated on a case-by-case basis. 
The absence of universality is probably the main cause why our understanding 
of Potts antiferromagnets is less advanced than that of their ferromagnetic 
(FM) counterparts. 
However, for the last decades, many new numerical and theoretical works 
have endeavoured to understand the AF regime. As a 
matter of fact, some sort of universality has been recovered in this regime.

In particular, one expects that for each lattice $\mathcal{L}$ there
exists a value $q_c(\mathcal{L})$ [possibly noninteger; 
see section~\ref{sec.potts}] such that for $q > q_c(\mathcal{L})$, 
the model has exponential decay of correlations at all temperatures 
including zero, while for $q = q_c(\mathcal{L})$ the model has a 
zero-temperature critical point. Finally, for $q < q_c(\mathcal{L})$ any 
behaviour is possible: often (though not always) the model has a phase 
transition at nonzero temperature, which may be of either first or second order
(e.g., the 3-state model on the triangular \cite{Adler_95},
or on the diced \cite{Kotecky-Salas-Sokal} lattices, respectively).  
In some other cases, the system has a critical point at $T=0$ and is 
disordered at any positive temperature, like the AF Ising model ($q=2$) on 
the triangular lattice, and in other cases, it is disordered at all 
temperatures $T\ge 0$, like the AF Ising model on the kagome lattice 
\cite{Suto_81a,Suto_81b}. The first task, for any lattice $\mathcal{L}$, 
is thus to determine the quantity $q_c(\mathcal{L})$.

Some AF models at $T=0$ have the remarkable property that they 
can be mapped exactly onto a `height' model (in general vector-valued) 
\cite{Henley_93,Kondev_96,Salas_98,Jacobsen_09}.
Since the height model must either be in a `smooth' (ordered)
or `rough' (massless) phase, the corresponding zero-temperature spin model 
must either be ordered or critical, never disordered. Moreover,  
Henley \cite{Henley_93} conjectured that all models that do \emph{not admit} 
a height representation are disordered at any temperature $T\ge 0$. 
When the height model is critical, the long-distance behaviour is that of 
a massless Gaussian model with some (\emph{a~priori} unknown) 
`stiffness tensor' ${\bf K} > 0$. The critical operators can be identified 
via the height mapping, and the corresponding critical exponents can be 
predicted in terms of the single parameter ${\bf K}$. Height representations 
thus provide a means for recovering a sort of universality for some 
(but not all) AF models, and for understanding their critical behaviour in 
terms of CFT \cite{DiFrancesco_97}.
All the nonuniversal details of the microscopic lattice structure
are encoded in the height representation and in the stiffness tensor ${\bf K}$.
Given these, everything can be understood in terms of
the universal behaviour of massless Gaussian fields.

Showing that a given model admits a height representation implies that this 
is an interesting model, as it has either a zero-temperature critical point,
or long-range order at $T=0$, implying that there is a finite-temperature 
critical point (belonging to some \emph{a priori} unknown
universality class). From now on, we will focus on two-dimensional (2D) 
$q$-state Potts AF models; but there are more models in the literature that 
admit height representations: the triangular-lattice spin-$s$ Ising 
antiferromagnet \cite{Zeng_Henley_97} (which is in the smooth phase for 
$s$ large enough),
the 4-state Potts model on the line graph of the square lattice
(which is exactly at the roughening transition, but this lattice is not planar) 
\cite{Kondev_95,Kondev_96}, 
a constrained 4-state AF Potts model on the square lattice 
\cite{Burton_Henley_97} (which also is in the smooth phase), 
a special six-vertex model \cite{Kondev_96}, several dimer 
\cite{Levitov_90,Thurston_90,Raghavan_97,Henley_97,Alet_05,Alet_06}
and other tiling models \cite{Kenyon_00,Ghosh_07,Korn_Pak_04,Jacobsen_07},  
and fully-packed loop models 
\cite{Bloete_Nienhuis_84,Kondev_94,Kondev_94b,Kondev_96,Jacobsen_Kondev_98} 
(see also \cite[and references therein]{Jacobsen_09}).

Prior to 2009, only a few 2D AF Potts models 
(i.e., with isotropic nearest-neighbour interactions and 
zero magnetic field) had been studied in detail. 
The four known models in this class admitting a height representation at
$T=0$, also displayed a zero-temperature critical point. These models were 
$q=2$ on the triangular lattice \cite{Stephenson_64,Bloete_Hilhorst_82}, 
$q=3$ on the square \cite{Nijs_82,Kolafa_84,Burton_Henley_97,Salas_98}
and kagome \cite{Huse_92,Kondev_96} lattices, and $q=4$ on the 
triangular lattice \cite{Henley_93,Moore_00}. 
Therefore, by that year, the existence of a zero-temperature critical point 
could be considered a rather uncommon phenomenon, and no 2D AF Potts model 
with a height representation and long-range order at $T=0$ was known in the 
literature.  

However, this scenario was challenged in \cite{Kotecky-Salas-Sokal}. The
authors noted that the very same height representation that was found for 
the 3-state Potts antiferromagnet on the square lattice, carries over 
unchanged to any plane quadrangulation. (A plane quadrangulation is a plane 
graph in which all faces are quadrilaterals or 4-cycles.) Therefore, it was 
natural to conjecture that $q_c=3$ for any plane quadrangulation. 
This conjecture was tested with the second simplest plane quadrangulation:
the diced lattice (i.e., the Laves lattice $[3,6,3,6]$ 
---using the notation of \cite{tilings}; see figure~\ref{fig_diced}(b)). 
Contrary to the previous expectations, it was rigorously 
proven that this model has long-range order at $T=0$. Furthermore, 
using high-precision Monte Carlo (MC) simulations, a finite-temperature 
phase transition was found. The numerical data strongly supports that this 
critical point belongs to the universality class of the \emph{FM} 
3-state Potts model. In summary, not only was the naive conjecture mentioned 
above false, but this work provided also the first example of an  
2D AF Potts model with both a height representation and long-range order 
at $T=0$. In particular, all this implies that $q_c(\text{diced})>3$ 
(numerical estimates from transfer matrices (TM) yield
$q_c(\text{diced}) \approx 3.45$ \cite{Jacobsen-Salas_unpub}).

In a similar way, Moore and Newman \cite{Moore_00} observed that the
same height representation found for the 4-state Potts AF on the triangular
lattice can also be applied without any modification to any Eulerian plane 
triangulation. (A triangulation is a plane graph in which all faces are 
triangles; it is Eulerian if moreover all the vertices have even degree.) 
Again, the natural conjecture is that $q_c=4$ for any Eulerian plane 
triangulation. And again, this conjecture turned out to be false: on one side, 
there are strong analytic arguments showing that on any Eulerian plane 
triangulation in which one sublattice consists entirely of vertices of 
degree 4, the 4-state Potts antiferromagnet has a finite-temperature phase
transition \cite{union-jack}, so that $q_c>4$ for this class of triangulations.
In addition, the universality class of this critical point could be predicted 
from the lattice structure. These findings were 
tested on two simple Eulerian plane triangulations: the union-jack and
the bisected hexagonal lattices  (i.e., the Laves $[4,8^2]$ and 
$[4,6,12]$ lattices, respectively; see \cite{tilings},
\cite[figure~2]{union-jack}). In both cases, MC simulations and TM 
computations found finite-temperature critical points belonging to the 
predicted universality classes \cite{union-jack}. In particular, 
$q_c(\text{union-jack})\approx 4.33$, and 
$q_c(\text{bisected-hexagonal})\approx 5.40$. This latter result was rather
surprising, as it implied not only a finite-temperature critical point
for $q=4$, but also for $q=5$. Again, examples of 2D AF $q$-state 
Potts models were found having a height representation and displaying an ordered phase
at low temperature.  

As a matter of fact, one can devise plane quadrangulations (e.g., 
$G'_n$ and $G''_n$) and plane triangulations (e.g., $H'''_n$) such that 
$q_c$ takes arbitrarily large values as $n$ increases 
\cite[figure~2, and tables~I and~II]{planar_AF_largeq}. Therefore, there
are infinitely many lattices for which there exists a height representation
(for $q=3$ or $q=4$ depending whether the lattice is a plane quadrangulation
or triangulation, respectively) with an ordered low-temperature phase.  
All these results imply a \emph{qualitative} change in the general picture 
about phase transitions for 2D AF Potts models: the class of these models with
a height representation and long-range order at $T=0$ contains actually 
\emph{infinitely many} elements. 

A natural question is to know whether the class of 2D AF Potts models with
a height representation and displaying a critical point at $T=0$ is finite 
or not. A good stating point is to investigate if there exists any 
useful condition on the class of plane quadrangulations telling us whether 
the corresponding 3-state AF Potts model at $T=0$ is critical or ordered. 
(One could have considered the same question for plane Eulerian triangulations 
and the 4-state AF Potts model. To our knowledge, this is still an open 
problem.) 

In this paper we shall propose a new criterion, involving graph duality, 
that appears to distinguish precisely whether the zero-temperature 
3-state AF Potts model on a quadrangulation is critical or ordered. 
(A summary of this work has appeared previously in \cite{Lv_17}.) 
This criterion depends on whether the quadrangulation belongs to one of two 
possible classes. In particular, we will show in section~\ref{sec.quad}, 
that there is a one-to-one correspondence between plane quadrangulations 
$\Gamma$ and pairs of dual plane graphs $(G,G^*)$. 
Then the class of planar quadrangulations can be split into two disjoint 
subclasses: those of self-dual type, if $G$ is self-dual (i.e., $G\simeq G^*$);
and those of non-self-dual type, otherwise 
(see definition~\ref{def.self_dual_Q}).  

Using arguments based on the existence of a height representation for these
models, we have arrived at the main result of this paper, that can be 
summarised as follows: 
 
\begin{conjecture} \label{conj.main}
For the three-state antiferromagnetic Potts model on a (periodic)
plane quadrangulation $\Gamma$, 
\begin{enumerate}
\item[(a)] If $\Gamma$ is a quadrangulation of self-dual type, then this model 
      has a zero-temperature critical point, so that $q_c(\Gamma) = 3$. The
      model is described by a CFT of central charge $c=1$.

\item[(b)] 
      If $\Gamma$ is a quadrangulation of non-self-dual type, this model has 
      long-range order at low temperature, and displays a
      finite-temperature phase transition, so that $q_c(\Gamma) > 3$. This 
      transition is second-order and lies in the universality class
      of the 3-state Potts ferromagnet, which is described by a CFT 
      of central charge $4/5$.
\end{enumerate}
\end{conjecture}

We have tested this conjecture on four (resp.\/ seven) quadrangulations of 
self-dual (resp.\/ non-self-dual) type using several complementary 
techniques: MC simulations, TM computations, and the method of 
critical polynomials (CP) \cite{JS_12a,JS_12b,JS_13,Jesper_14,JS_16}. 
These numerical results agree, \emph{without exception}, with 
conjecture~\ref{conj.main}. 
As a side result, we have computed some critical exponent for the 
3-state AF Potts model on several quadrangulations of self-dual type not 
considered before in the literature.

It is worth noticing that there are infinitely many self-dual periodic 
planar lattices \cite{Ashley_91,Okeeffe_92,Okeeffe_96,Servatius_98,Wierman_06,%
Scullard_06,Ziff_12}. Thus, conjecture~\ref{conj.main}(a) implies that 
there are also infinitely many 3-state AF Potts models displaying 
a zero-temperature critical point. 
This observation entails another \emph{qualitative} change in the general 
picture about phase transitions in this kind of models: the class of 2D AF 
Potts models with a height representation and a zero-temperature critical 
point also contains \emph{infinitely} many elements.
Finally, there is a close connection between our results and
those by Delfino and Tartaglia \cite{Delfino_17}. They have obtained 
renormalisation-group (RG) fixed points of CFT invariant under the symmetric 
group $S_q$. The Potts models in conjecture~\ref{conj.main}(a) provide a 
lattice realisation of their solution~I (see \cite[table~I]{Delfino_17}).
 
On the other hand, there are also infinitely many quadrangulations of 
non-self-dual type (see e.g., the families $G'_n$ and $G''_n$ of 
\cite{planar_AF_largeq}). Thus, conjecture~\ref{conj.main}(b) implies 
that there are also infinitely many 3-state AF Potts models displaying 
long-range order at $T=0$ and undergoing a finite-temperature 
phase transition. Although this transition has been found in all the 
cases considered here to be of second order, we cannot rule out the
possibility that in some cases it might be of first order 
(see section~\ref{sec.disc}). 

In summary, we have recovered some sort of universality for the 3-state AF 
Potts model on the \emph{whole} class of (periodic) plane quadrangulations.
The phase diagram of this model will depend only on whether the quadrangulation 
$\Gamma$ is of self-dual type or not. 

As a side result, we have found strong empirical evidence that the 
Wang--Swendsen--Koteck\'y (WSK) algorithm \cite{WSK_89,WSK_90} 
for the zero-temperature 3-state AF Potts model on quadrangulations 
of self-dual type has no critical slowing down (CSD), even though this is a
critical point for all these models. In this case, the MC simulations
were carried over quadrangulations of this type with toroidal boundary 
conditions with sizes such that the resulting graphs were bipartite. This
condition is essential for the WSK Markov-chain to be irreducible 
(or ergodic) \cite{Burton_Henley_97,Sokal_Ferreira,Mohar}. 
The absence of CDS for this algorithm was first observed on the square lattice 
\cite{Sokal_Ferreira,Salas_98}. 

On the other hand, for the 3-state Potts AF model on five quadrangulations 
of non-self-dual type, we have found that the WSK algorithm does present
CSD. Furthermore, the estimated dynamic critical exponents 
$z_{\text{int},\mathcal{M}_\text{stagg}^2}$ and 
$z_{\text{int},\mathcal{M}_\text{u}^2}$ agree well within errors with the 
corresponding dynamic critical exponent 
$z_{\text{int},\mathcal{M}^2}$ for the Swendsen--Wang (SW) algorithm 
\cite{SW_87} for the FM 3-state Potts model on the square 
lattice \cite{Salas_Sokal_97,Garoni_11}. 

Therefore, we propose the following:

\begin{conjecture} \label{conj.wsk}
Consider the Wang--Swendsen--Koteck\'y algorithm for the three-state 
antiferromagnetic Potts model on a (periodic) bipartite quadrangulation 
$\Gamma$ embedded in a torus. Then: 
\begin{enumerate}
\item[(a)] If $\Gamma$ is a quadrangulation of self-dual type, this 
      algorithm has no critical slowing down at the zero-temperature 
      critical point. 

\item[(b)]
      If $\Gamma$ is a quadrangulation of non-self-dual type and 
      the finite-temperature phase-transition point of this model is of 
      second order, this algorithm has critical slowing down at the critical 
      temperature. This algorithm belongs to the same dynamic universality
      class as the Swendsen--Wang algorithm for the 3-state ferromagnetic 
      Potts model.  
\end{enumerate}
\end{conjecture}

It is worth noticing that conjecture~\ref{conj.wsk}(a) provides 
\emph{infinitely} many critical models for which the WSK algorithm has no CSD. 
Before this work (and \cite{Lv_17}), the square-lattice case was the only 
known example with this uncommon (but very desirable) dynamic property. 
Finally, conjecture~\ref{conj.wsk}(b) seems `natural' when compared to 
conjecture~\ref{conj.main}(b). 

This paper is organised as follows. In section~\ref{sec.setup} we provide the
necessary background we need to make this paper as self-contained as possible. 
Our numerical results are described in section~\ref{sec.results}. Finally,
we discuss the physical interpretation of our numerical findings in 
section~\ref{sec.disc}. In appendix~\ref{sec.app}, we show how
to compute the geometric factor for two of the above quadrangulations. 

%
%
\section{Basic setup} \label{sec.setup}

This section is devoted to explaining the main definitions we will use 
afterwards.
In particular, in section~\ref{sec.potts}, we will review basic facts about 
the $q$-state Potts model. Section~\ref{sec.quad} will deal with the
class of lattices we are interested in; namely, plane quadrangulations.
Finally, in section~\ref{sec.height_rep}, we will work out the height 
representation for the zero-temperature 3-state Potts antiferromagnet on 
any plane quadrangulation $\Gamma$. This representation will
lead us to make some predictions about the critical nature of these Potts
models based solely on whether $\Gamma$ is of self-dual type or not.

%
%
\subsection{The Potts model} \label{sec.potts}

The $q$-state Potts model \cite{Potts_52,Wu_82,Wu_84} can be defined on any 
undirected graph $G=(V,E)$ with vertex set $V$ and edge set $E$. On each 
vertex of the graph $x\in V$, we place a spin 
$\sigma_x \in \Omega = \{0,1,\ldots,q-1\}$,
where $q\ge 2$ is a positive integer. Each spin has $q$ possible values or 
`colours', and interacts with those spins located on the neighbouring 
vertices with some coupling constant $J\in\R$. 
The partition function of this model is written as
\begin{equation}
Z_G(q,J) \;=\; \sum\limits_{\sigma\colon V\to \Omega} 
    \exp \left( J \sum\limits_{ \{x,y\}\in E} \delta_{\sigma_x,\sigma_y}
         \right)  \,,
\label{def.Z.potts}
\end{equation}
where the first sum is over all possible colourings of the spins of the system,
and the second one (inside the exponential) is over all edges of the graph,
with $\delta_{\sigma_x,\sigma_y}$ being the Kronecker delta. The coupling 
constant is proportional to the inverse of the temperature $T$, and its sign
defines the regime of the model: if $J>0$ (resp.\/ $J<0$) the model
is in the FM (resp.\/ AF) regime.  

Even though we will deal with $q=3$ in this paper, it is useful to
rewrite \eqref{def.Z.potts} in the Fortuin--Kasteleyn (FK) representation
\cite{FK_69,FK_72}:
\begin{equation}
Z_G(q,v) \;=\; \sum\limits_{F \subseteq E} v^{|F|}\, q^{k(F)} \,, 
\label{def.Z.FK}
\end{equation}
where the sum is over all spanning subgraphs $(V,F)$ of $G=(V,E)$, $k(F)$ is
the number of connected components of $(V,F)$, and  
the temperature-like parameter $v$ is defined as
\begin{equation}
v \;=\; e^{J} - 1 \,.
\end{equation}
Therefore, in the FM regime $v\ge 0$; in the AF one, $v\in[-1,0)$, and the 
model is unphysical for $v< -1$ (complex $J$). This variable $v$ will be used 
in the plots shown in section~\ref{sec.results}. One important 
consequence of \eqref{def.Z.FK} is that $Z_G$ is a polynomial in both $q$ 
and $v$, therefore we can promote both variables outside their original 
physical range. In particular, $q,v$ can be any real, or even complex 
numbers. This representation gives \emph{some} sense to the $q$-state Potts
model for noninteger values of $q$: for instance, for the hexagonal lattice,
we expect that $q_c(\text{hexagonal})=(3+\sqrt{5})/2$ 
\cite[and references therein]{Salas_Sokal_97a}.

In Statistical Mechanics one is mainly interested in the thermodynamic limit,
in which $G$ is a graph that `tends to infinity' in a suitable way. In 
order to achieve this, we first define the \emph{free energy per vertex} 
for a finite graph:
\begin{equation}
f_G(q,v) \;=\; \frac{1}{|V|} \, \log Z_G(q,v) \,.
\label{def.f_G}
\end{equation}
We then define a suitable sequence of graphs $(G_n)$ with $G_n=(V_n,E_n)$ 
which is usually a subset of an infinite periodic lattice $\mathcal{G}$ 
with some boundary conditions (free, cylindrical, toroidal, etc). 
The thermodynamic limit of the free energy is defined as the limit
\begin{equation}
f(\mathcal{G};q,v) \;=\; \lim_{n\to\infty} \frac{1}{|V_n|} \, 
                         \log Z_{G_n}(q,v) \,,
\label{def.f}
\end{equation}
if that limit exists. It is well known that for $q\in\N$ and $v>-1$ [using the
spin representation \eqref{def.Z.potts}] or for $q>0,v\ge 0$ [using the 
FK representation \eqref{def.Z.FK}] this limit exists, and it is a continuous 
function of $v$; but it might fail to be analytic at some points (where phase
transitions occur). For other values of the parameters $q,v$ the very existence 
and/or uniqueness of the above limit is a difficult question. 

%
%
\subsection{Plane quadrangulations} \label{sec.quad}

In this section we will describe how to build a general plane 
quadrangulation. This construction was already briefly outlined in 
\cite{union-jack}.

For any (finite or infinite) graph $G=(V,E)$ embedded in the plane,
the dual graph $G^* = (V^*,E^*)$ is defined by placing a vertex in each 
face of $G$ and drawing an edge $e^*$ across each edge $e$ of $G$.
Since $G^{**} = G$, we refer to the pair $(G,G^*)$ as a \emph{dual pair}.
A graph $G$ is called \emph{self-dual} if $G$ is isomorphic to $G^*$.

%
%
\begin{figure}[htb]
\begin{center}
\begin{tabular}{cc}
\includegraphics[width=200pt]{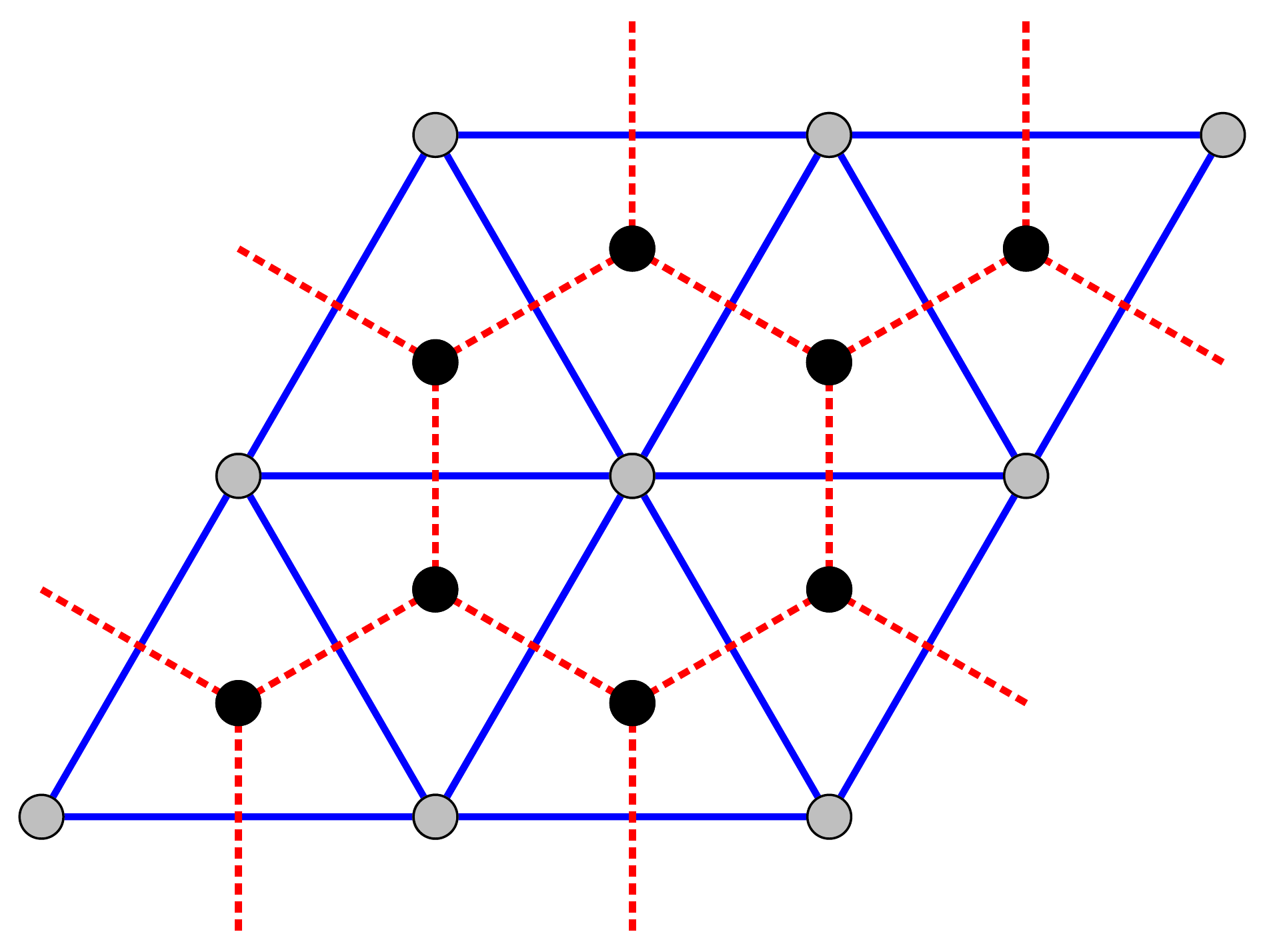} &
\includegraphics[width=200pt]{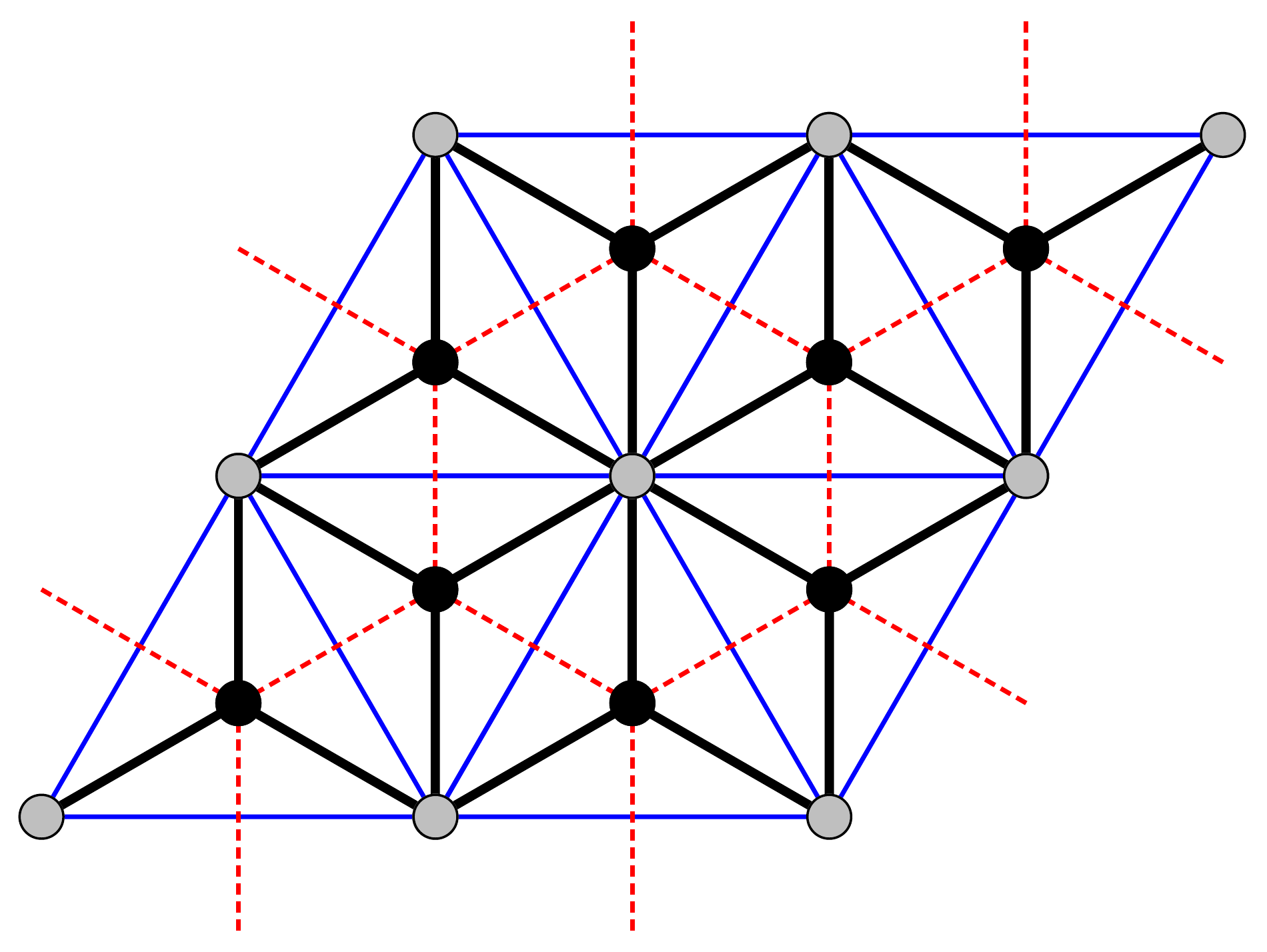} \\[2mm]
(a) & (b) 
\end{tabular}
\end{center}
\vspace*{-6mm}
\caption{\label{fig_diced}
   Building the diced lattice from the dual pair $(G_0,G_1)=$
   (triangular, hexagonal).
   (a) We show a triangular lattice $G_0=(V_0,E_0)$ with gray vertices 
       and solid (blue) edges. We also depict its dual hexagonal 
       lattice $G_1=(V_1,E_1) = G_0^*$ with black vertices and dashed 
       (red) edges. 
   (b) We show the corresponding quadrangulation $\Gamma = 
       \mathcal{Q}(G_0)=$ diced lattice, with vertex set $V_0 \cup V_1$
       (depicted as black and gray vertices), and edge set depicted as 
       solid thick black lines. [The edges of the underlying triangular
       and hexagonal lattices are also depicted as in panel~(a). They are the diagonals of the quadrangles.] 
}
\end{figure}

Let us consider a connected plane graph $G=(V,E)$ and its dual graph 
$G^* =(V^*,E^*)$. See figure~\ref{fig_diced}(a) for $G=$ triangular 
lattice and $G^*=$ hexagonal lattice (see \cite[figure~1]{union-jack} for 
a more general example). We then define the new graph 
$\Gamma = (V\cup V^*,E')$, such that its vertex set is $V\cup V^*$, and its
edge set $E'$ contains all edges $e=\{x,y\}\in E'$, whenever 
$x\in V$ belongs to the boundary of the face of $G$ that contains the dual 
vertex $y\in V^*$ (see figure~\ref{fig_diced}(b) for $\Gamma=$ diced 
lattice). This new graph $\Gamma$ is a plane quadrangulation: each face of 
$\Gamma$ contains exactly one pair of diametrically opposite vertices of $V$, 
corresponding to an edge $e\in E$; and another pair of diametrically 
opposite vertices of $V^*$, which correspond to the dual edge $e^*\in E^*$. 
As a matter of fact, $\Gamma$ is the dual of the medial graph 
$\mathcal{M}(G) = \mathcal{M}(G^*)$. (In particular, the diced lattice of 
figure~\ref{fig_diced}(b) is the dual of the kagome lattice, which is the
medial graph of either the triangular and the hexagonal lattice.) 

Indeed, there is a one-to-one correspondence between plane 
quadrangulations $\Gamma$ and dual pairs of plane graphs $(G_0,G_1)$. 
If we consider a connected plane quadrangulation $\Gamma = (V,E)$, 
then $\Gamma$ is bipartite, and therefore $V = V_0 \cup V_1$. We can define 
two sublattices $G_0 = (V_0,E_0)$ and $G_1 = (V_1,E_1)$ by 
drawing edges across the diagonals of the quadrilateral faces, as in
figure~\ref{fig_diced}(b). Those edges connecting vertices of $V_0$ belong to 
$E_0$, and conversely, those connecting vertices of $V_1$ belong to $E_1$.
Finally, it is easy to see that $G_0$ and $G_1$ form a dual pair. 

Finally, as $G_0$ and $G_1$ play the same role in the above constructions,
both pairs $(G_0,G_1)$ and $(G_1,G_0)$ lead to the same quadrangulation 
$\Gamma$. For simplicity, we will denote the quadrangulation arising 
from the dual pair $(G_0,G_1)$ as $\mathcal{Q}(G_0)=\mathcal{Q}(G_1)$. 
Let us formalise these ideas in the following: 

\begin{definition} \label{def.self_dual_Q}
The plane quadrangulation $\Gamma = \mathcal{Q}(G_0)$ associated to the
dual pair ($G_0,G_1)$ is of \textbf{self-dual type}
if $G_0$ is self-dual, and of \textbf{non-self-dual type} otherwise.
\end{definition} 

For instance, the square lattice [$=\mathcal{Q}(\text{square})$] is a 
quadrangulation of self-dual type (both $G_0$ and $G_1$ are themselves 
square lattices), while the diced lattice [$=\mathcal{Q}(\text{triangular})$] 
is a quadrangulation of non-self-dual type (the sublattices are the triangular 
and hexagonal lattices). 

\begin{figure}[htb]
\begin{center}
\begin{tabular}{ccc}
\includegraphics[width=130pt]{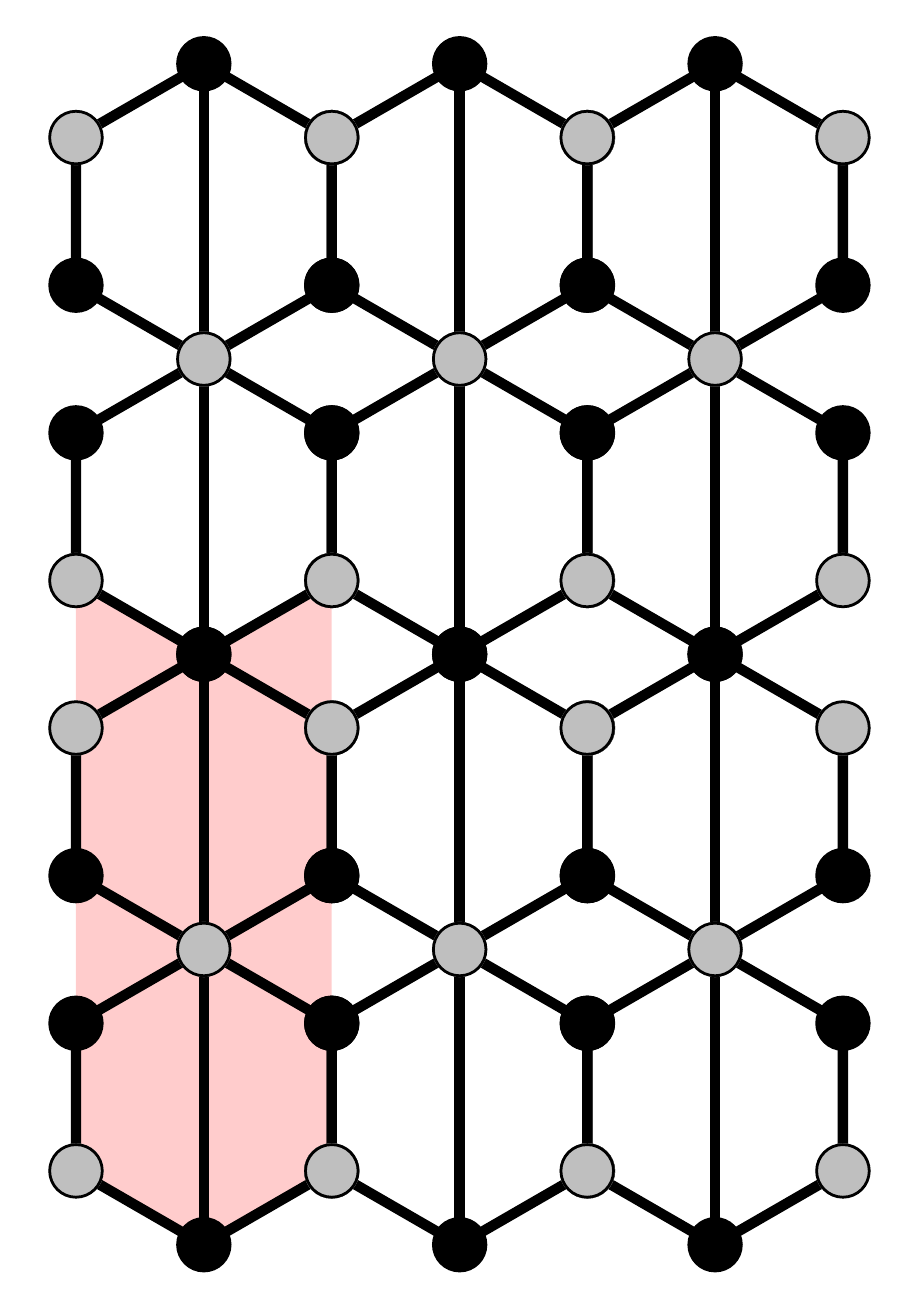} &
\includegraphics[width=140pt]{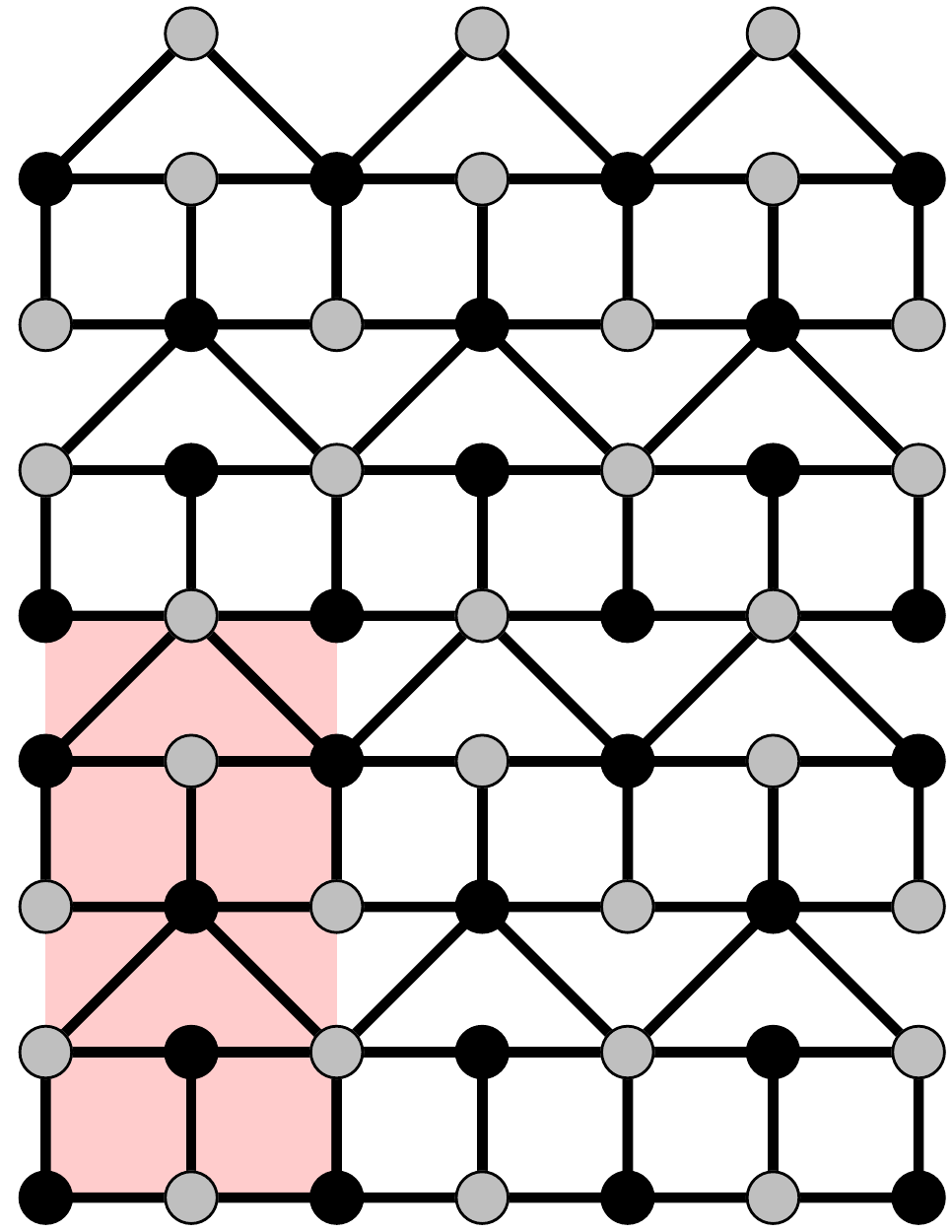}  &  
\includegraphics[width=140pt]{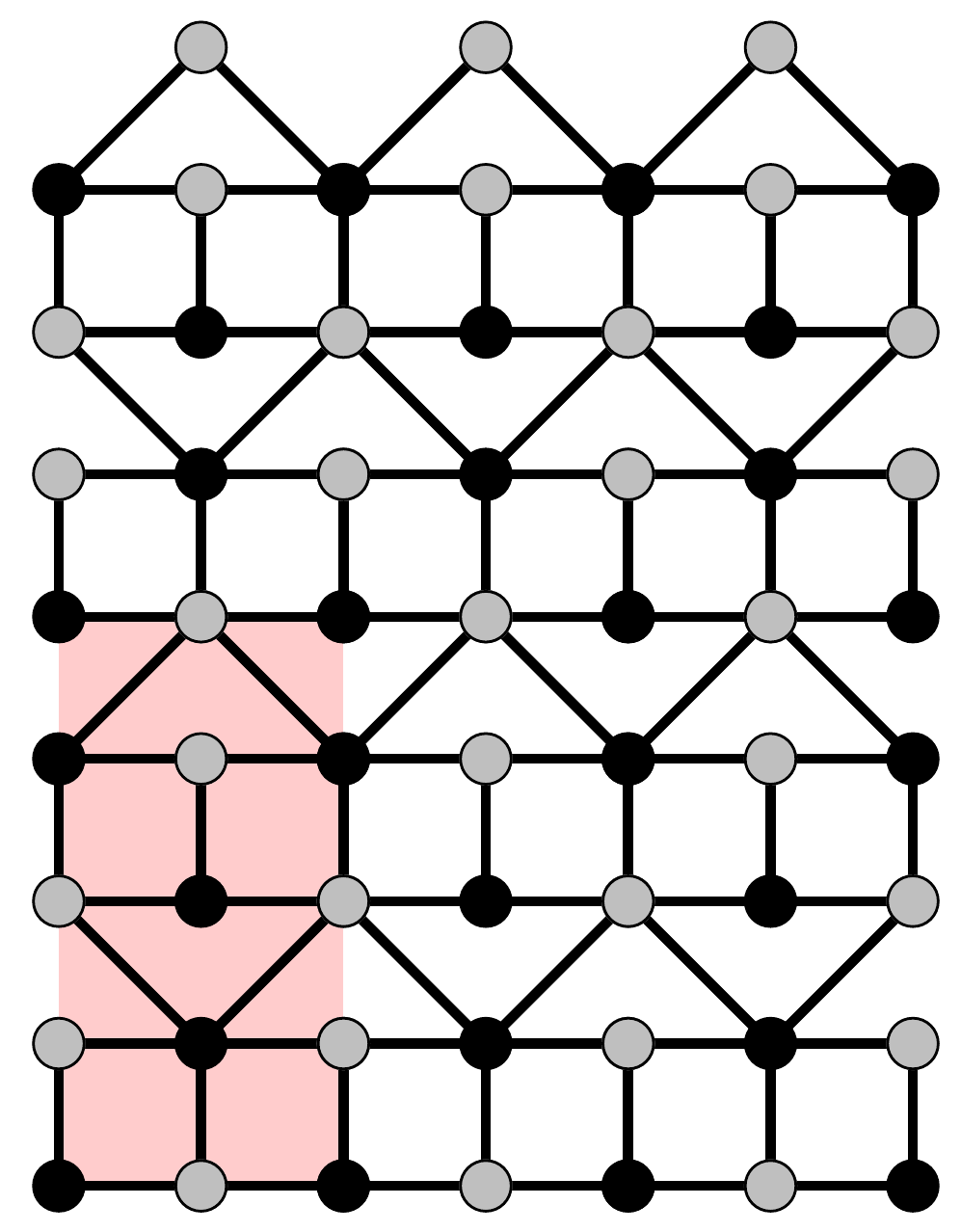} \\[1mm] 
(a) & (b)  & (c) \\[4mm]
\multicolumn{3}{c}{%
\includegraphics[width=300pt]{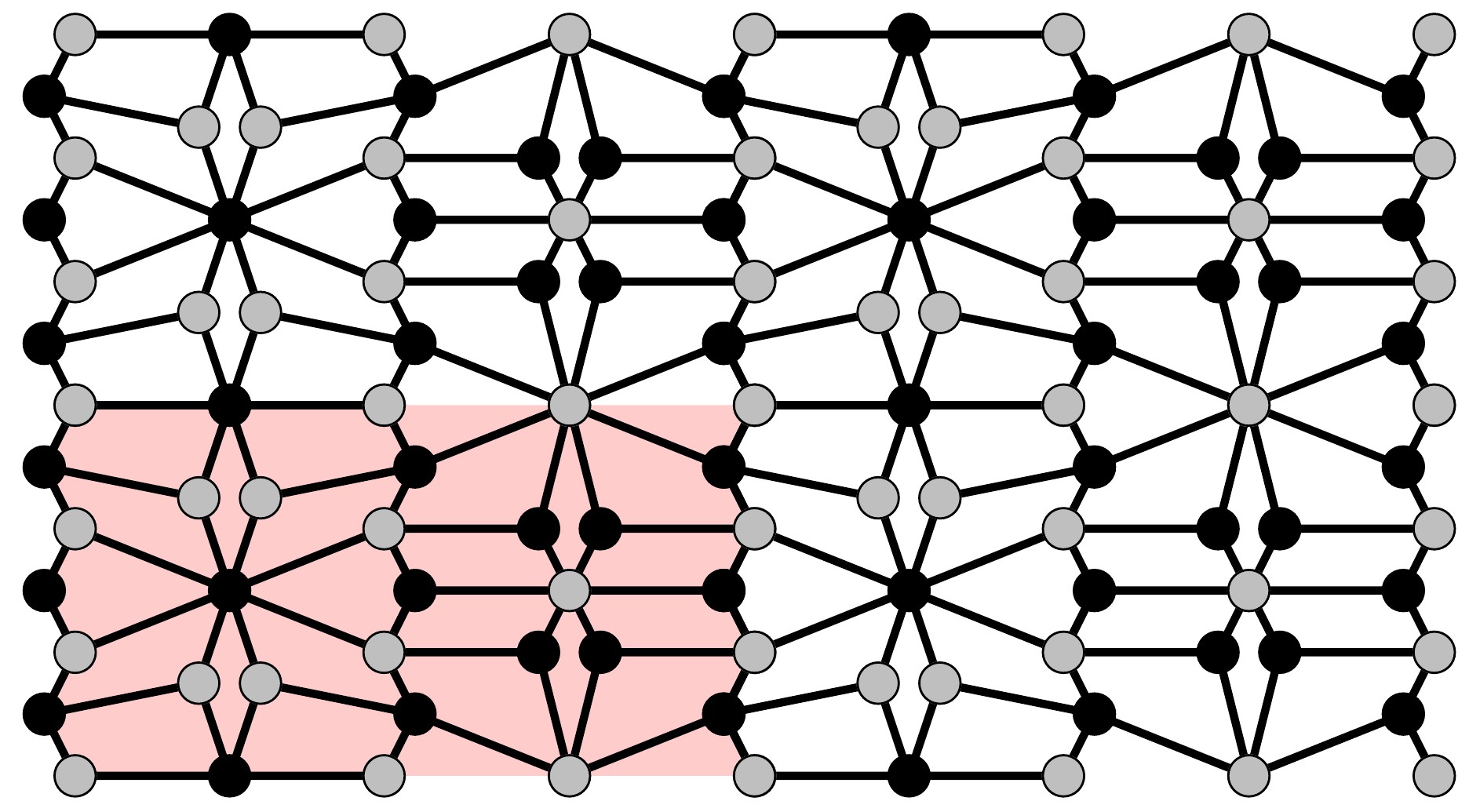}} \\[1mm]
\multicolumn{3}{c}{\;\;\;(d)} \\[1mm]
\end{tabular}
\end{center}
\vspace*{-6mm}
\caption{\label{fig_Q_selfdual}
   The four plane quadrangulations of self-dual type considered in this paper: 
   (a) $\mathcal{Q}(\text{hextri})$, 
   (b) $\mathcal{Q}(\text{house})$, 
   (c) $\mathcal{Q}(\text{martini-B})$, and 
   (d) $\mathcal{Q}(\text{cmm-pmm})$. The pink area in each panel  
   shows the corresponding unit cell. We have depicted finite pieces
   of such lattices with free boundary conditions and sizes $3\times 2$ (unit 
   cells), except for (d) which has size $2\times 2$ (see text). 
}
\end{figure}

In this paper, we will restrict attention to periodic plane lattices.
It is well known (and obvious) that the square lattice is self-dual;
what seems to be less well known is that there exist \emph{infinitely many}
self-dual periodic plane lattices
\cite{Ashley_91,Okeeffe_92,Okeeffe_96,Servatius_98,Wierman_06,Scullard_06,%
Ziff_12},
including the `hextri' \cite[figures~1 and~10]{Okeeffe_92}%
\cite[figure~16]{Servatius_98} \cite[figure~1b]{Wierman_06}, 
the `house' \cite[figure~2]{Okeeffe_92}, the martini-B 
\cite[figure~8]{Scullard_06}, and the `cmm-pmm' 
\cite[figure~29]{Servatius_98} lattices. 
In particular, from each of these lattices we can construct the corresponding
quadrangulation of self-dual type. Those are depicted in 
figure~\ref{fig_Q_selfdual}. 
This figure shows that all these quadrangulation of self-dual type
$\Gamma$ can be regarded as a square Bravais lattice $\mathcal{B}_\Gamma$ 
with a basis formed by $n_\Gamma$ vertices. In particular,
$n_{\mathcal{Q}(\text{hextri})}=6$,
$n_{\mathcal{Q}(\text{house})}=n_{\mathcal{Q}(\text{martini-B})}=8$, and 
$n_{\mathcal{Q}(\text{cmm-pmm})}=24$. We say that a quadrangulation $\Gamma$ 
has size $L\times N$ when the underlying square Bravais lattice 
$\mathcal{B}_\Gamma$ has size $L\times N$.

In addition, we will consider seven quadrangulations of non-self-dual type: 
\begin{itemize}
\item $\mathcal{Q}(\text{diced})$ = dual ruby. This quadrangulation is the 
      Laves lattice $[3,4,6,4]$; see \cite[figure~2.7.1]{tilings}. 
\item $\mathcal{Q}(\text{martini})$, where the martini lattice is 
      depicted in \cite[figure~1]{Scullard_06}. This is a 2-homeohedral 
      tiling of valence $3$ in the notation of ref.~\cite[p.~186]{tilings}: 
      this is a cubic graph with vertices of type $(3,9^2)$ and $(9^3)$ in the 
      notation of ref.~\cite{tilings}. 
\item $\mathcal{Q}(\text{ruby})$, where the ruby lattice is the 
      Archimedean lattice $(3,4,6,4)$ \cite[figure~2.1.5]{tilings}.
\item $\mathcal{Q}(\text{asanoha})$, where the asanoha lattice is the 
      Laves lattice $[3,12^2]$; see \cite[figure~2.7.1]{tilings}. 
\end{itemize}

%
%

\begin{itemize}
\item $\mathcal{Q}(\text{cross})$, where the cross lattice is the 
      Archimedean lattice $(4,6,12)$. The dual cross lattice is also called
      the bisected hexagonal lattice \cite[figure~2(b)]{union-jack} and 
      corresponds to the Laves lattice $[4,6,12]$; see 
      \cite[figure~2.7.1]{tilings}.
\item $G''_2 = \mathcal{Q}(\text{union-jack})$ \cite[figure~2(a)]{union-jack},
      where the union-jack lattice is the Laves lattice $[4,8^2]$ 
      (see \cite[figure~2.7.1]{tilings}), and the dual of the 
      four-eight lattice.
\item $G''_3 = \mathcal{Q}(\text{decorated four-eight})$ 
      \cite[figure~2, central panel]{planar_AF_largeq}. The four-eight 
      lattice is the Archimedean lattice $(4,8^2)$ \cite[figure~2.1.5]{tilings}.
      The decorated four-eight lattice is obtained from the former one  
      by subdividing once those edges shared by two octogonal faces.
      Its dual is the lattice $G'_2$ 
      \cite[figure~2, left panel]{planar_AF_largeq}.
\end{itemize}

%
%
\begin{figure}[htb]
\begin{center}
\begin{tabular}{cc}
\includegraphics[width=200pt]{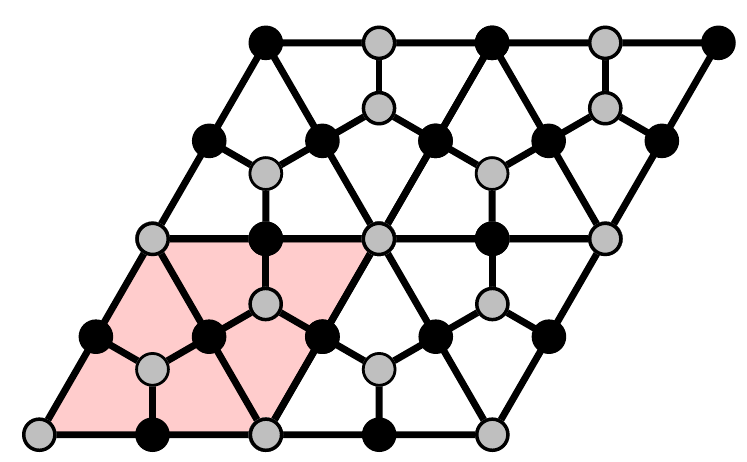} & 
\includegraphics[width=200pt]{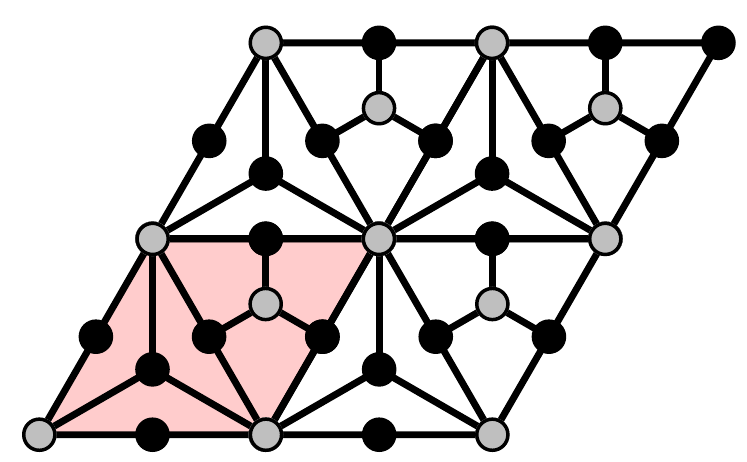} \\[2mm] 
\hspace*{-2cm}(a) &\hspace*{-2cm} (b) \\[2mm] 
\multicolumn{2}{c}{%
\includegraphics[width=200pt]{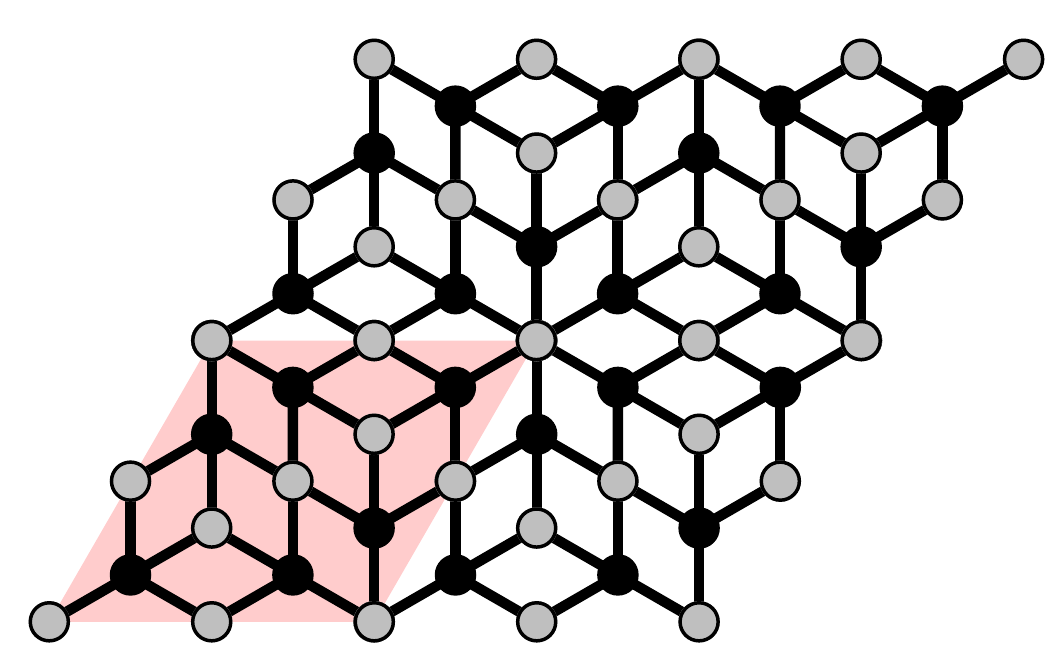}} \\[2mm] 
\multicolumn{2}{c}{\hspace*{-2cm}(c)}\\[2mm] 
\end{tabular}
\end{center}
\vspace*{-5mm}
\caption{\label{fig_Q_non-selfdual}
   Three quadrangulations of non-self-dual type considered in this paper with 
   an underlying triangular Bravais lattice: 
   (a) $\mathcal{Q}(\text{diced}) =$ dual ruby, 
   (b) $\mathcal{Q}(\text{martini})$, and  
   (c) $\mathcal{Q}(\text{ruby})$. 
   The pink areas show the corresponding unit cells. We have depicted finite 
   pieces of such lattices with free boundary conditions and sizes $2\times 2$
   (unit cells).
}
\end{figure}

%
%
\begin{figure}[htb]
\begin{center}
\begin{tabular}{cc}
\hspace*{-0.5cm} 
\includegraphics[width=230pt]{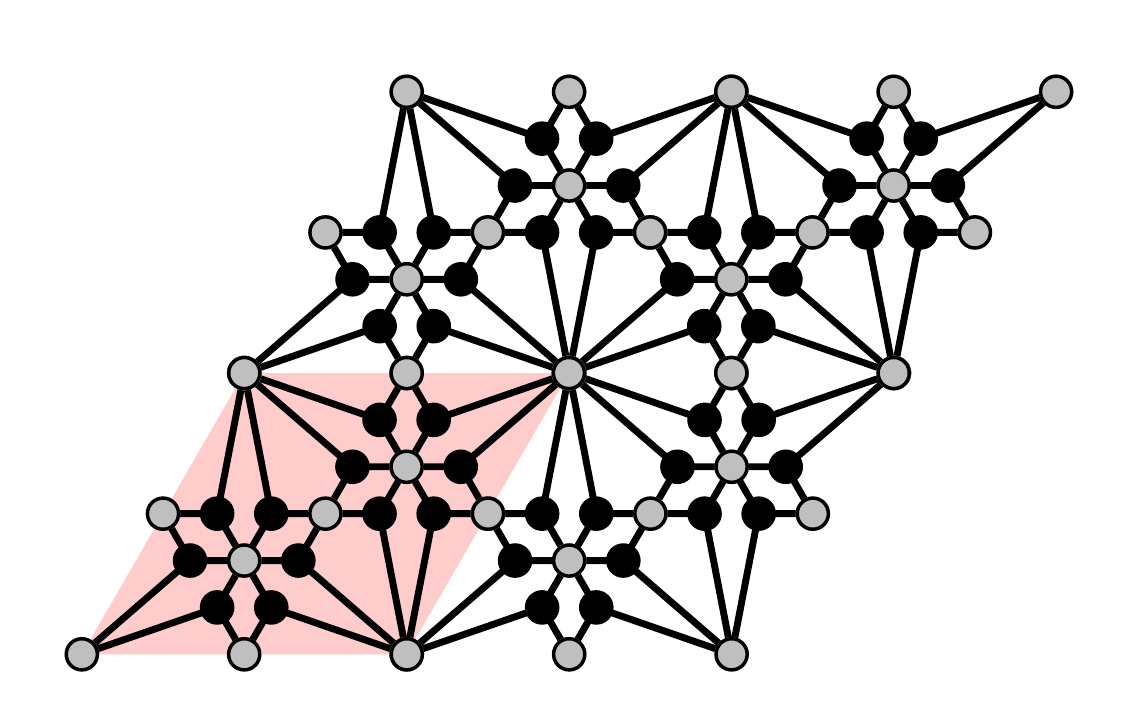} & \hspace*{-0.9cm} 
\includegraphics[width=230pt]{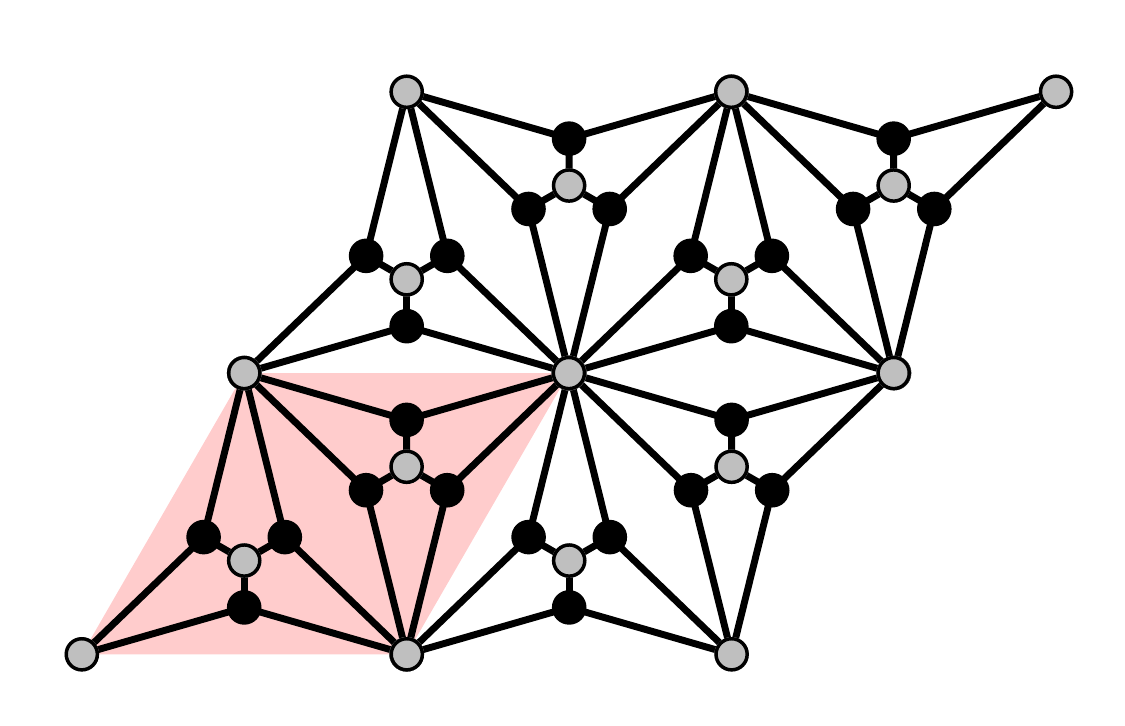} \\[2mm]  
\hspace*{-2.7cm} (a) & \hspace*{-3.1cm} (b) \\[2mm] 
\hspace*{-2.7cm} \includegraphics[width=150pt]{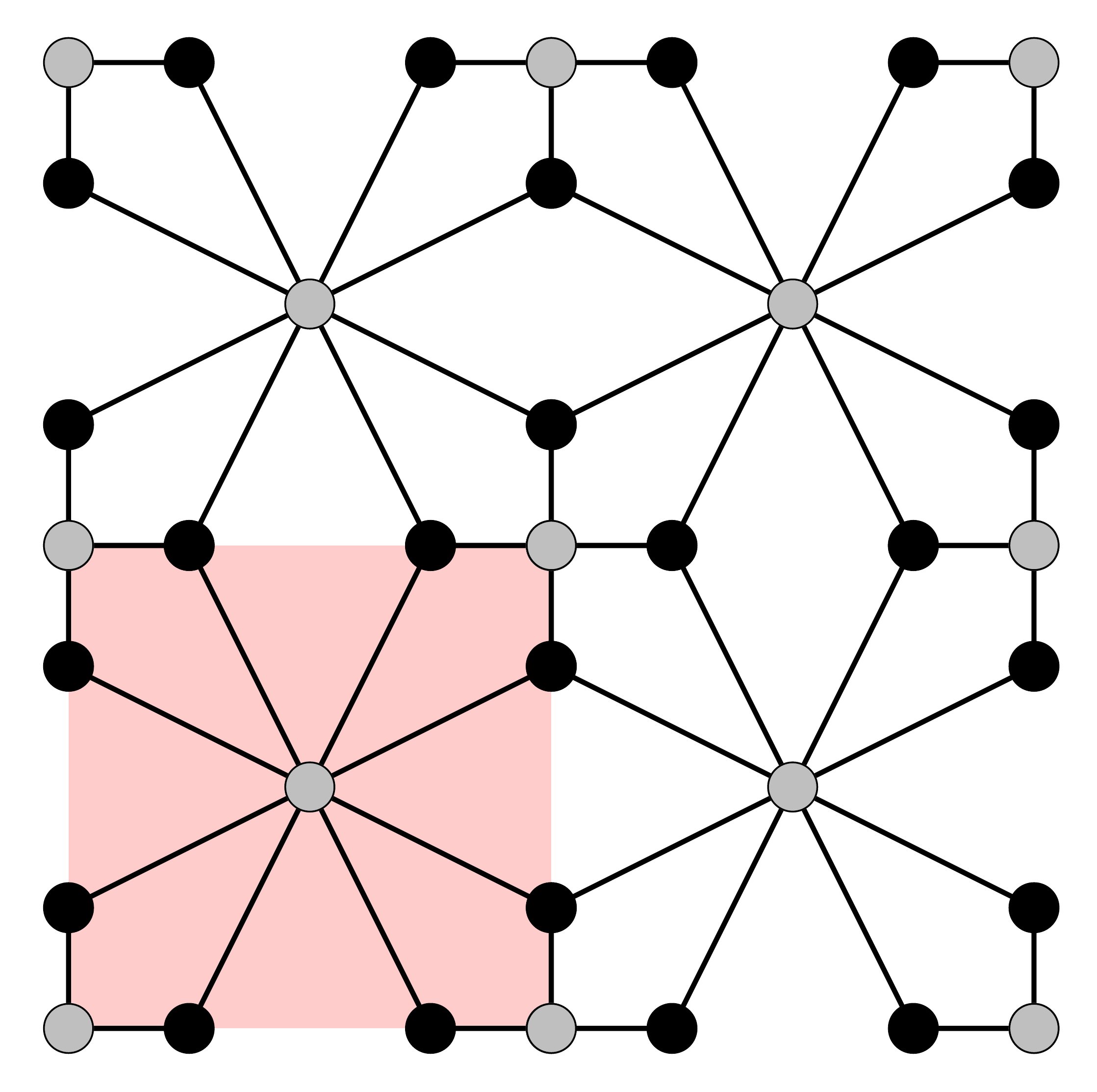}  & \hspace*{-2.9cm} 
\includegraphics[width=150pt]{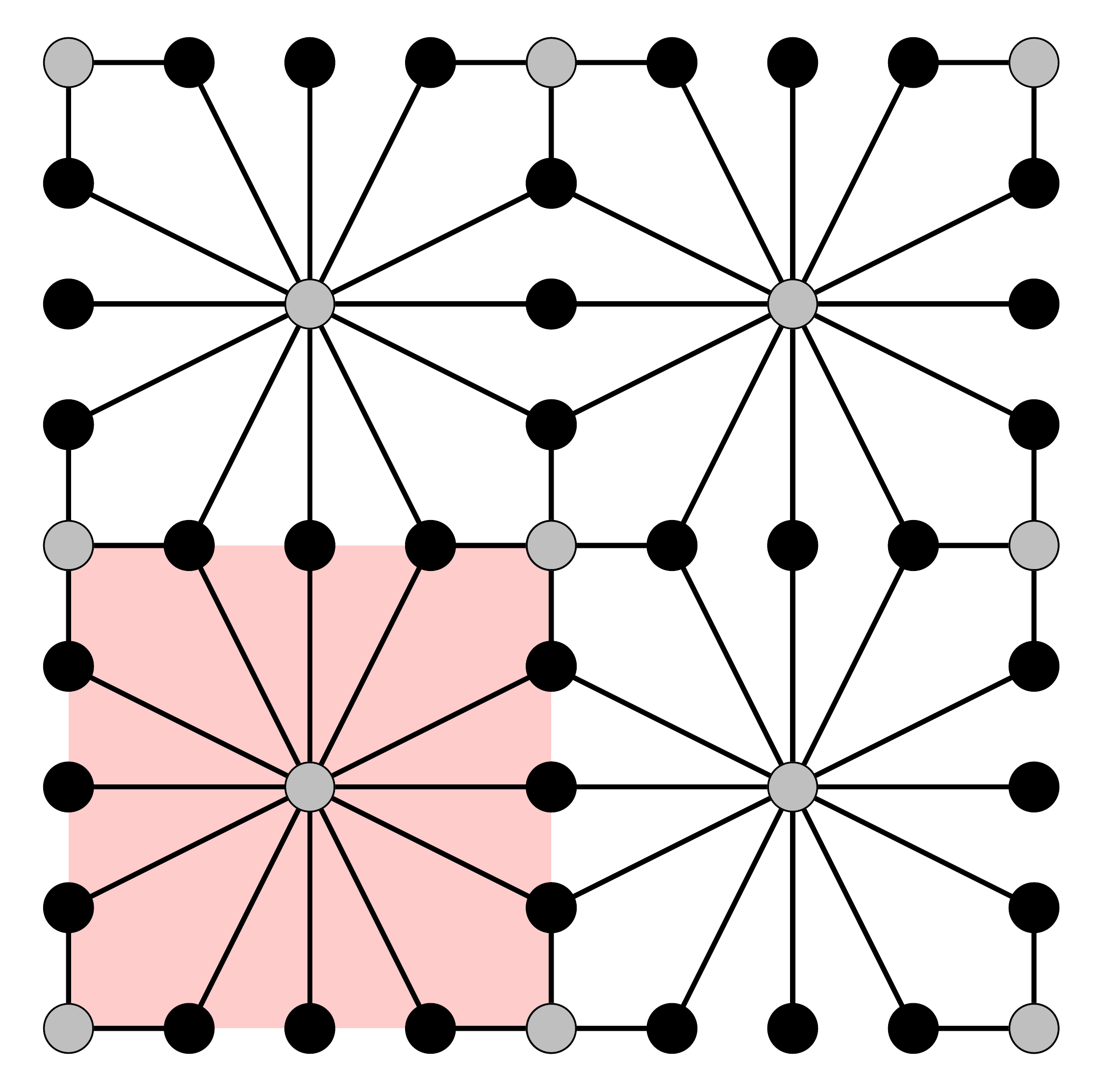}  \\[2mm]  
\hspace*{-2.7cm} (c) & \hspace*{-2.9cm}  (d) \\[2mm] 
\end{tabular}
\end{center}
\vspace*{-5mm}
\caption{\label{fig_Q_non-selfdual2}
   Quadrangulations of non-self-dual type considered in this paper. 
   We show two of them with an underlying triangular Bravais lattice: 
   (a) $\mathcal{Q}(\text{cross})$, and  
   (b) $\mathcal{Q}(\text{asanoha})$. 
   We also show another two quadrangulations with an underlying square 
   Bravais lattice: (c) $G''_2$, and (d) $G''_3$.
   The lattices are depicted as in figure~\ref{fig_Q_non-selfdual}. 
}
\end{figure}

Notice that this list encompasses several qualitatively distinct cases: either 
$G_0$ or $G_1$ is a plane triangulation (i.e., asanoha, dual cross, 
dual martini, and union-jack), or a plane quadrangulation (i.e., diced, dual 
ruby, and $G'_2$). These seven quadrangulations are depicted in 
figures~\ref{fig_Q_non-selfdual}--\ref{fig_Q_non-selfdual2}.

Concerning their translational invariance, there are five lattices with
an underlying triangular Bravais lattice with a basis formed by 
$n_{\mathcal{Q}(\text{diced})}= n_{\mathcal{Q}(\text{martini})}=6$,
$n_{\mathcal{Q}(\text{asanoha})}=9$, 
$n_{\mathcal{Q}(\text{ruby})}=12$, and
$n_{\mathcal{Q}(\text{cross})}=18$. 
The other two quadrangulations have an underlying square Bravais lattice with 
a basis formed by $n_{G''_2}=6$, and $n_{G''_3}=8$. Again, we say that a 
quadrangulation $\Gamma$ has size $L\times N$ when the corresponding 
underlying square Bravais lattice $\mathcal{B}_\Gamma$ has size $L\times N$.

\medskip

\noindent
{\bf Remark.} The above construction of a plane quadrangulation $\Gamma$ 
associated to a pair of dual plane graphs $(G_0,G_1)$ can be extended 
to any closed orientable surface. If $G_0$ is a finite connected graph 
embedded in a closed orientable surface $S$, then one can construct its dual 
in an analogous way, so that the dual graph $G_1$ is also embedded in $S$. 
Then, the quadrangulation $\Gamma$ can be built as before, and it is also a 
graph embedded in $S$. In this way, for any dual pair $(G_0,G_1)$ there is 
a unique quadrangulation $\Gamma$ embedded in the same surface. However, the
inverse relation is \emph{false}: there are quadrangulations in a closed
orientable surface that do \emph{not} correspond to a dual pair. 
Simple examples of these quadrangulations are square lattices embedded 
in a torus with sizes 
$(2L+1)\times 2N$, $2L\times (2N+1)$ and $(2L+1)\times (2N+1)$ for any 
$L,N\in\N$. The reason is that these graphs are \emph{not} bipartite. 
This observation leads to the following statement: there is a one-to-one 
correspondence between \emph{bipartite} quadrangulations embedded in a 
closed orientable surface $S$ and dual pairs $(G_0,G_1)$ embedded also in $S$. 
We will be interested in section~\ref{sec.MC} in bipartite quadrangulations 
embedded in a torus.   

%
%
\subsection{Height representation} \label{sec.height_rep}

In this section we will discuss the height representation for the
AF 3-state Potts model on a quadrangulation. In particular, we will
extend the well-known height representation for the square lattice 
\cite{Nijs_82,Burton_Henley_97,Salas_98} to a general plane quadrangulation.
This extension is not however, the most general case discussed in 
\cite{Henley_93,Kondev_96,Salas_98}. We refrain from repeating this general setup
here to keep the exposition short. 

We consider a finite subset of a plane quadrangulation $\Gamma=(V,E)$ with 
\emph{free} boundary conditions (to ensure that everything is well defined). 
This quadrangulation is associated to the dual pair $(G_0,G_1)$, as explained
in the previous section with $G_0=(V_0,E_0)$ and $G_1=(V_1,E_1)$. 
We define the zero-temperature 3-state AF Potts model as usual: to each 
vertex $x\in V$, we assign a spin variable $\sigma_x \in \{0,1,2\}$, such 
that two neighbouring spins cannot have the same value: i.e., 
if $\{x,y\}\in E$, then $\sigma_x \neq \sigma_y$. Indeed, as $\Gamma$ is
bipartite, there are many possible ground states (= proper colourings of 
$\Gamma$) without frustration.  

The first step consists in defining a height rule, so we can map the 
spin configuration to a height configuration. In our case, the 
\emph{microscopic} height variables $h(x)$ live on the vertices of $\Gamma$,  
and they are defined as follows:
\begin{itemize}
 \item Pick, without loss of generality, any vertex $0\in V_0$. We will 
       call this distinguished vertex the `origin'. We fix the height at 
       the origin as $h(0)\equiv 0,4,2 \pmod{6}$ according to whether 
       $\sigma_0 = 0,1,2$. This choice ensures that
\begin{equation}
h(0) \;\equiv\; \sigma_0 \pmod{3} \,, \qquad h(0) \;\equiv\; 0 \pmod{2}\,.
\label{def_height_origin}
\end{equation} 
 \item The increment in height in going from vertex $x$ to a neighbour 
       vertex $y$ is given by 
\begin{equation}
\label{def_height_rule}
  h(x) - h(y) \;\equiv\; \sigma_x - \sigma_y \pmod{3} \,, \qquad  
  h(x) - h(y) \;=\; \pm 1 \,.                       
\end{equation}
\end{itemize}
This rule is well defined, as the change $\Delta h$ around any quadrilateral 
face is zero. (If four numbers $\pm 1$ add up to $0 \pmod{3}$, they must be 
two $+1$ and two $-1$, so their sum is zero.) The above rules 
\eqref{def_height_origin}/\eqref{def_height_rule} fix the height 
$h(x) \bmod 6$ at any vertex $x\in V= V_0\cup V_1$: 
\begin{equation}
\label{def_heights}
  h(x) \;\equiv\; \sigma_x \pmod{3} \,, \qquad  
  h(x) \;\equiv\; \begin{cases} 0 \pmod{2} & \text{if $x \in V_0$,} \\
                                1 \pmod{2} & \text{if $x \in V_1$,} 
                  \end{cases}
\end{equation}
once we know the height at the origin $h(0)$, and the sublattice $x$ belongs 
to. Then the \emph{height lattice} (i.e., the subset of $\R^D$ 
where the height variables live) is $\Z$ with $D=1$.  

The second step is to determine the \emph{ideal states}: they are 
in general families of proper colourings of $\Gamma$ whose
corresponding height configurations are macroscopically flat and  
maximize the entropy density. These ideal states depend on the type
of quadrangulation we are considering:

\begin{itemize}

\item If $\Gamma$ is a plane \emph{quadrangulation of self-dual type}, then 
      both sublattices are equivalent $G_0 \simeq G_1$. Therefore, there 
      are six ideal states such that one sublattice is ordered (i.e, all spins 
      take a constant value), while the other sublattice is 
      disordered (i.e., the spins take randomly the other two values). 
      The system has to choose which sublattice is the ordered one, and then 
      it has to choose which value is taken by all the spins in such sublattice.
      If $\Gamma$ is associated to the dual pair $(G_0,G_1)$ we will label 
      our ideal states as $a/bc$ (resp.\/ $ab/c$) with $a,b,c=0,1,2$, if the  
      $G_0$ (resp.\/ $G_1$) sublattice is ordered. 

\item If $\Gamma$ is a plane \emph{quadrangulation of non-self-dual type}, 
      then $G_0$ and $G_1$ are no longer equivalent ($G_0 \not\simeq G_1$). 
      Let us first consider the most often case, where one sublattice has 
      less vertices (or a smaller vertex density) than the other one. 
      Without loss of generality, if we assume that $|V_0| < |V_1|$, then the    
      ordered sublattice will be $G_0$, and the disordered one will be
      $G_1$. Therefore, we expect three ideal states of the form $a/bc$. 
      The case for which $|V_0| = |V_1|$ needs more work, and will be 
      considered at the end of this section. 

\end{itemize} 

In both cases, these ideal states are macroscopically flat: the microscopic 
height at the vertices of the ordered sublattice is constant $h_\circ$, 
and on the vertices of the other sublattices, the height takes randomly 
the values $h_\circ \pm 1$. We can therefore label each ideal state by its 
average height $\<h(x)\> = h_\circ$. 

The above ideal states also maximize the zero-temperature entropy density 
$W(\Gamma;q)$ in the thermodynamic limit (see \eqref{def.f}):
\begin{equation} 
W(\Gamma;q) \;=\; e^{f(\Gamma;q,-1)} \;=\; 
                  \lim_{n\to\infty} Z_{\Gamma_n}(q,-1)^{1/|U_n|}\,,
\label{def_W}
\end{equation}
for a suitable sequence $(\Gamma_n)$ of graphs $\Gamma_n=(U_n,F_n)$ tending, 
as $n\to\infty$, to some infinite periodic quadrangulation $\Gamma$. 
We can compare the entropy of ideal states to the entropy of three-colourings 
{\em tout court}\/:

\begin{itemize}

\item Let $\Gamma_n=(V_n,E_n)$ be a plane \emph{quadrangulation of self-dual 
      type} with $|V_n|=n$. The ideal-state entropy density is given by 
\begin{equation}
W(\Gamma;3) \;=\; \lim_{n\to\infty} \left(6 \times 2^{n/2}\right)^{1/n} 
      \;=\; \sqrt{2} \;\approx\; 1.41421\,.
\end{equation}
      This result is not far from (and smaller than) Lieb's exact result for 
      the square lattice $W(\text{square},3) = (4/3)^{3/2} \approx 1.53960$ 
      \cite{Lieb_67a,Lieb_67b}.

\item Let $\Gamma_n=(V_n,E_n)$ be a plane \emph{quadrangulation of 
      non-self-dual type}, with $|V_n|=n$, and associated to the dual pair 
      $(G_{0,n},G_{1,n})$ with $G_{j,n} = (V_{j,n},E_{j,n})$ for $j=0,1$, and 
      $|V_{0,n}|+|V_{1,n}|=n$. The ideal-state entropy density is given by 
\begin{equation}
W(\Gamma;3) \;=\; \lim_{n\to\infty} (3 \times 2^{|V_{1,n}|})^{1/n} \;=\; 
      \lim_{n\to\infty} 2^{|V_{1,n}|/n}\,.
\end{equation} 
      In the diced case, $|V_{0,n}|=n/3$ and $|V_{1,n}|=2n/3$, so 
      $W(\text{diced},3) = 2^{2/3} \approx 1.5874$. This value is close to
      (and smaller than) the estimated value by Chen \emph{et al.} 
      \cite{Chen_11}: $W(\text{diced},3) \approx e^{0.473839} \approx 1.6062$.

\end{itemize} 

The examples discussed above show that the naive ideal-state entropy density 
is slightly smaller that the true entropy density for the square and diced 
lattices, respectively. But this ideal-state picture captures 
the relevant physics of the corresponding models, as we shall see below. 

\begin{figure}[htb]
\begin{center}
\includegraphics[width=300pt]{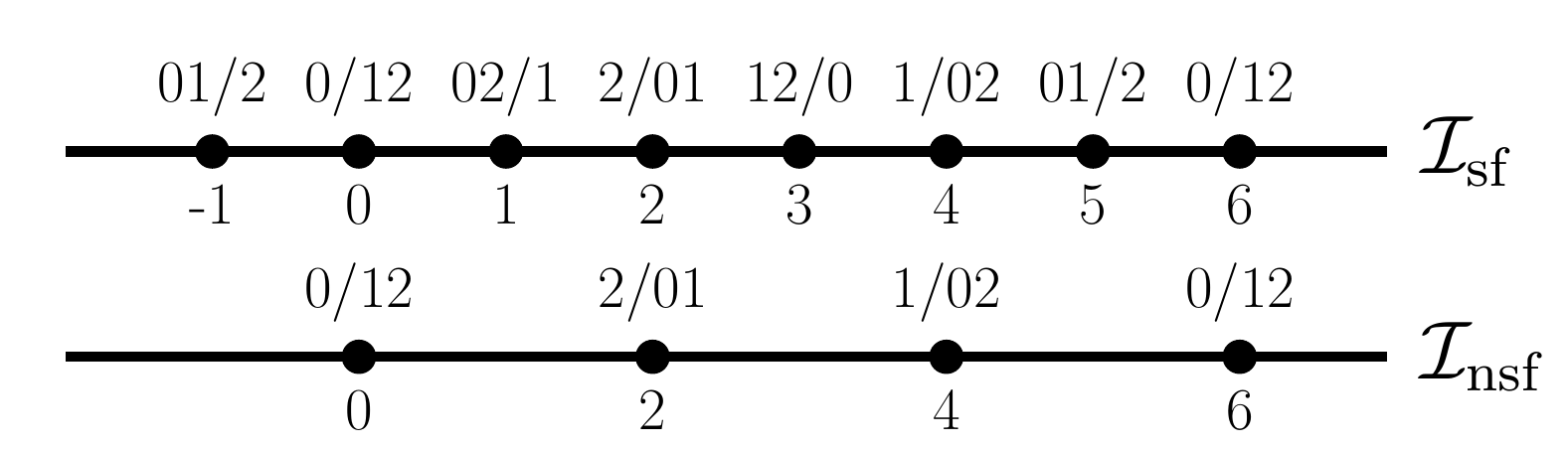} 
\end{center}
\vspace*{-5mm}
\caption{\label{fig_ideal_states}
 Ideal-state lattice for the zero-temperature 3-state AF Potts model 
 on a plane quadrangulation of self-dual type (top figure labeled 
 $\mathcal{I}_\text{sf}$), and for the same model on a quadrangulation of 
 non-self-dual type (bottom figure labeled $\mathcal{I}_\text{nsf}$). 
 The dots depict the ideal states: the numbers below (resp.\/ above) each dot 
 represent the average height $h_\circ$ (resp.\/ the spin structure) of the 
 corresponding ideal state (see text).
}
\end{figure}

It follows from \eqref{def_heights} that there is a one-to-one correspondence 
between ideal states and heights ${\rm mod} \ 6$. The \emph{ideal-state lattice} 
$\mathcal{I}$ (which is the set of all average heights of ideal states) 
for the quadrangulations of self-dual type is $\mathcal{I}_\text{sf}=\Z$; but 
for quadrangulations of non-self-dual type is 
$\mathcal{I}_\text{nsf}=2\Z$ (see figure~\ref{fig_ideal_states}). 
 
%
%

Notice that a given ideal state corresponds to infinitely many different 
average heights $h\in \mathcal{I}$. In particular, there
is a one-to-one correspondence between ideal states and the
coset $\mathcal{I}/\mathcal{R}$, where $\mathcal{R}$ is the 
\emph{repeat lattice}. This is a subgroup of $\Z$ such that if 
$h\in \mathcal{I}$ is a given ideal state, then $h+\mathcal{R}$ corresponds
to the same ideal state. In our case, it is obvious that
$\mathcal{R}=6\Z$ for any plane quadrangulation, as the heights are 
defined modulo $6$ (see figure~\ref{fig_ideal_states}).   
 
Up to this point the arguments are rigorous. Now we guess that typical 
configurations of the spin model are built from relatively large domains 
(where on each of these domains, the spin configuration corresponds to 
small fluctuations around one of the above ideal states), separated by
relatively narrow interfaces. If we then define suitable 
\emph{coarse-grained} height variables $\bar{h}(x)$, we
expect that they will take values in or near the ideal-state lattice
$\mathcal{I}$, except at the boundary between domains. The long-wavelength
behaviour of such a coarse-grained model is believed to be controlled
by the effective coarse-grained Hamiltonian (or Euclidean action)
\begin{equation}
H \;=\; \int \left[ \frac{K}{2} | \nabla \bar{h}(x) |^2 + 
                    V_\text{lock}( \bar{h}(x) ) \right] \, {\rm d}^2 x \,,
\label{def_hamiltonian_height_model}
\end{equation}
where we have taken into account that $D=1$. The first term in the integral 
takes into account the entropy of small fluctuations around the ideal states; 
and the \emph{locking potential} $V_\text{lock}$ favours the heights to 
take their values in $\mathcal{I}$. The coupling constant (or \emph{stiffness})
$K$ controls the physics of this model: there is a critical value $K_c$ such 
that if $K<K_c$ (resp.\/ $K>K_c$), $V_\text{lock}$ is irrelevant 
(resp.\/ relevant) in the RG sense. Therefore, if $K<K_c$ the height model 
is in the rough phase, and its long-wavelength behaviour is described by a 
massless Gaussian model with height correlations diverging logarithmically 
with distance. In this case, the corresponding spin model is critical, and it
is described by a CFT with central charge $c=1$. 
On the contrary, if $K>K_c$, the height model is in its smooth phase, 
exhibiting long-range order and bounded fluctuations around the ordered
state. Then, the corresponding spin model describes small fluctuations around
one of the ideal states. At $K=K_c$ the system is at the roughening transition,
and the spin system is also critical.

If $K< K_c$ this approach also predicts the number of critical observables
and their corresponding critical exponents in terms of the unique free 
parameter $K$. In particular, the correlation functions of local operators 
of the coarse-grained heights should have the periodicity of the repeat lattice
$\mathcal{R}$. Therefore, the Fourier transform of such correlators should
contain only wavevectors belonging to the reciprocal of the repeat lattice
\begin{equation}
\mathcal{R}^o \;=\; \{ G \in \R \colon G \cdot a \in 2\pi \Z\,, \forall a\in 
\mathcal{R} \} \,.
\end{equation}
In particular, we have three candidates for critical observables 
in this model 
\begin{itemize}
  \item The staggered magnetisation with $G=\pm \pi/3$.
  \item The uniform   magnetisation with $G=\pm 2\pi/3$.
  \item The staggered polarisation  with $G=\pm \pi$ (see 
        \cite{Burton_Henley_97,Salas_98} for details). 
\end{itemize}  

Given the wavelength $G$, the corresponding correlator will decrease
algebraically with distance like $\sim |x|^{-\eta_G}$ with 
\begin{equation}
\eta_G  \;=\; \frac{G^2}{2\pi K} \,.
\label{def_etaG}
\end{equation} 
In particular, \emph{all} critical exponents depend on the single parameter
$K$. The usual critical-exponents ratios can be computed by using 
\eqref{def_etaG} and the usual scaling relation
\begin{equation}
\left. \frac{\gamma}{\nu}\right|_G \;=\; 2 - \eta_G \;=\; 
                                         2 - \frac{G^2}{2\pi K} \,.
\label{def_gamma_over_nu}
\end{equation}

Notice that the value of $K_c$ is obtained by making the locking potential
exactly marginal: i.e., $\eta = 4$. This potential has the periodicity of the 
ideal-state lattice $\mathcal{I}$, so its Fourier transform can only 
contain wavevectors $G$ belonging to the reciprocal of the ideal-state lattice 
$\mathcal{I}^\circ$. Then the critical stiffness $K_c$ is given by  
\begin{equation}
K_c \;=\; \frac{ a_{\mathcal{I}^\circ}^2}{8 \, \pi} \;=\; 
\begin{cases} 
\dfrac{\pi}{2} & \text{ if $\Gamma$ is of self-dual type,} \\[5mm] 
\dfrac{\pi}{8} & \text{ if $\Gamma$ is of non-self-dual type.}
\end{cases} 
\label{def_Kc}
\end{equation}
as $\mathcal{I}_\text{sf}^\circ = 2\pi \Z$ for quadrangulations of self-dual type, 
and $\mathcal{I}_\text{nsf}^\circ = \pi \Z$ for quadrangulations of non-self-dual 
type. Note that the critical stiffness for the latter case is \emph{four times
smaller} than for quadrangulations of self-dual type.  

Let us now discuss the behavior of the 3-state AF Potts model on a plane
quadrangulation depending of its type. This discussion will lead naturally
to conjecture~\ref{conj.main} stated in the Introduction.

\subsubsection*{Quadrangulations of self-dual type}

In the square-lattice case, from the known result for the staggered 
magnetisation $\eta_\text{stag}=1/3$ \cite{Nijs_82,Park_Widom_89}, 
the value $K=\pi/6$ follows from \eqref{def_etaG}, and the other exponents 
for this model can be computed by using 
\eqref{def_etaG}/\eqref{def_gamma_over_nu}.  
Notice that in this case, $K < K_c =\pi/2$ \eqref{def_Kc}. Therefore this spin 
model has a zero-temperature critical point whose large-distance behaviour 
is governed by a CFT with central charge $c=1$. In addition, the three 
observables mentioned above are relevant; although only the first two have a
diverging susceptibility. 

We expect that for the 3-state AF Potts model on any plane quadrangulation of
self-dual type all the above conclusions should hold by using standard 
universality arguments: the symmetry and ground-state degeneracy are exactly 
the same (i.e., six ideal states, each of them with a sublattice 
ferromagnetically ordered). Moreover, the ideal-state and repeat lattices are 
identical to the square-lattice case. Therefore, $q_c(\Gamma)=3$ for any 
quadrangulation $\Gamma$ of self-dual type.  
However, as there is a free parameter $K$, the critical exponents may be (and 
in fact are) lattice dependent through the unique free parameter $K$.
Therefore, the stiffness $K$ encodes the microscopic properties of the lattice  
for this class of models.

In this sense, we recover some sort of universality for the whole family of 
plane periodic quadrangulations of self-dual type. The only difference is that
the critical exponents depend on $K$, rather than being the same for all 
elements of this family. 
This discussion motivates part (a) of conjecture~\ref{conj.main}.  

\subsubsection*{Quadrangulations of non-self-dual type}

If $\Gamma$ is of non-self-dual type, then the two sublattices are not
equivalent. Let us assume in this preliminary discussion that there is 
one sublattice ($G_0$) that contains less vertices than the other one $(G_1)$
(or, equivalently, that the fraction of vertices belonging to $G_0$
is smaller than that of $G_1$). Then, as we discussed earlier, all the spins 
on $G_0$ will take the same value, while those on $G_1$ will take randomly the
other two values. Therefore, we have only three ideal states (each of them with
$G_0$ ferromagnetically ordered).  
 
The diced lattice \cite{Kotecky-Salas-Sokal} [depicted in 
figure~\ref{fig_diced}(b)] is the simplest case: the triangular sublattice 
$G_0$ contains one third of the vertices, and each of them has degree six,
while $G_1$ is an hexagonal sublattice with two thirds of the vertices, 
each of them of degree three. (The average degree is four, as it should be.)  
Then $G_0$ is the ordered sublattice in the three ideal states. 
In fact, it has been rigorously proven that the 3-state Potts antiferromagnet
has long-range order at $T=0$ \cite{Kotecky-Salas-Sokal} with three ordered 
states (one for each ideal state).
Furthermore, it has been shown numerically via MC simulations, that 
there is a finite-temperature critical point belonging to the 
3-state \emph{FM} Potts-model universality class. 

If we consider the 3-state AF Potts model on any plane periodic quadrangulation
of non-self-dual type, we find that the symmetry and ground-state degeneracy 
are the same as for the diced lattice. Therefore, using the same universality
arguments as for the self-dual case, we expect that the ground state is
ordered with three distinct and coexisting phases. Therefore, there should be 
a phase-transition at some (lattice-dependent) finite temperature 
$v_c(\Gamma)$. The existence of such a transition implies that 
$q_c(\Gamma) > 3$ for any quadrangulation $\Gamma$ belonging to this family.   

The fact that the critical stiffness $K_c=\pi/8 \approx 0.39269908$ 
\eqref{def_Kc} is four
times smaller than for the self-dual case, implies that for the non-self-dual
case, it is, in principle, more difficult to fall in the critical case 
$K < K_c=\pi/8$. In particular, the staggered polarisation operator mentioned 
above has wavevector $G=\pm \pi$, which coincides with the wavevector of 
the locking potential; therefore it is no longer a relevant operator, but 
a marginal one. (Actually, the stiffness for all  
quadrangulations of self-dual type displayed in 
table~\ref{table_gamma_self_dual} are larger than this critical value.) 

The nature of the predicted finite-temperature phase transition is 
\emph{a priori} not clear:
it might be either a first- or a second-order critical point. Again, if this
finite-temperature phase transition is of second order, then universality 
predicts that it must belong to the universality class of the 3-state FM 
Potts model. (At $T=0$, the ground-state of both models 
has the same $\Z_3$ symmetry and degeneracy.) 
Moreover, the previous results for the diced lattice and the 
new ones for the other seven lattices considered in this
paper agree well with this scenario.  
This discussion motivates part (b) of conjecture~\ref{conj.main}.  

\medskip

\noindent
{\bf Remark.} The above arguments assume that the quadrangulation $\Gamma$
corresponds to a dual pair $(G_0,G_1)$ such that $|V_0|<|V_1|$. However,
there are quadrangulations of non-self-dual type that satisfy 
$|V_0|=|V_1|$ [e.g. $\Gamma = \mathcal{Q}(\text{diced})$ and 
$\mathcal{Q}(\text{ruby})$]. In these cases, there is no {\em a priori} 
reason why only three out of the initial six 
candidates are actually the ideal states of the system. 

If one of the sublattices (say, $G_0$) has a sub-sublattice $G_{0,0}$ with 
vertices with a degree larger than the degrees of the vertices on the other 
sub-sublattices of $G_0$ and $G_1$, then we expect that $G_0$ should
be the `ordered' lattice. In fact, the numerical study of 
$\mathcal{Q}(\text{diced})$ and $\mathcal{Q}(\text{ruby})$ provides additional
support to this argument: the sublattice $G_0$ that is more ordered is 
indeed the one containing the vertices with larger degree. 
However, in this case the sublattice $G_0$ and its sub-sublattice $G_{0,0}$ 
are not completely ordered, contrary to what happened to $G_0$ when 
$|V_0|<|V_1|$. 

In other words, even though the `naive' entropy density of the two sets 
of ideal states is the same, the \emph{true} entropy density is not, as one 
has to consider also the fluctuations around these ideal states. And this true 
entropy density is larger for the sublattice containing a sub-sublattice 
with vertices of larger degree.

Finally, notice that if we have a quadrangulation of non-self-dual type 
$\Gamma$ associated to a dual pair $(G_0,G_1)$ satisfying $|V_0|=|V_1|$,
then the quadrangulation $\mathcal{Q}(\Gamma)$ associated to a new dual
pair $(G'_0,G'_1)$ will also satisfy $|V'_0|=|V'_1|$. This observation 
uses the fact that the average degree on a quadrangulation is $4$, and  
provides additional examples to further test this subtle point in our
arguments. 

%
%
\section{Numerical results} \label{sec.results}

In this section we will describe our numerical tests to confirm or disprove
conjectures~\ref{conj.main} and~\ref{conj.wsk}. We will consider the four
quadrangulations of self-dual type shown in figure~\ref{fig_Q_selfdual} 
(namely, $\mathcal{Q}(\text{hextri})$, $\mathcal{Q}(\text{house})$,
$\mathcal{Q}(\text{martini-B})$, and $\mathcal{Q}(\text{cmm-pmm})$),
and the seven quadrangulations of non-self-dual type shown in 
figures~\ref{fig_Q_non-selfdual}--\ref{fig_Q_non-selfdual2} (namely,  
$\mathcal{Q}(\text{diced})$, $\mathcal{Q}(\text{martini})$, 
$\mathcal{Q}(\text{ruby})$, $\mathcal{Q}(\text{asanoha})$,
$\mathcal{Q}(\text{cross})$, $G''_2$ and $G''_3$). 
In addition, we also have two already well-known cases: the square 
\cite{Burton_Henley_97,Salas_98} and the diced
\cite{Kotecky-Salas-Sokal} lattices. 

Most of our numerical work consist in high-precision MC
simulations of the 3-state Potts AF model on these seven lattices; but we 
will also use other techniques: TM and CP. All these methods will be 
described in the following sections.

%
%
\subsection{Monte--Carlo simulations} \label{sec.MC}

We have made extensive MC simulations for the eleven lattices mentioned
above using the Wang--Swendsen--Koteck\'y (WSK) cluster algorithm 
\cite{WSK_89,WSK_90}. To avoid surface effects, we have considered the 
Potts model on finite subsets of each lattice with periodic boundary 
conditions. Although WSK is irreducible (or ergodic) on any graph $G$ 
at any non-zero temperature, periodic boundary conditions may cause troubles 
with the ergodicity of this algorithm at zero temperature. 
In general, at $T=0$, the Potts-model probability distribution
becomes the uniform measure over the set of \emph{proper} $q$-colourings of 
$G$. While for planar graphs, ergodicity of WSK can be proven for any 
$q>\chi(G)$, where $\chi(G)$ is the chromatic number of $G$ 
\cite[Corollary~4.5]{Mohar}, 
for non-planar graphs the ergodicity of WSK can be guaranteed only for 
$q\ge \Delta+1$, where $\Delta$ is the maximum degree of $G$ 
\cite[Corollary~2.5]{Mohar}. However, we can always choose a finite
subset of any quadrangulation with periodic boundary conditions in such a 
way that the graph is still bipartite. This observation is crucial, as  
it is well-known \cite{Burton_Henley_97,Sokal_Ferreira,Mohar} that WSK 
is ergodic at zero temperature for any \emph{bipartite} graph and 
any number of states $q\ge 2$. Therefore, we can use WSK safely even at
zero temperature for all the quadrangulations depicted in
figures~\ref{fig_Q_selfdual}--\ref{fig_Q_non-selfdual2}. (However, the 
ergodicity of WSK at $T=0$ cannot be taken for granted for 
non-bipartite lattices and certain values of the number of states $q$: e.g., 
the triangular lattice for $q=4$ \cite{Mohar_tri}, or the kagome lattice 
for $q=3$ \cite{Mohar_kag}.)
 
As mentioned in section~\ref{sec.quad}, it is clear from 
figures~\ref{fig_Q_selfdual}--\ref{fig_Q_non-selfdual2} that
all the quadrangulations considered in this paper can be regarded as a 
Bravais lattice $\mathcal{B}_\Gamma$ with a basis formed by 
$n_\Gamma$ vertices. 

Dealing with a Bravais lattice with a non-trivial basis is not hard, but 
it involves some extra work. To our knowledge, in the literature 
one can find only simple cases \cite{Campostrini_98,MC_hc}, like the
hexagonal lattice. Therefore, we will explain how to deal with a generic 
\emph{bipartite} quadrangulation 
$\Gamma = \mathcal{Q}(G_0) = (V_\Gamma,E_\Gamma)$ associated to a dual pair 
$(G_0,G_1)$. Let us assume that the sublattice $G_0$ (resp.\/ $G_1$) is the 
even (resp.\/ odd) sublattice with vertex set $V_0$ (resp.\/ $V_1$). Indeed,
$V_\Gamma = V_0 \cup V_1$.  

In order to measure some observables (see below), we need to embed the graph
$\Gamma$ in a torus, and assign to each vertex $x \in V_\Gamma$ a vector  
$\bm{x} \in\R^2$. Then, each vertex $x\in V$ is associated to a vector that
can be written as:
\begin{equation}
\bm{x} \;=\; \sum\limits_{j=1}^2 x'_j \, \bm{\eta}_j  + 
             \sum\limits_{j=1}^{n_\Gamma -1} \epsilon_j \, \bm{\mu}_j 
   \;=\; \bm{x}' + \sum\limits_{j=1}^{n_\Gamma -1} \epsilon_j \, \bm{\mu}_j\,,
\quad x'_1,x'_2 = 1,\ldots,L \,, 
\label{def_x}
\end{equation} 
where the \emph{unit} vectors $\bm{\eta}_1,\bm{\eta}_2\in\R^2$ span the
Bravais lattice $\mathcal{B}_\Gamma$, which has dimensions $L\times L$ 
(unit cells) with periodic boundary conditions. Then $\Gamma$ has 
$|V_\Gamma| = |V_0|+|V_1| = n_\Gamma L^2$ vertices. The non-trivial
content of the basis is given in \eqref{def_x} by the $n_\Gamma-1$ vectors 
$\{\bm{\mu}_j\}$. If we assume that one of the vertices of the basis 
(i.e., $x'$ with vector $\bm{x}'$) belongs to the Bravais lattice 
$\mathcal{B}_\Gamma$, then the vectors $\bm{\mu}_j$ give the
position of the non-trivial vertices of the basis with respect to $\bm{x}'$.  
Finally, the $\epsilon_j$ are the components of an $(n_\Gamma -1)$-dimensional 
vector $\bm{\epsilon}=(\epsilon_1,\epsilon_2,\ldots,\epsilon_{n_\Gamma -1})$, 
that (according to the preceding discussion) can take only $n_\Gamma$ values:
it can be either the zero vector $\bm{\epsilon}=\bm{0}$ (when the vertex 
$x$ belongs to $\mathcal{B}_\Gamma$), or any of the 
unit vectors of the standard basis of $\R^{n_\Gamma -1}$: i.e.,  
$\bm{\epsilon}=(0,\ldots,0,1,0,\ldots,0)$, when $x$ is a non-trivial 
vertex of the basis.  
 
We define the sublattice magnetisations $\bm{\mathcal{M}}_i$ (for $i=0,1$) as: 
\begin{equation}
\bm{\mathcal{M}}_i \;=\; \sum\limits_{x \in V_i} \bm{\sigma}_x \;=\;  
                         \sum\limits_{x \in V_i} \sum\limits_{\alpha=1}^q
    \bm{e}^{(\alpha)} \, \delta_{\sigma_x,\alpha} \,, 
\label{def_magnetization_i}
\end{equation}
where $\bm{\sigma}_x \in \R^{q-1}$, and the unit vectors $\bm{e}^{(\alpha)}$
satisfy  
\begin{equation}
\bm{e}^{(\alpha)} \cdot \bm{e}^{(\beta)} \;=\; 
                \frac{q \delta_{\alpha,\beta} -1}{q-1} \,.
\label{eq_e_alpha}
\end{equation}

As explained in section~\ref{sec.height_rep}, we should compute both the
staggered and the uniform susceptibilities. As the graph $\Gamma$ is 
bipartite, the natural staggering should be the one that assigns a $+1$
(resp.\/ $-1$) to all vertices in $V_0$ (resp.\/ $V_1$). This is obviously  
motivated by the Ising case. In our case, the staggered 
$\bm{\mathcal{M}}_\text{stagg}$ an uniform $\bm{\mathcal{M}}_\text{u}$
magnetisations are given by 
\begin{subequations}
\label{def_Mags}
\begin{align}
\bm{\mathcal{M}}_\text{stagg} &\;=\; \bm{\mathcal{M}}_0 - \bm{\mathcal{M}}_1\,,
\label{def_Mstagg} \\
\bm{\mathcal{M}}_\text{u}     &\;=\; \bm{\mathcal{M}}_0 + \bm{\mathcal{M}}_1\,, 
\label{def_Mu}
\end{align}
\end{subequations}
Then, the formulas for $\bm{\mathcal{M}}_\text{stagg}^2$ and 
$\bm{\mathcal{M}}_\text{u}^2$ follow from 
\eqref{def_magnetization_i}/\eqref{eq_e_alpha}: 
\begin{subequations}
\label{def_mag2}
\begin{align}
\bm{\mathcal{M}}_\text{stagg}^2 &\;=\; 
\frac{q}{q-1} \, \sum\limits_{\alpha=1}^q \left[ 
\sum\limits_{x\in V_0} \delta_{\sigma_x,\alpha} - 
\sum\limits_{x\in V_1} \delta_{\sigma_x,\alpha} \right]^2 - 
\frac{(|V_0|-|V_1|)^2}{q-1}  \label{def_Mstagg2} \\ 
\bm{\mathcal{M}}_\text{u}^2   &\;=\; 
\frac{q}{q-1} \, \sum\limits_{\alpha=1}^q 
\left( \sum\limits_{x\in V} \delta_{\sigma_x,\alpha} \right)^2 - 
\frac{|V_\Gamma|^2}{q-1} \label{def_Mu2}
\end{align}
\end{subequations}

The corresponding susceptibilities are defined as follows:
\begin{equation}
\chi_\text{stagg} \;=\; \frac{1}{|V_\Gamma|} \, 
    \< \bm{\mathcal{M}}_\text{stagg}^2 \> \,, \quad 
\chi_\text{u} \;=\; \frac{1}{|V_\Gamma|} \, 
    \< \bm{\mathcal{M}}_\text{u}^2 \> \,, 
\label{def_sus}
\end{equation} 
and the corresponding dimensionless Binder-like ratios are given by
\begin{equation}
R_\text{stagg} \;=\; \frac{ \< (\bm{\mathcal{M}}_\text{stagg}^2)^2 \>}
                          { \<  \bm{\mathcal{M}}_\text{stagg}^2 \>^2 }\,, \quad
R_\text{u} \;=\; \frac{ \< (\bm{\mathcal{M}}_\text{u}^2)^2 \>}
                      { \<  \bm{\mathcal{M}}_\text{u}^2 \>^2 }\,. 
\label{def_R}
\end{equation}

The number of MC steps we have run for each simulation [i.e. for each 
triplet $(\Gamma,q,v)$] is in the range $6\times 10^6$--$10^8$, although 
most of them consist in $\approx 10^7$ MC steps. In all cases, we have 
discarded between 10\% and 20\% of the total number of MC steps to allow
the system to attain thermodynamic equilibrium. The error bars in our MC 
estimates were computed using the jackknife method
(see e.g., \cite[Section 5.7.5, and references therein]{Landau_09}). With 
our statistics, we have been able to obtain error bars of relative size 
$\lesssim 0.1\%$. 

The number of MC steps discarded at the beginning of the 
simulations is also large enough for the system to reach its thermodynamic 
equilibrium. We have measured the integrated autocorrelation times 
$\tau_\text{int}$ for four quadrangulations of non-self-dual type: 
$\mathcal{Q}(\text{diced})$,
$\mathcal{Q}(\text{martini})$, $\mathcal{Q}(\text{ruby})$, and $G''_3$.  
This autocorrelation time is roughly speaking the number of MC steps between
two statistically independent spin configurations \cite{Sokal_book},
once the Markov-chain MC has reached equilibrium.
In all MC simulations performed on these lattices, the worst case corresponds 
to the $\mathcal{Q}(\text{martini})$ lattice with linear size $L=512$, giving 
$\tau_\text{int}\lesssim 90$. This means that for an average simulation of 
$10^7$ MC steps, the number of discarded steps is at least  
$\gtrsim 1.1\times 10^4\, \tau_\text{int}$ for all cases. This is more than
enough to get rid of any initialisation bias. 

For quadrangulations of self-dual type, the situation is even better: we 
have found that the WSK algorithm \emph{at} the zero-temperature critical point 
does not suffer from CSD. This means that  
$\tau_\text{int}$ is (for any of these latices) uniformly bounded in the 
lattice linear size $L$. In our case, we have found that
$\tau_\text{int} \lesssim 8$ for the four lattices considered in 
section~\ref{sec.MC_selfdual}. This result also agrees with the previous
computation for the square-lattice 3-state AF Potts model: 
$\tau_\text{int} \lesssim 8$ uniformly in $L$ \cite{Salas_98}. 
 
%
%
\subsubsection{Quadrangulations of self-dual type} \label{sec.MC_selfdual}

In this section we will report the numerical results for the four 
quadrangulations of self-dual type that we have simulated: 
$\mathcal{Q}(\text{hextri})$, $\mathcal{Q}(\text{house})$, 
$\mathcal{Q}(\text{martini-B})$, and 
$\mathcal{Q}(\text{cmm-pmm})$ (see figure~\ref{fig_Q_selfdual}).

For each lattice, we have first computed the Binder ratios \eqref{def_R}, 
and found that all the curves $R_\text{a}(v;L)$ for 
$\text{a}\in\{\text{stagg},\text{u}\}$, nicely overlap 
at $v=-1$, confirming part~(1) of conjecture~\ref{conj.main}: the AF 
3-state Potts model on a quadrangulation of self-dual type is critical 
at zero temperature $v_c=-1$. As in this case the critical temperature is 
known, we have fitted the zero-temperature susceptibilities to the 
standard power-law {\em Ansatz}
\begin{equation}
\chi_\text{a}(-1;L) \;=\; L^{ (\gamma/\nu)_\text{a} }\, \left[ A + 
   B \, L^{-\omega} + \cdots \right] \,, \qquad 
   \text{a}\in\{\text{stagg},\text{u}\}\,.  
\label{def_sus_Ansatz}
\end{equation} 
As a precaution against subdominant finite-size-scaling (FSS) corrections, 
we have systematically
varied the minimum value $L_\text{min}$ of the data included in the fit. 
In this way we obtain the results displayed in 
table~\ref{table_gamma_self_dual}.

\begin{figure}[htb]
\begin{center}
\begin{tabular}{cc}
\includegraphics[width=200pt]{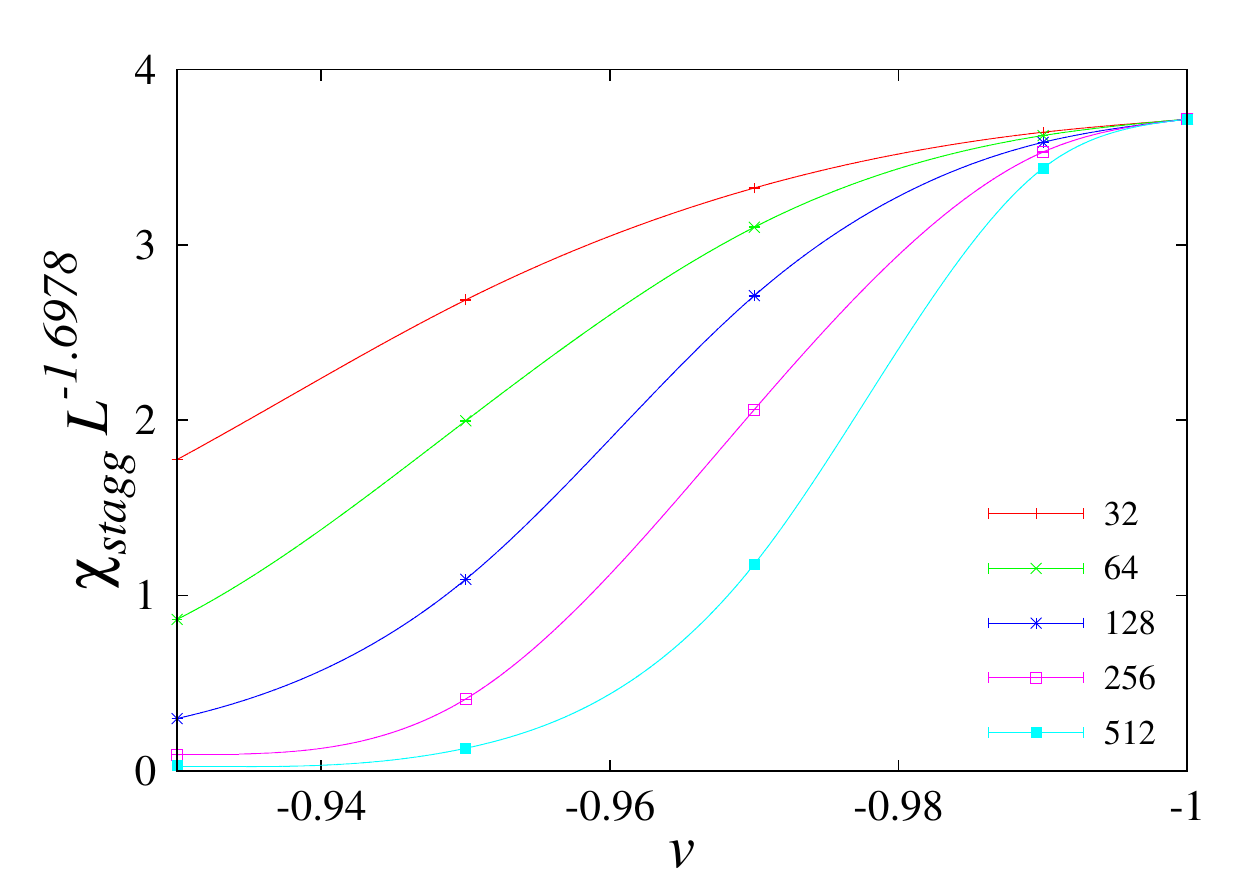} & 
\includegraphics[width=200pt]{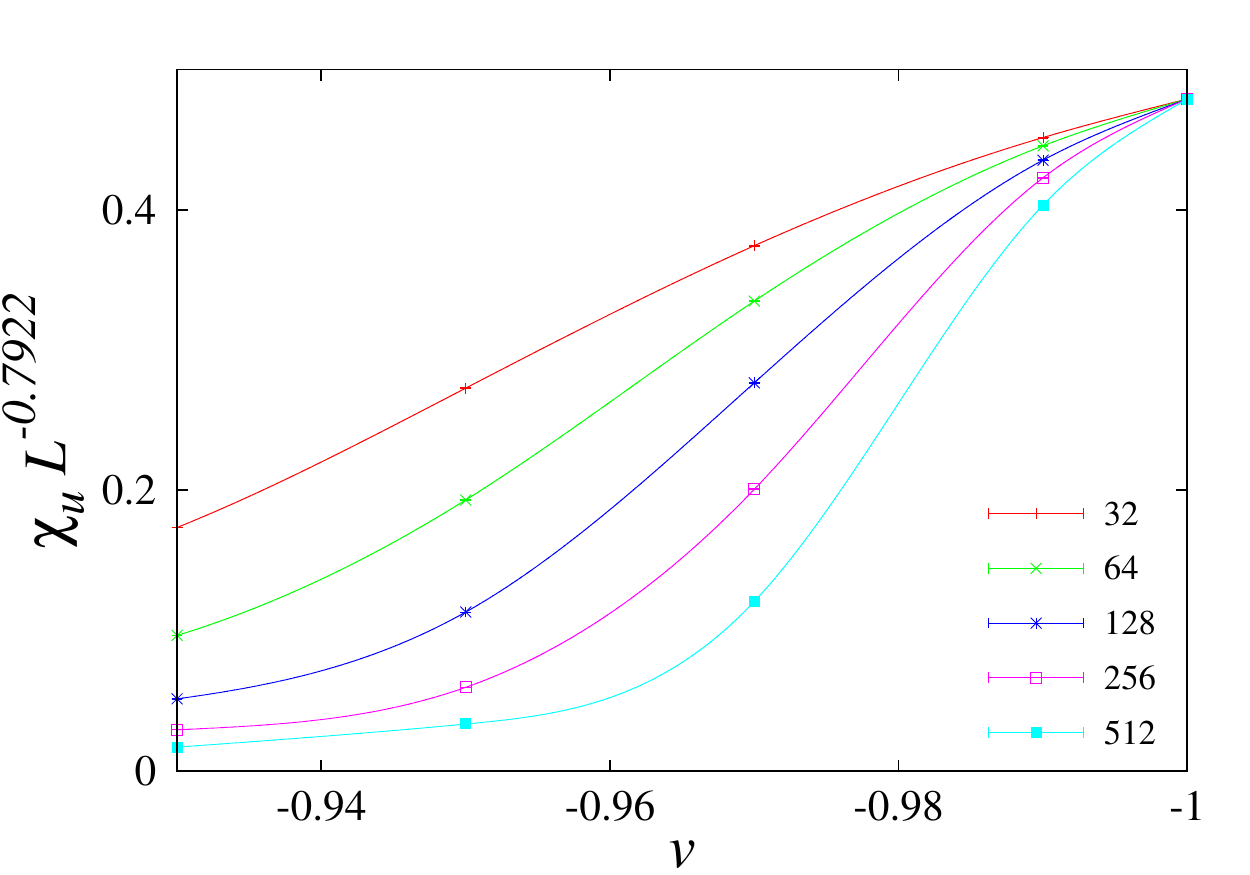} \\[2mm]  
\quad\; (a) &\quad\;  (b) \\
\end{tabular}
\end{center}
\vspace*{-5mm}
\caption{\label{fig_house_sus}
   Scaled staggered $\chi_\text{stagg}$ (a) and uniform $\chi_\text{u}$ (b) 
   susceptibilities (see~\eqref{def_sus}) as a function of the 
   temperature-like parameter $v$ for a $\mathcal{Q}(\text{house})$ lattice 
   of dimensions $L\times L$ and periodic boundary conditions. 
   We show the data points for $L=32$ (red), $L=64$ (green), 
   $L=128$ (navy blue), $L=256$ (pink), and $L=512$ (cyan).
   For each value of $L\le 256$, we have also 
   drawn a spline-interpolation curve joining the corresponding points to 
   guide the eye. 
   The error bars of the points are smaller than the corresponding symbols. 
}
\end{figure}

If we plot the scaled susceptibilities 
$\chi_\text{a} \, L^{-(\gamma/\nu)_\text{a}}$ with 
$\text{a}\in\{\text{stagg},\text{u}\}$ for the 
$\mathcal{Q}(\text{house})$ lattice, we obtain the plots displayed 
in figure~\ref{fig_house_sus}. It is clear that the curves for 
different values of the linear size $L$ overlap at the critical point 
$v_c=-1$, as expected. As the corresponding figures for the other three 
lattices are very similar to those shown in figure~\ref{fig_house_sus},
we refrain from showing them. (Notice that the plot of the scaled
staggered susceptibility for the lattice $\mathcal{Q}(\text{hextri})$ 
has already appeared in \cite[figure~3]{Lv_17}.)

%
%

%
%
\def\kkk{\phantom{1}}
\begin{table}[htb]
\centering
\begin{tabular}{lrrrc}
\hline\hline
\multicolumn{1}{c}{$\Gamma$} &
\multicolumn{1}{c}{$(\gamma/\nu)_\text{stagg}$} & 
\multicolumn{1}{c}{$(\gamma/\nu)_\text{u}$} & 
\multicolumn{1}{c}{$K$} &  
\multicolumn{1}{l}{$\Delta$} \\ 
\hline
$\mathcal{Q}(\text{cmm-pmm})$   & $1.71762(9)$    & $0.8691(5)$ & 
                                  $0.6177(6)$ &8\\ 
$\mathcal{Q}(\text{hextri})$    & $1.7024(3)\kkk$ & $0.8096(9)$ & 
                                  $0.5865(6)$ &6\\ 
$\mathcal{Q}(\text{house})$     & $1.6978(3)\kkk$ & $0.7922(4)$ & 
                                  $0.5778(8)$ &5\\
$\mathcal{Q}(\text{martini-B})$ & $1.6882(3)\kkk$ & $0.7557(9)$ & 
                                  $0.5609(6)$ &5\\ 
\text{square}                   & $5/3 \approx 1.66667\,\,\,\,\kkk$& 
                                  $2/3 \approx 0.66667\,\,\,\,$& 
                                  $\pi/6\approx 0.52360\,\,\,\,$& 4 \\ 
\hline\hline
\end{tabular}
\caption{Critical exponents $(\gamma/\nu)_\text{stagg}$ and 
$(\gamma/\nu)_\text{u}$, and the estimated value of the stiffness $K$ for 
the zero-temperature AF 3-state Potts model on the quadrangulations of 
self-dual type $\Gamma$ studied in this paper. We also show the maximum 
degree $\Delta$ for each of these lattices. 
These results have already appeared in \cite[table~I]{Lv_17}. We include 
for comparison the exact values for the square lattice.
}
\label{table_gamma_self_dual}
\end{table}

As both critical exponents $(\gamma/\nu)_\text{stagg}$ and 
$(\gamma/\nu)_\text{u}$ depend on a single parameter, i.e., the stiffness
$K$ (see~\eqref{def_etaG}/\eqref{def_gamma_over_nu}), we can estimate 
$K$ by using our numerical estimates for the former:  
\begin{equation}
\left. \frac{\gamma}{\nu}\right|_\text{stagg} \;=\; 2 - \frac{\pi}{18K}\,, 
\qquad 
\left. \frac{\gamma}{\nu}\right|_\text{u} \;=\; 2 - \frac{2\pi}{9K}\,.  
\end{equation}
The results of such fits are displayed on the fourth column of 
table~\ref{table_gamma_self_dual}. Indeed, all of them are smaller
than the critical stiffness $K_c =\pi/2\approx 1.5708$. 

As explained in section~\ref{sec.height_rep}, we expect that the stiffness
$K$ will depend on the lattice structure, and this feature is observed 
in table~\ref{table_gamma_self_dual}. It is worth noticing that 
a potential counterexample to conjecture~\ref{conj.main}(a) will be provided 
by a quadrangulation of self-dual type $\Gamma$ for which $K>K_c=\pi/2$. From
table~\ref{table_gamma_self_dual}, we see that the estimated values for $K$ 
do not vary much from that of the square lattice $K=\pi/6$, and all of them
are roughly speaking $2.5$ times smaller than the critical value
$K_c=\pi/2 \approx 1.570796$. A closer inspection of  
table~\ref{table_gamma_self_dual} reveals that the stiffness $K$ seems to
grow \emph{weakly} with the maximum degree of the lattice $\Delta$. 
Therefore, for all plane quadrangulations of self-dual type with maximum 
degree $\Delta \lesssim 8$, we expect that $K < K_c$. On the other hand, 
potential counterexamples to conjecture~\ref{conj.main}(a) might be found 
when $\Delta$ becomes very large, if these quadrangulations do exist at all. 
As a matter of fact, we chose the rather complicated 
quadrangulation $\mathcal{Q}(\text{cmm-pmm})$ with $\Delta=8$ to test this 
weak dependency of the stiffness $K$ on $\Delta$.

The dynamic behaviour of the WSK algorithm at the zero-temperature
critical point found for these four systems is worth discussing in detail.
Once a Markov-chain MC has reached its equilibrium state, then, for any local 
observable $\mathcal{O}$, one can compute its integrated autocorrelation
time $\tau_{\text{int},\mathcal{O}}$ \cite{Sokal_book}. The integrated
autocorrelation time $\tau_\text{int}$ is just the maximum over all 
observables of $\tau_{\text{int},\mathcal{O}}$. Therefore, each 
$\tau_{\text{int},\mathcal{O}}$ is a lower bound for the true value 
$\tau_\text{int}$. We have measured the autocorrelation times 
for the critical observables $\bm{\mathcal{M}}_\text{stagg}^2$
and $\bm{\mathcal{M}}_\text{stagg}^2$ (see table~\ref{table_tauint_sd}).  

%
%
\begin{table}[htb]
\centering
\begin{tabular}{r|rr|rr|rr|rr}
\hline\hline
$\Gamma$ & 
\multicolumn{2}{c|}{$\mathcal{Q}(\text{martini-B})$} & 
\multicolumn{2}{c|}{$\mathcal{Q}(\text{house})$} & 
\multicolumn{2}{c|}{$\mathcal{Q}(\text{hextri})$} & 
\multicolumn{2}{c} {$\mathcal{Q}(\text{cmm-pmm})$} \\
$\Delta$ &  
\multicolumn{2}{c|}{$5$} & 
\multicolumn{2}{c|}{$5$} & 
\multicolumn{2}{c|}{$6$} & 
\multicolumn{2}{c} {$8$} \\
\hline 
 & & & & & & & & \\[-4mm]
\multicolumn{1}{r|}{$L$} &  
\multicolumn{1}{c}{$\bm{\mathcal{M}}_\text{stagg}^2$} & 
\multicolumn{1}{c|}{$\bm{\mathcal{M}}_\text{u}^2$} & 
\multicolumn{1}{c}{$\bm{\mathcal{M}}_\text{stagg}^2$} & 
\multicolumn{1}{c|}{$\bm{\mathcal{M}}_\text{u}^2$} & 
\multicolumn{1}{c}{$\bm{\mathcal{M}}_\text{stagg}^2$} & 
\multicolumn{1}{c|}{$\bm{\mathcal{M}}_\text{u}^2$} & 
\multicolumn{1}{c}{$\bm{\mathcal{M}}_\text{stagg}^2$} & 
\multicolumn{1}{c}{$\bm{\mathcal{M}}_\text{u}^2$} \\[1mm] 
\hline
32 &        &        &        &        &        &        & 6.2(1)  & 7.2(1)\\
64 & 5.0(1) & 4.9(1) & 5.0(1) & 5.3(1) & 5.6(1) & 6.3(1) & 6.3(1)  & 7.6(1)\\
128& 5.0(1) & 5.0(1) & 5.1(1) & 5.4(1) & 5.7(1) & 6.5(1) & 6.5(1)  & 8.0(1)\\
256& 5.0(1) & 4.9(1) & 5.1(1) & 5.4(1) & 5.7(1) & 6.5(1) & 6.3(2)  & 7.9(2)\\
384&        &        &        &        &        &        & 6.2(2)  & 7.7(3)\\
512& 4.9(1) & 5.0(2) & 5.0(1) & 5.3(2) & 5.6(1) & 6.5(1) &         &       \\
\hline\hline
\end{tabular}
\caption{Integrated autocorrelation times of the WSK algorithm for the the
zero-temperature critical point of the AF 3-state Potts model on the 
quadrangulations $\Gamma$ of self-dual type considered in this work. 
For each lattice $\Gamma$, we show in the second row its maximum degree 
$\Delta$. We also display the integrated autocorrelation times 
$\tau_{\text{int},\mathcal{O}}$ for the critical observables 
$\mathcal{O}=\bm{\mathcal{M}}_\text{stagg}^2,\bm{\mathcal{M}}_\text{u}^2$ 
as a function of the linear size $L$ (in unit cells). 
The missing entries were not measured. The data for the square 
lattice can be found in \cite{Salas_98}.
}
\label{table_tauint_sd}
\end{table}

It is clear from this table that in all cases, 
$\tau_{\text{int},\bm{\mathcal{M}}_\text{stagg}^2} \lesssim 
 \tau_{\text{int},\bm{\mathcal{M}}_\text{u}^2} \lesssim 8$ uniformly in
$L$, at least for the range of sizes considered in this paper 
$32\le L \le 512$. For each lattice shown in table~\ref{table_tauint_sd}, we
see that both autocorrelation times 
$\tau_{\text{int},\bm{\mathcal{M}}_\text{stagg}^2}$ and 
$\tau_{\text{int},\bm{\mathcal{M}}_\text{u}^2}$ are roughly independent of 
$L$ within errors. This observation gives a strong support to our
conjecture that these integrated autocorrelation times are uniformly bounded
in $L$; i.e., the absence of CSD for the WSK algorithm on this class of 
AF models.  
For the square-lattice case \cite{Salas_98}, another
critical observable (the staggered polarisation) was considered, whose integrated 
autocorrelation time was also uniformly bounded in $L$, but this bound was 
larger than the corresponding magnetisation-square bounds. 
This example illustrates that we cannot rule out the possibility that there 
is another observable whose integrated autocorrelation time is larger than 
those quoted in table~\ref{table_tauint_sd}; but we expect that this 
autocorrelation time will be also uniformly bounded in $L$. If this is
true, then the conclusions discussed above still hold, although the 
uniform bound on $\tau_\text{int}$ will be larger. 

Finally, let us stress the empirical observation that the bounds displayed
in table~\ref{table_tauint_sd} for 
$\tau_{\text{int},\bm{\mathcal{M}}_\text{u}^2}$
seem to (weakly) grow with the maximum degree $\Delta$ of the quadrangulation. 
Notice that for the square lattice with $\Delta=4$, the bound is 
$\tau_{\text{int},\bm{\mathcal{M}}_\text{u}^2} \lesssim 5$ \cite{Salas_98}. 

%
%
\subsubsection{Quadrangulations of non-self-dual type} 
\label{sec.MC_no_selfdual}

\begin{figure}[htb]
\begin{center}
\begin{tabular}{cc}
\includegraphics[width=200pt]{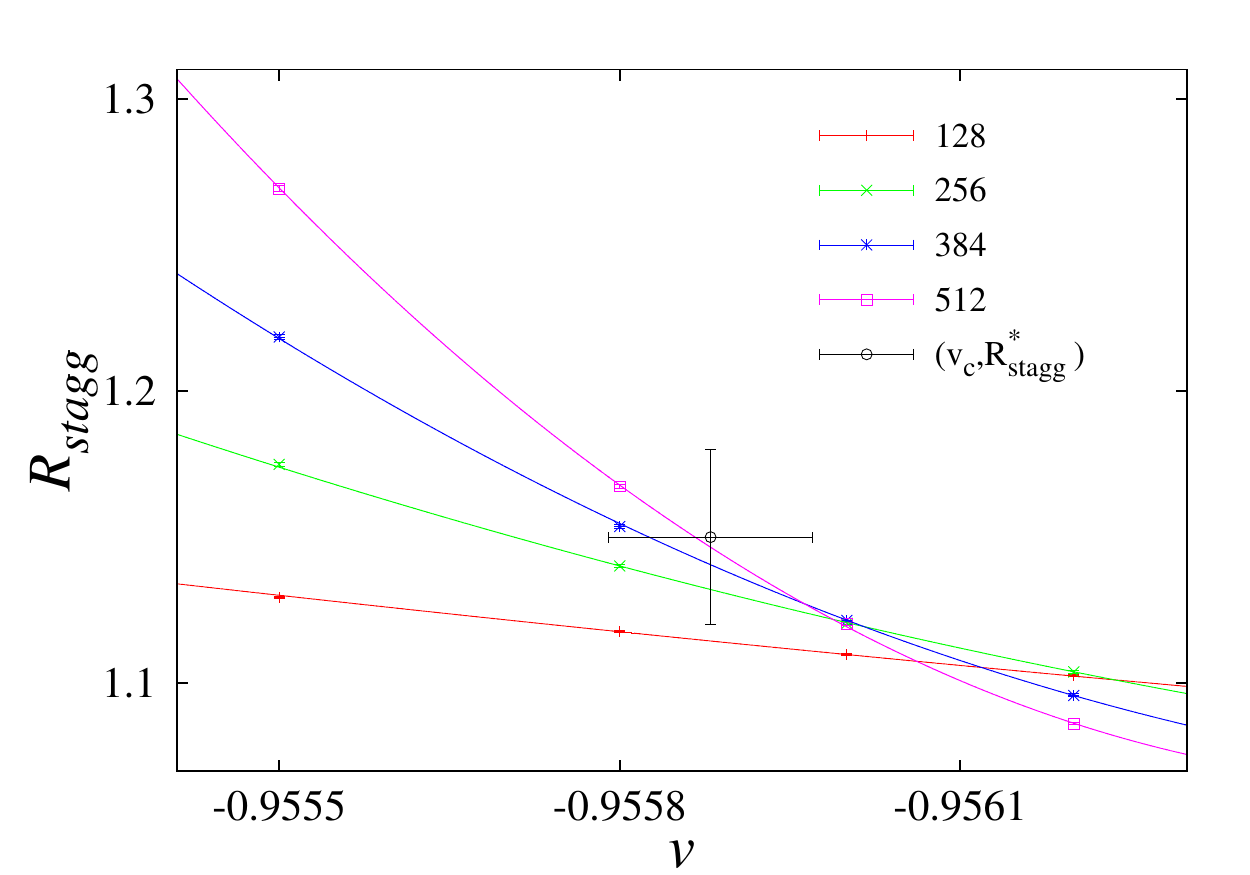} & 
\includegraphics[width=200pt]{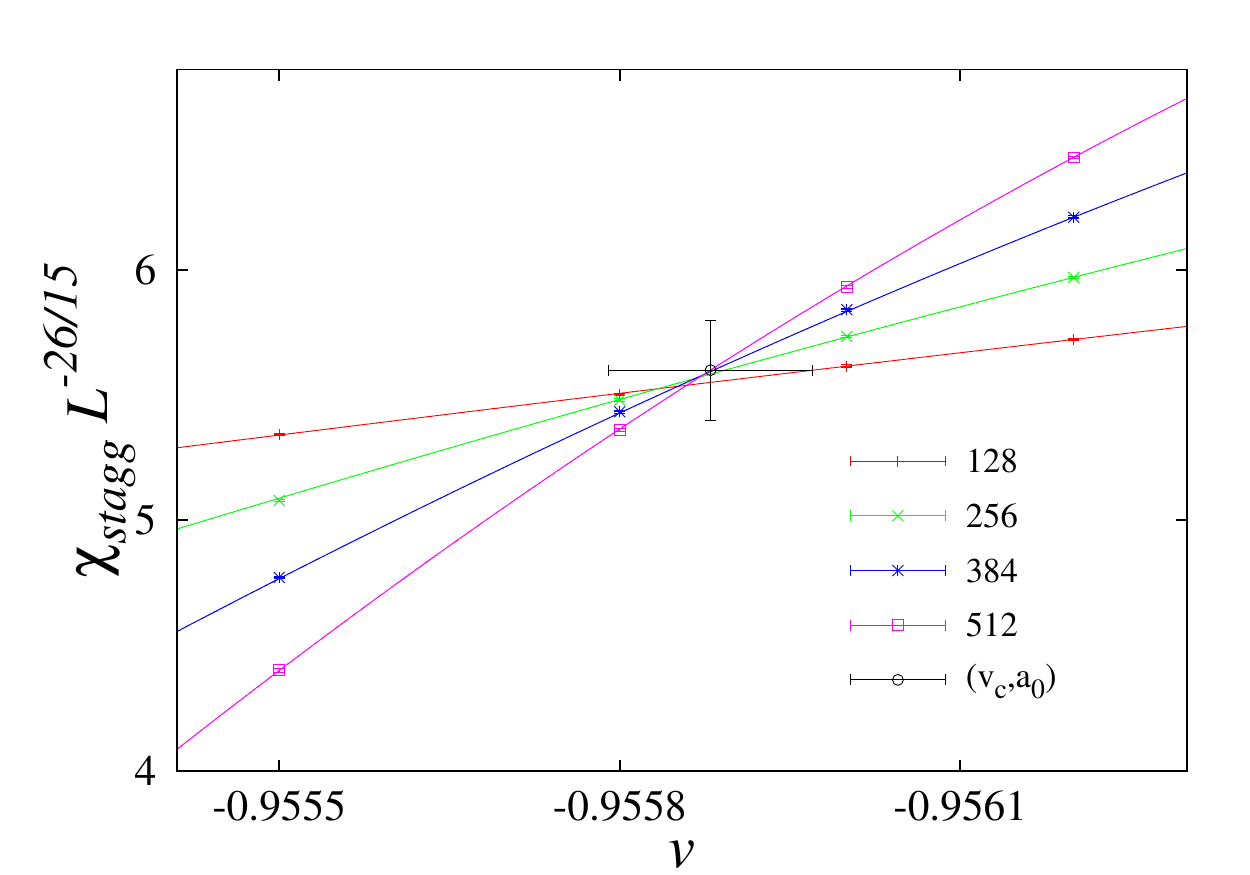} \\[2mm]  
\quad\; (a) &\quad\;  (b) \\ 
\end{tabular}
\end{center}
\vspace*{-5mm}
\caption{\label{fig_Qdiced_sus}
   Binder cumulant $R_\text{stagg}$ (a) and scaled staggered susceptibility
   $L^{-26/15} \chi_\text{stagg}$ (b) as a function of the 
   temperature-like parameter $v$ for a $\mathcal{Q}(\text{ruby})$ lattice
   of dimensions $L\times L$ (unit cells) and periodic boundary conditions.
   We depict the data points for $L=128$ (red), $L=256$ (green), 
   $L=384$ (navy blue), and $L=512$ (pink). 
   The error bars of the points are smaller than the corresponding symbols. 
   The curves correspond to our preferred fit 
   \eqref{def_Ansatz_R}/\eqref{def_Ansatz_susB} with 
   both $1/\nu$ and $\gamma/\nu$ fixed to the exact values for the 
   3-state Potts ferromagnet. 
}
\end{figure}

In this section we will report the numerical results for the seven 
quadrangulations of non-self-dual type that we have simulated  
(see figures~\ref{fig_Q_non-selfdual}--\ref{fig_Q_non-selfdual2}). 
In four cases (namely, $\mathcal{Q}(\text{diced})$, 
$\mathcal{Q}(\text{martini-B})$, $\mathcal{Q}(\text{ruby})$, and 
$G''_3$), we have measured both the static and dynamic observables. For the
other three cases (i.e., $\mathcal{Q}(\text{cross})$, 
$\mathcal{Q}(\text{asanoha})$,
and $G''_2$), we have focused on the static observables. Concerning the
static observables, we have found that the staggered and uniform observables
behave in the same way. Therefore, we will consider the former, and omit
details about the latter. Moreover, as the plots for all lattices are
quite similar, we will refrain from showing all the plots, and display 
only those
for $\mathcal{Q}(\text{ruby})$ in figure~\ref{fig_Qdiced_sus}. (The plot 
of the scaled staggered susceptibility for the lattice 
$\mathcal{Q}(\text{diced})$ has already appeared in \cite[figure~4]{Lv_17}.)  

%
%

First, we have computed the Binder ratio 
$R_\text{stagg}$ (see~\eqref{def_R}). In all cases, we find that the curves
$R_\text{stagg}(v;L)$ cross at a non-zero value of the temperature.  
This common crossing point determines the position of the critical value
$-1< v_c < 0$ for each lattice. This critical value $v_c$, as well as the
critical exponent $1/\nu$, and the FSS correction exponent $\omega$
can be estimated by using {\em Ans\"atze} of the form  
\begin{equation}
R_\text{stagg}(v;L)  \;=\; R_\text{stagg}^*  + a_1\, (v -v_c)\, L^{1/\nu}  +
  a_2\, (v -v_c)^2\, L^{2/\nu} + b\, L^{-\omega} + \cdots 
  \,,
\label{def_Ansatz_R}
\end{equation} 
by omitting various subsets of terms, and varying the smallest value 
$L_\text{min}$ of the data included in the fit. In this way we obtained
stable fits for the parameters and their error bars. In all cases, the 
estimates for $1/\nu$ agree well within statistical errors with the expected 
value. We then redid the fits, with $1/\nu$ \emph{fixed} to its expected value; 
in this way we obtained better estimates for $v_c$. 
In figure~\ref{fig_Qdiced_sus}(a) we show the Binder cumulant 
$R_\text{stagg}(v;L)$ for the quadrangulation $\mathcal{Q}(\text{ruby})$. 
We display our measurements for values of $v$ close enough to the critical 
value $v_c$, and $L=32,64,128,256,512$. Our preferred estimate for the
point $(v_c,R_\text{stagg}^*)$ is also shown.  

In the same way, we fitted the staggered susceptibility $\chi_\text{stagg}$
(see~\eqref{def_sus}). In this case, the form of our {\em Ans\"atze} should be
\begin{multline}
\chi_\text{stagg}(v;L)  \;=\; L^{(\gamma/\nu)_\text{stagg}} \, \Bigl[ 
  a_0  + a_1\,  (v -v_c)\,  L^{1/\nu}  + a_2\,  (v_c -v)^2\,  L^{2/\nu} \\ 
  + b\,  L^{-\omega} + \cdots \Bigr] \,. 
\label{def_Ansatz_susB}
\end{multline}
Again, we checked that the estimates for both $1/\nu$ and $\gamma/\nu$ agree 
well with the expected values within statistical errors. Then, we redid the 
fits with (some of) these parameters fixed to their expected values to obtain 
more precise estimates for $v_c$.
In figure~\ref{fig_Qdiced_sus}(a) we show the scaled staggered susceptibility
$\chi_\text{stagg}(v;L) \, L^{-26/15}$ for the quadrangulation 
$\mathcal{Q}(\text{ruby})$. We display our numerical data for values of $v$ 
sufficiently close to the critical value $v_c$, and $L=32,64,128,256,512$. Our 
preferred estimate for the point $(v_c,\chi_\text{stagg}^* L^{-26/15})$ 
is also shown. Note that the critical exponent 
$(\gamma/\nu)_\text{stagg} = 26/15$ is the one expected 
for the FM 3-state Potts model, and it is \emph{not} the 
one estimated using \eqref{def_Ansatz_susB}. 

The estimates of $1/\nu$ and $v_c$ obtained from 
\eqref{def_Ansatz_R}/\eqref{def_Ansatz_susB} are combined, and the 
final result is displayed in table~\ref{table_results_non_self_dual}. In the
same table we will show the value of $R_\text{stagg}^*$ and 
$(\gamma/\nu)_\text{stagg}$ obtained in \eqref{def_Ansatz_R} and 
\eqref{def_Ansatz_susB}, respectively. The final error bars are obtained by
combining the statistical errors and the systematic errors. These ones 
are computed by comparing different estimates coming from all good fits 
obtained in the above described procedure.

%
%
\begin{table}[htb]
\centering
\begin{tabular}{llrrl}
\hline\hline
\multicolumn{1}{c}{$\Gamma$} &
\multicolumn{1}{c}{$v_c$} & 
\multicolumn{1}{c}{$(\gamma/\nu)_\text{stagg}$} & 
\multicolumn{1}{c}{$\nu$} & 
\multicolumn{1}{c}{$R_\text{stagg}^*$} \\   
\hline
$\mathcal{Q}(\text{ruby})$    & $-0.95588(9)$ & $1.737(5)$             &
                                                 $0.80(5)$  & $1.15(3)$ \\ 
$\mathcal{Q}(\text{diced})$   & $-0.94075(12)$ & $1.737(6)$             &
                                                 $0.83(7)$  & $1.17(2)$ \\ 
\text{diced}                  & $-0.860599(4)$ & $1.737(4)$             &
                                                 $0.81(2)$  & $1.170(7)$\\ 
$G''_2$                       & $-0.82278(4)$ & $1.737(5)$             &
                                                 $0.82(2)$  & $1.17(1)$\\ 
$\mathcal{Q}(\text{cross})$   & $-0.80057(4)$  & $1.736(7)$             &
                                                 $0.85(3)$  & $1.19(2)$ \\
$\mathcal{Q}(\text{martini})$ & $-0.77454(6)$  & $1.735(4)$             &
                                                 $0.83(3)$  & $1.16(1)$ \\
$G''_3$                       & $-0.72278(2)$  & $1.736(4)$             &
                                                 $0.82(2)$  & $1.17(1)$ \\ 
$\mathcal{Q}(\text{asanoha})$ & $-0.72033(3)$  & $1.735(4)$             &
                                                 $0.83(2)$  & $1.16(2)$ \\
\hline
Prediction                    &                & $26/15 \approx 1.73333\,\,$ &
                                                 $5/6   \approx 0.8333\,\,$ & 
                                                 $1.1711(5)$ \\
\hline\hline
\end{tabular}
\caption{Critical temperature $v_c$, critical exponents 
$(\gamma/\nu)_\text{stagg}$ and $\nu$, and critical value of the Binder
cumulant $R_\text{stagg}^*$ for the AF 3-state Potts model  
on quadrangulations of non-self-dual type $\Gamma$. 
We also include for comparison the results for the diced lattice 
\cite{Kotecky-Salas-Sokal}, and the last line (labeled `Prediction')
shows the corresponding values for the FM 3-state Potts model 
\cite{Salas_Sokal_97,Garoni_11}. 
The results for $\mathcal{Q}(\text{ruby})$, $\mathcal{Q}(\text{diced})$, 
$\mathcal{Q}(\text{martini})$, and $G_3''$ have already appeared in 
\cite[table~III]{Lv_17}. 
}
\label{table_results_non_self_dual}
\end{table}

If we look at the estimates displayed in 
table~\ref{table_results_non_self_dual}, we find a rather good agreement with
the prediction of conjecture~\ref{conj.main}(b). Moreover, as all MC 
simulations are statistically independent, we can test the hypothesis that
the values for $(\gamma/\nu)_\text{stagg}$, $\nu$, and $R_\text{stagg}^*$ 
are constant among this class of quadrangulations. A fit to a constant
reveals that
\begin{subequations}
\label{def_res_non_self_dual}
\begin{align}
\left.\frac{\gamma}{\nu}\right|_\text{stagg} &\;=\; 1.736(2)\,, \quad
    \chi^2 \;=\; 0.29\,, \quad \text{DF} \;=\; 7 \,, 
    \quad \text{CL} \;=\; 99.9\%\, \\
\nu &\;=\; 0.823(9)\,, \quad 
    \chi^2 \;=\; 1.68\,, \quad \text{DF} \;=\; 7 \,, 
    \quad \text{CL} \;=\; 97.6\%\, \\
R_\text{stagg}^* &\;=\; 1.168(4)\,, \quad 
    \chi^2 \;=\; 2.53\,, \quad \text{DF} \;=\; 7 \,, 
    \quad \text{CL} \;=\; 92.4\%\, 
\end{align}
\end{subequations}
where DF is the number of degrees of freedom, and CL is the
confidence level of the fit. The estimates agree within less than 1.5 
standard deviations from those on the row labeled `Prediction' in 
table~\ref{table_results_non_self_dual}. Therefore, we have found that
for all the quadrangulations of non-self-dual type considered in this work,
the phase transition separating the low- and high-temperature phases is 
of second order, and that this finite-temperature critical point belongs 
to the universality class of the FM 3-state Potts model (i.e., identical 
critical exponents and universal amplitudes). Therefore, all
cases studied in this section are in full agreement with 
conjecture~\ref{conj.main}(b).  

\begin{figure}[htb]
\begin{center}
\begin{tabular}{cc}
\includegraphics[width=150pt]{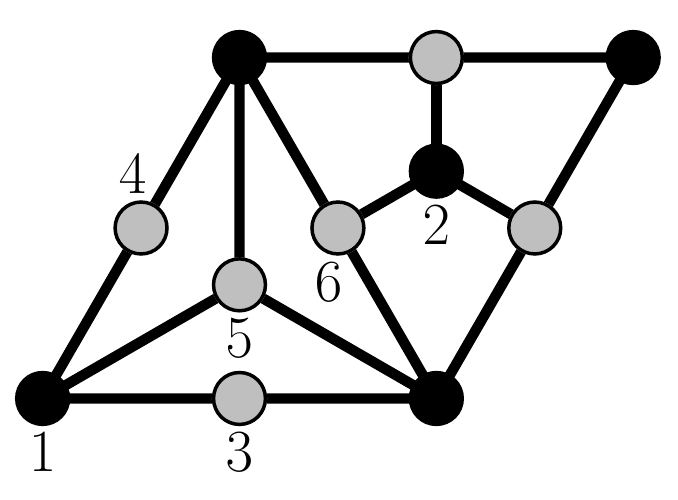} &  
\includegraphics[width=100pt]{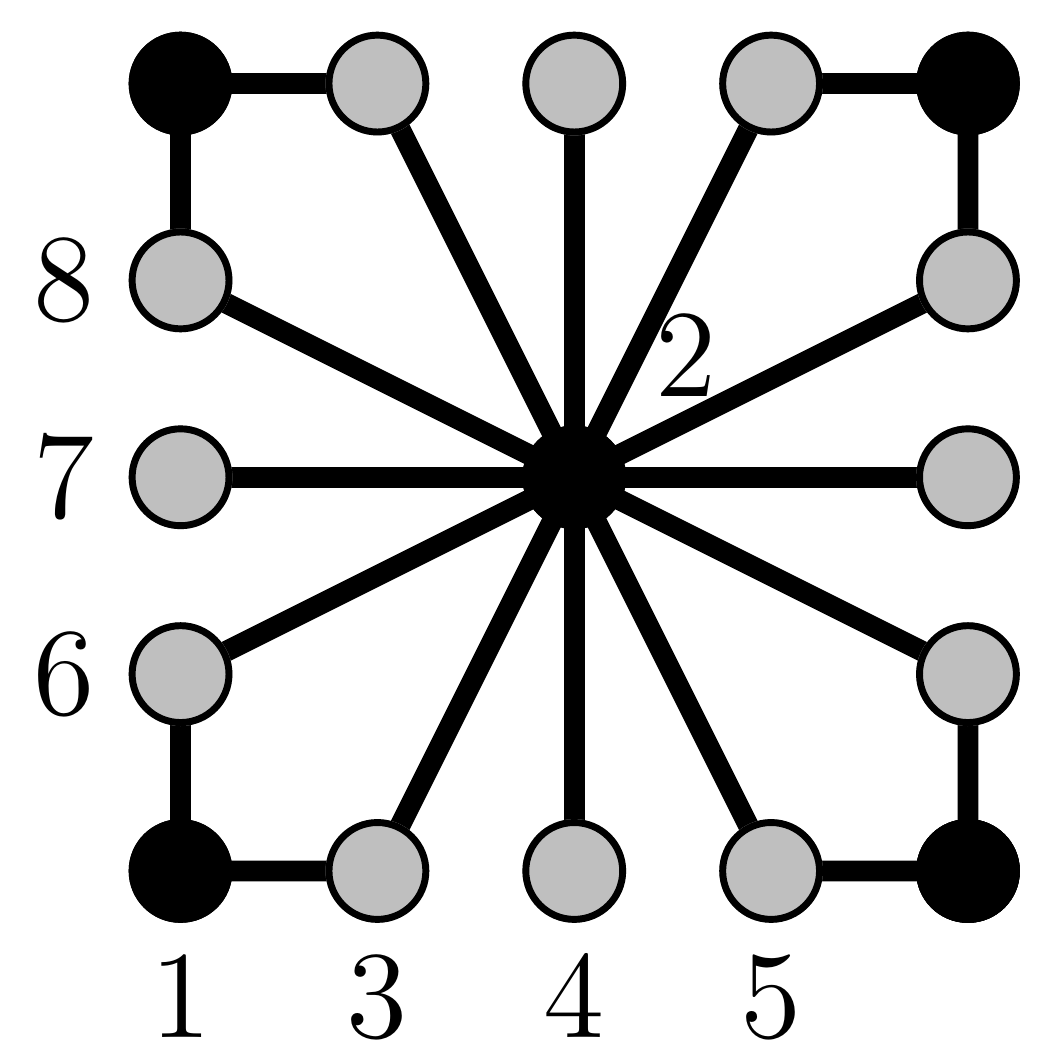} \\
 \hspace*{-1.6cm}(a)  & \,(b) \\[2mm] 
\includegraphics[width=150pt]{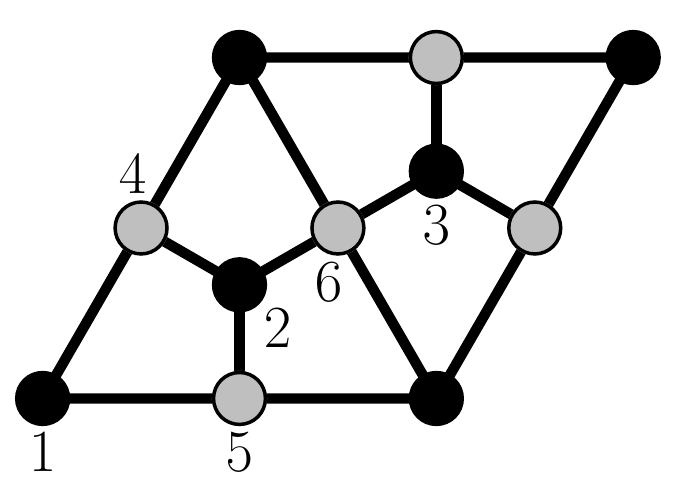} & 
\includegraphics[width=150pt]{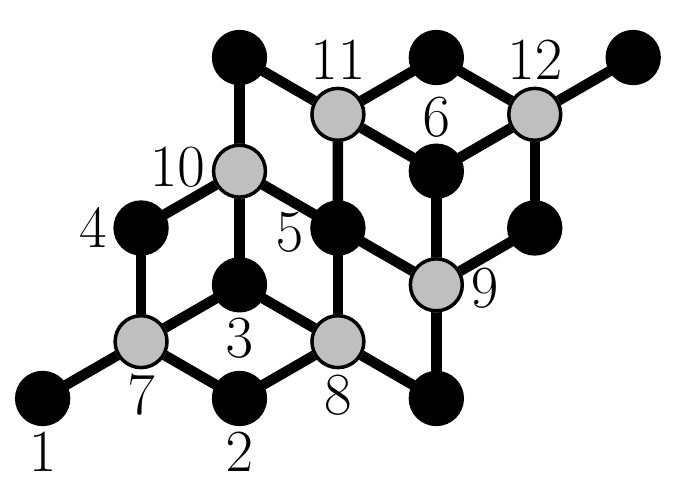} \\ 
 \hspace*{-1.6cm}(c)   & \hspace*{-1.6cm} (d)
\end{tabular}
\end{center}
\vspace*{-5mm}
\caption{\label{fig_unit_cells}
   Unit cells for some of the quadrangulations $\Gamma$ of non-self-dual type 
   depicted in figures~\ref{fig_Q_non-selfdual} and~\ref{fig_Q_non-selfdual2}:
   (a) $\mathcal{Q}(\text{martini})$; 
   (b) $G''_3$;
   (c) $\mathcal{Q}(\text{diced})$; and  
   (d) $\mathcal{Q}(\text{ruby})$. 
   Each quadrangulation $\Gamma$ is associated to a dual pair $(G_0,G_1)$.
   The black (resp.\/ gray) vertices belong to $G_0$ (resp.\/ $G_1$). 
   In each unit cell, the numbers label the different sub-sublattices it
   contains. 
}
\end{figure}

Finally, let us discuss in detail what happens for the special cases 
$\mathcal{Q}(\text{diced})$ and $\mathcal{Q}(\text{ruby})$. 
Please recall that for these two lattices, both sublattices $G_0$ and $G_1$
have the same number of vertices, and the general argument leading to 
conjecture~\ref{conj.main}(b) does not work in a straightforward sense. 
We are now interested in investigating the typical ordering in each sublattice.
To achieve this goal, we will consider all possible sub-sublattices of each of
four quadrangulations of non-self-dual type studied in this section (namely,
$\mathcal{Q}(\text{diced})$, $\mathcal{Q}(\text{ruby})$, 
$\mathcal{Q}(\text{martini})$, and $G''_3$), and measure their magnetisation. 
These sub-sublattices are depicted in figure~\ref{fig_unit_cells}. 

%
%

The vertex set of each one of these four lattices $\Gamma$ can be partitioned
into $n_\Gamma$ disjoint sets $V_s$ (see section~\ref{sec.quad}). Then, 
for each sub-sublattice $s$, we can define the vector magnetisation
\begin{equation}
\bm{\mathcal{M}}^{(s)} \;=\; \sum\limits_{\alpha=1}^3 m_\alpha^{(s)} \, 
   \bm{e}^{(\alpha)} \,,
\label{def_mag_sublattice}
\end{equation} 
where the unit vectors $\bm{e}^{(\alpha)}$ are given in 
\eqref{eq_e_alpha}, and 
\begin{equation}
m_\alpha^{(s)} \;=\; \frac{1}{|V_s|} \sum\limits_{x\in V_s} 
      \delta_{\sigma_x,\alpha} 
\end{equation}
is the fraction of the spins in $V_s$ taking the value $\alpha \in\{1,2,3\}$. 
We then define the mean squared magnetisation density (MSMD) 
for each sub-sublattice $s$ as:
\begin{equation}
 M^2_s \;=\; \< \bm{\mathcal{M}}^{(s)}  \cdot \bm{\mathcal{M}}^{(s)} \>\,.
\label{def_sus_sublattice}
\end{equation} 
This quantity measures the FM order of the spins within each sub-sublattice 
$s$. If the spins in $s$ are completely FM ordered (i.e., they take a 
common value), then $M^2_s =1$; if the spins in $s$ take randomly two 
distinct values, then $M^2_s =1/4$; and if the spins in $s$ 
are completely uncorrelated, then $M^2_s=0$. 

The unit cell of the lattice $\mathcal{Q}(\text{martini})$ is shown in 
figure~\ref{fig_unit_cells}(a). The black (resp.\/ gray) dots
represent the vertices in $G_0$ (resp.\/ $G_1=\text{martini}$), and the 
labels refer to its six different sub-sublattices.
The sublattice $G_0$ (resp.\/ $G_1$) contains $|V_0|=2 L^2$ (resp.\/ 
$|V_1|=4L^2$) vertices, where $L$ is the linear size of the underlying 
triangular lattice. 
The sublattice with less vertices $G_0$ contains two sub-sublattices: 
one with degree-9 vertices (labeled~1), and another with degree-3 vertices 
(labeled~2). The corresponding  MSMD are $M_1^2 \approx 0.97$ and 
$M_2^2 \approx 0.48$. $G_1$ can be decomposed into four sublattices 
(each of them with degree-3 vertices and labeled~3, 4, 5, and 6). For all of 
them, $M_s^2 \approx 0.25$. 

The unit cell of the lattice $G''_3$ is depicted in 
figure~\ref{fig_unit_cells}(b). In this case $G_0=G'_2$ contains $2L^2$ 
vertices, while $G_1$ (= decorated 4--8 lattice) has $6L^2$ vertices, 
where $L$ is the linear size of the underlying square lattice. 
$G_0$ can be split into two sub-sublattices: 
one with degree-12 vertices (labeled~2) and $M_2^2\approx 0.99$, and another
with degree-4 vertices (labeled~1) and $M_1^2\approx 0.69$. Moreover,
$G_1$ can also be split into two degree-2 sublattices (labeled~4 and 7) and
four degree-3 sublattices (labeled 3, 4, 6, and 8). For all of them,
$M^2_s \approx 0.25$. 

In both lattices, $|V_0| < |V_1|$, and the behavior of the MSMD for each 
sub-sublattice agrees qualitatively with our conclusions based on the height 
representation: the spins living on the sublattice with less vertices are 
FM ordered, while on the other sublattice, the spins take randomly the other 
two values. Notice however, that the ordering in $G_0$ is not uniform: the 
sub-sublattice with vertices of higher degree is almost completely ordered, 
while the other sub-sublattice is only partially ordered (i.e., 
$1/4 < 0.48 \lesssim  M_a^2  \lesssim 0.69 < 1$). On the other hand, all the
sub-sublattices in $G_1$ take two random values $M_a^2 \approx 1/4$. This 
difference with respect to the scenario presented in 
section~\ref{sec.height_rep} is probably due to fluctuations around the 
three ideal states (which were ignored in section~\ref{sec.height_rep}).

Let us now consider the other two cases for which $|V_0| = |V_1|$. 
The unit cell of the lattice $\mathcal{Q}(\text{diced})$ is shown in 
figure~\ref{fig_unit_cells}(c). In this case, both sublattices
have $3L^2$ vertices. The diced sublattice $G_0$ contains a 
degree-6 sublattice (labeled~1), and two degree-3 sublattices 
(labeled 2 and 3). The corresponding MSMD are $M_1^2\approx 0.49$ and 
$M_2^2 \approx M_3^2 \approx 0.28$, respectively. The kagome sublattice $G_1$ 
contains three degree-4 sublattices (labeled 4 to 6) and for all of them, 
we obtain $M^2_s \approx 0.20$. 

Finally, the unit cell of the lattice $\mathcal{Q}(\text{ruby})$ is shown in
figure~\ref{fig_unit_cells}(d). In this case both sublattices have 
$6L^2$ vertices. The $G_0 = \mathcal{Q}(\text{diced})$ sublattice  
contains a degree-6 sub-sublattice (labeled~1), three degree-4 sub-sublattices
(labeled 2, 4, and 5), and two degree-3 sub-sublattices (labeled 3 and 6).
The corresponding MSMD are $M_1^2\approx 0.47$, 
$M_2^2 \approx 0.32$, and $M_3^2 \approx 0.22$ (all sub-sublattices with the
same degree have approximately the same MSMD). The ruby sublattice $G_1$ has
six degree-4 sub-sublattices (labeled 7 to 12) with $M_7^2 \approx 0.20$. 

We observe that, when $|V_0| = |V_1|$, the same conclusions hold qualitatively;
but there are significant quantitative changes. First of all, the most ordered
sub-sublattice is that of $G_0$ with the largest degree (as in the generic 
case); but now it is only partially FM ordered: 
$1/4 < 0.48 \approx  M_a^2 < 1$. Secondly, the other sub-sublattices are 
less ordered: $M_a^2$ is in the range $0.22$--$0.28$, which is roughly 
speaking similar to $1/4$. This means that the spins in those sub-sublattices 
basically take two values at random. Finally, all sub-sublattices of $G_1$ 
are less ordered than for the generic case: $M_a^2 \approx 0.20 < 1/4$. 
Again these differences could be attributed to fluctuations around the 
three ideal states; but for this particular case ($|V_0| = |V_1|$) these
fluctuations look larger than for the generic one ($|V_0| < |V_1|$).
In conclusion, we see that, even for the most involved cases, the sublattice 
$G_0$ containing the sub-sublattice of largest degree is the one that is more 
ordered; while the other sublattice $G_1$ is disordered. These results 
give a stronger empirical support to conjecture~\ref{conj.main}(b). 

%
%
\begin{table}[htb]
\centering
\begin{tabular}{r|rr|rr|rr|rr}
\hline\hline
$\Gamma$ & 
\multicolumn{2}{c|}{$\mathcal{Q}(\text{diced})$} & 
\multicolumn{2}{c|}{$\mathcal{Q}(\text{martini})$} & 
\multicolumn{2}{c|}{$G''_3$} & 
\multicolumn{2}{c} {$\mathcal{Q}(\text{ruby})$} \\
\hline 
$v$ &  
\multicolumn{2}{c|}{$-0.94071$} & 
\multicolumn{2}{c|}{$-0.77454$} & 
\multicolumn{2}{c|}{$-0.72278$} & 
\multicolumn{2}{c} {$-0.95588$} \\
 & & & & & & & & \\[-4mm]
\multicolumn{1}{r|}{$L$} &  
\multicolumn{1}{c}{$\bm{\mathcal{M}}_\text{stagg}^2$} & 
\multicolumn{1}{c|}{$\bm{\mathcal{M}}_\text{u}^2$} & 
\multicolumn{1}{c}{$\bm{\mathcal{M}}_\text{stagg}^2$} & 
\multicolumn{1}{c|}{$\bm{\mathcal{M}}_\text{u}^2$} & 
\multicolumn{1}{c}{$\bm{\mathcal{M}}_\text{stagg}^2$} & 
\multicolumn{1}{c|}{$\bm{\mathcal{M}}_\text{u}^2$} & 
\multicolumn{1}{c}{$\bm{\mathcal{M}}_\text{stagg}^2$} & 
\multicolumn{1}{c}{$\bm{\mathcal{M}}_\text{u}^2$} \\[1mm] 
\hline
16 &        &        &        &        &        &        & 9.3(1) &10.2(1)\\
24 &        &        &        &        &19(3)   &18.3(1) &        &       \\
32 &11.7(1) &13.3(1) &21.1(1) & 15.4(1)&27(7)   &22.0(2) &11.6(1) &13.6(1)\\
40 &        &        &        &        &31(7)   &24.9(2) &        &       \\
48 &        &        &        &        &28(14)  &27.6(3) &        &       \\
56 &        &        &        &        &21(8)   &30.3(2) &        &       \\
64 &14.6(1) &16.7(1) &30.1(2) & 27.0(2)&49(33)  &32.3(4) &14.1(1) &16.4(1)\\
72 &        &        &        &        &36(22)  &34.0(4) &        &       \\
128&18.8(1) &20.9(2) &42.7(4) & 40.3(4)&57(29)  &46.1(5) &17.6(3) &19.9(3)\\
256&25.2(2) &27.1(2) &59.5(9) & 57.2(8)&        &65(1)   &22.8(4) &24.8(4)\\
384&        &        &        &        &        &        &27.5(5) &29.4(6)\\
512&33.3(9) &35.9(7) &88(3)   & 79(3)  &        &87(3)   &31(2)   &33(2)  \\
\hline\hline
\end{tabular}
\caption{Integrated autocorrelation times of the WSK algorithm for the the
finite-temperature critical point of the AF 3-state Potts model on four of the 
quadrangulations $\Gamma$ of non-self-dual type considered in this work. 
For each lattice $\Gamma$, we show the integrated autocorrelation times 
$\tau_{\text{int},\mathcal{O}}$ for the observables 
$\mathcal{O}=\bm{\mathcal{M}}_\text{stagg}^2,\bm{\mathcal{M}}_\text{u}^2$ 
as a function of the linear size $L$ (in unit cells). The second row (labeled
`$v$') shows the temperature used to measure these estimates.  
The missing entries were not measured or could not be estimated (like the 
$L=256,512$ entries in the first column for $G''_3$). 
}
\label{table_tauint_nsd}
\end{table}

Now let us take a look at the dynamic behaviour of the WSK algorithm for the
AF 3-state Potts model on the four quadrangulations considered above:
$\mathcal{Q}(\text{martini})$, $G''_3$, $\mathcal{Q}(\text{diced})$ and
$\mathcal{Q}(\text{ruby})$. We have measured the integrated autocorrelation
times $\tau_{\text{int},\mathcal{O}}$ for the operators 
$\mathcal{O}=\bm{\mathcal{M}}_\text{stagg}^2,\bm{\mathcal{M}}_\text{u}^2$.
The raw data is displayed in table~\ref{table_tauint_nsd}. The second row of
this table shows the temperature $v$ used in the MC simulations to estimate
these autocorrelation times. Notice that this temperature is our preferred  
estimate for $v_c$, except for the lattice $\mathcal{Q}(\text{diced})$. In
this case $v=-0.94071$, which differs from $v_c=-0.94075(12)$ by just one 
third of one standard deviation for the latter value.

Table~\ref{table_tauint_nsd} shows that the autocorrelation time for 
$\bm{\mathcal{M}}_\text{stagg}^2$ (resp.\/ $\bm{\mathcal{M}}_\text{u}^2$)
is greater than the other one for $\mathcal{Q}(\text{martini})$ and
$G''_3$ (resp.\/ $\mathcal{Q}(\text{diced})$ and $\mathcal{Q}(\text{ruby})$).
It seems that for those generic quadrangulations of non-self-dual type
(with $|V_0| < |V_1|$) the slowest mode is $\bm{\mathcal{M}}_\text{stagg}^2$,
while for those quadrangulations with $|V_0| = |V_1|$, the slowest mode 
is $\bm{\mathcal{M}}_\text{u}^2$. 

We have fitted the data of table~\ref{table_tauint_nsd} to a power-law 
\emph{Ansatz}: 
\begin{equation}
\tau_{\text{int},\mathcal{O}} \;=\; A_\mathcal{O}  + B_\mathcal{O} \, 
L^{z_{\text{int},\mathcal{O}}} \,,
\label{def_Ansatz_z}
\end{equation}
where $z_{\text{int},\mathcal{O}}$ is the corresponding dynamical critical
exponent. Again, as a precaution against FSS artifacts, we have systematically
varied the minimum value $L_\text{min}$ of the data included in the fit.
In some cases, a better and more stable fit is obtained by fixing 
$A_\mathcal{O}=0$. In this way we obtain the results displayed in
table~\ref{table_z_non_self_dual}. 
For completeness, we have also added data
for the diced lattice. In this case, we have used (unpublished) dynamic data 
corresponding to \cite{Kotecky-Salas-Sokal}, and the analysis was performed 
by using the FSS \emph{Ansatz} \eqref{def_Ansatz_susB} adapted to
\eqref{def_Ansatz_z}. 

%
%
\def\kk{$\phantom{1}$}
\begin{table}[htb]
\centering
\begin{tabular}{lllrlrl}
\hline\hline
\multicolumn{1}{c}{$\Gamma$} & 
\multicolumn{1}{c}{$\mathcal{O}$}  &  
\multicolumn{1}{c}{$z_{\text{int},\mathcal{O}}$}  &  
\multicolumn{1}{c}{$L_\text{min}$} &  
\multicolumn{1}{c}{$\chi^2$} &  
\multicolumn{1}{c}{DF} &  
\multicolumn{1}{c}{CL} \\ 
\hline \\[-4mm]
$\mathcal{Q}(\text{diced})$ &  
$\bm{\mathcal{M}}_\text{stagg}^2$ & $0.55(3)$ & 
$32$ & \kk$1.59$ & $2$ & $45\%$ \\[1mm] 
 & $\bm{\mathcal{M}}_\text{u}^2$ & $0.45(4)$ & 
$32$ & \kk$2.17$ & $2$ & $34\%$ \\ 
\hline \\[-4mm]  
$\mathcal{Q}(\text{martini})$& 
$\bm{\mathcal{M}}_\text{stagg}^2$ & $0.49(3)$ & 
$32$ & \kk$1.78$ & $2$ &  $41\%$ \\[1mm]  
 & $\bm{\mathcal{M}}_\text{u}^2$& $0.50(2)$ & 
$128^\dagger$ & \kk$0.32$ & $1$ & $57\%$ \\ 
\hline \\[-4mm] 
$G''_3$ & 
$\bm{\mathcal{M}}_\text{stagg}^2$ & $1.0(6)$  & 
$48^\dagger$ & \kk$0.88$ & $3$ & $83\%$ \\[1mm]  
 & $\bm{\mathcal{M}}_\text{u}^2$ & $0.48(2)$ & 
$128^\dagger$ & \kk$1.21$ & $1$ & $27\%$ \\ 
\hline \\[-4mm] 
$\mathcal{Q}(\text{ruby})$  & 
$\bm{\mathcal{M}}_\text{stagg}^2$ & $0.46(6)$ & 
$256^\dagger$ & \kk$0.033$ & $1$ & $86\%$ \\[1mm] 
 & $\bm{\mathcal{M}}_\text{u}^2$ & $0.42(6)$ & 
$256^\dagger$ & \kk$0.0054$ & $1$ & $94\%$ \\ 
\hline \\[-4mm] 
$\text{diced}$ \cite{Kotecky-Salas-Sokal} & 
$\bm{\mathcal{M}}_\text{stagg}^2$ & $0.48(2)$ & 
$48$ & $15.82$ & $16$ & $47\%$ \\[1mm] 
 & $\bm{\mathcal{M}}_\text{u}^2$ & $0.47(7)$ & 
$192$ & \kk$3.71$ & $8$ & $88\%$ \\ 
\hline\hline
\end{tabular}
\caption{Dynamic critical exponents 
$z_{\text{int},\bm{\mathcal{M}}_\text{stagg}^2}$ and 
$z_{\text{int},\bm{\mathcal{M}}_\text{u}^2}$ for the WSK algorithm for the
AF 3-state Potts model on quadrangulations of non-self-dual type $\Gamma$. 
The estimates for these critical exponents have been obtained using the
\emph{Ansatz} \eqref{def_Ansatz_z} using data with $\ge L_\text{min}$. 
For each estimate, we show the corresponding values of $\chi^2$, the 
number of degrees of freedom (DF) and the confidence level (CL) of the fit.
The symbol ${}^\dagger$ in the value of $L_\text{min}$ means that the fit 
has been done with $A_\mathcal{O}=0$. 
For the diced lattice, we have used 
(unpublished) dynamic data from \cite{Kotecky-Salas-Sokal}, and an FSS
\emph{Ansatz} similar to \eqref{def_Ansatz_susB}/\eqref{def_Ansatz_z}
for the corresponding autocorrelation times. 
}
\label{table_z_non_self_dual}
\end{table}

A closer look at this table reveals that all dynamic critical exponents are 
roughly consistent within errors. A fit to a constant shows that this 
observation is true:  
\begin{subequations}
\label{res_z_non_self_dual}
\begin{align}
z_{\text{int}, \bm{\mathcal{M}^2_\text{stagg}}} &\;=\; 0.50(1)\,, \quad
    \chi^2 \;=\; 4.98\,, \quad \text{DF} \;=\; 4 \,, 
    \quad \text{CL} \;=\; 47\%\,, \\
z_{\text{int}, \bm{\mathcal{M}^2_\text{u}}} &\;=\; 0.48(2)\,, \quad
    \chi^2 \;=\; 3.02\,, \quad \text{DF} \;=\; 4 \,, 
    \quad \text{CL} \;=\; 56\%\,.
\end{align}
\end{subequations}
Finally, these two values are also consistent within $1$--$1.5$ 
standard deviations. If these exponents are in fact equal,
then our preferred estimate (taking into account the statistical 
non-independence of the two estimates) would be
\begin{equation}
z_{\text{int}} \;=\; 0.49(2)\,.  
\label{res_z_OK}
\end{equation}
This value is close to the estimate of the dynamic critical exponent for the 
SW \cite{SW_87} algorithm for the square-lattice FM 3-state Potts model 
$z_{\text{int},\bm{\mathcal{M}^2}}=0.475(6)$ \cite{Salas_Sokal_97,Garoni_11}.

%
%
\subsection{Transfer matrices} \label{sec.TM}

In this section we will show some results using the TM approach in the 
FK representation for the quadrangulations $\mathcal{Q}(\text{hextri})$ and
$\mathcal{Q}(\text{house})$ of self-dual type. 
In particular, our goal is to show numerically that 
$(q,v)=(3,-1)$ is a critical point with central charge $c=1$, in agreement  
with conjecture~\ref{conj.main}(a). Therefore, we will focus on the 
chromatic polynomial case $v=-1$, and see how the free energy behaves as
$q$ is varied in an interval around $q=3$. 
We will consider strip graphs of this lattice with
cylindrical boundary conditions; i.e., periodic (resp.\/ free) 
boundary conditions in the transverse (resp.\/ longitudinal) direction. 
The TM for a strip graph of width $L$, and the chromatic polynomial 
$Z_{L\times N}(q,-1)$ for a strip graph of size $L\times N$ can be obtained 
in terms of join $\mathsf{J}_{ij}$ and detach $\mathsf{D}_j$ operators 
(acting on a suitable connectivity space), as it is shown in detail, e.g., 
in \cite{transfer1,transfer2,JSS_tri}. 

We have chosen cylindrical boundary conditions because they are easier 
to deal with, so they allow us to study strip graphs with larger widths 
$L\le 14$; but the limiting curves $\mathcal{B}_L$ (in the complex $q$-axis) 
along which the chromatic zeros accumulate in the limit $N\to \infty$, 
may or may not cross the real $q$ axis. For instance, for $L=4,12$, 
the finite-size estimates 
$q_c(L)$ for the parameter $q_c(\mathcal{Q}(\text{hextri}))$ do not exist.
(Toroidal boundary conditions \cite{JS_torus} are expected to provide
estimates $q_c(L)$ for any $L$ large enough, and these estimates are also
expected to converge faster to their thermodynamic limit $q_c$; but are
technically more difficult to handle and the range of amenable widths 
is therefore smaller: in particular $L\le 8$ for 
$\mathcal{Q}(\text{hextri})$.)   
 
%
%
\subsubsection[Transfer matrix for Q(hextri)]%
  {Transfer matrix for $\bm{\mathcal{Q}(\text{hextri})}$} 
\label{sec.TM_Qhextri}

\begin{figure}[htb]
\begin{center}
\begin{tabular}{cc}
\includegraphics[width=200pt]{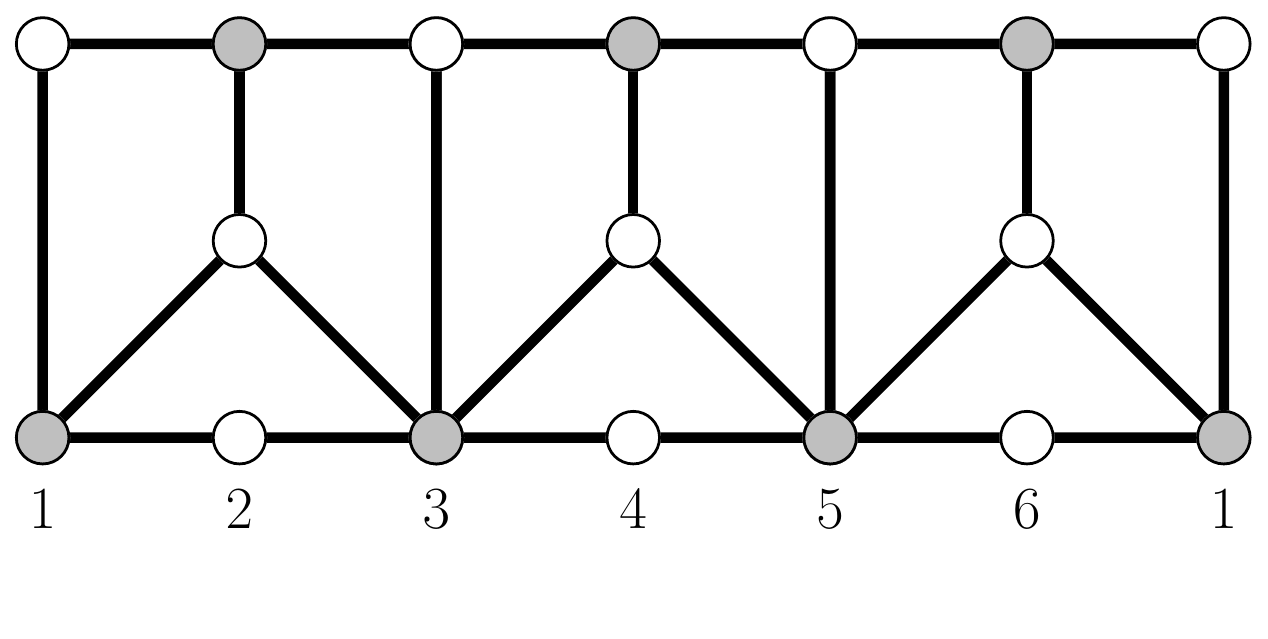} & 
\quad \includegraphics[width=100pt]{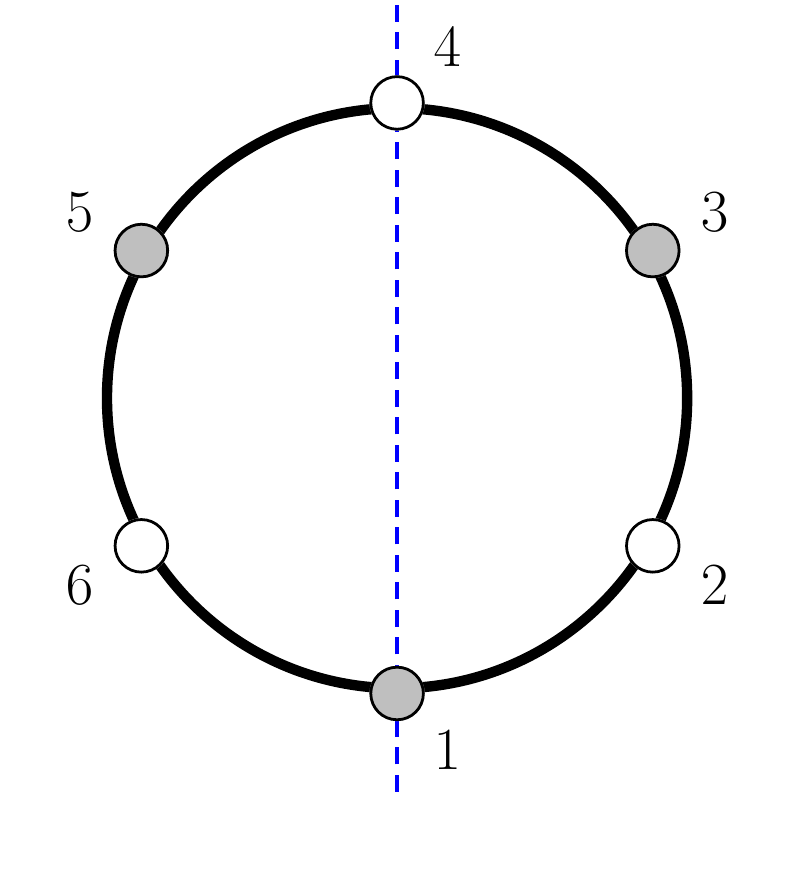} \\[2mm]  
\quad\; (a) &\quad\;  (b) \\ 
\end{tabular}
\end{center}
\vspace*{-0.5cm}
\caption{\label{fig_qhextri_tm}
   (a) TM construction for a layer of a strip graph of the 
   $\mathcal{Q}(\text{hextri})$ lattice of width $L=6$ and cylindrical  
   boundary conditions. There are two classes of 
   nonequivalent vertices, depicted respectively, as white and gray dots.
   (b) Another representation of the bottom row of panel (a) so that the
   periodic boundary conditions in the transverse direction are evident. 
   The strip graph is symmetric with respect to the vertical dashed line 
   going through vertices $1$ and $1+L/2=4$. 
}
\end{figure}

Let us start with the TM description of the $\mathcal{Q}(\text{hextri})$
lattice (see figure~\ref{fig_qhextri_tm}). Notice that for this lattice,  
the width $L$ should be an even integer, and there are two different classes 
of vertices (depicted as white and gray dots in figure~\ref{fig_qhextri_tm}). 
The building of the chromatic-polynomial TM for this strip graph with 
cylindrical boundary conditions is shown in figure~\ref{fig_qhextri_tm}(a). 
This matrix can be written simply as 
\begin{equation}
  \mathsf{T} \;=\; \mathsf{V} \cdot \mathsf{H} \cdot \mathsf{V}_0^\text{(even)}
     \cdot \mathsf{H} \,,
\label{def_tm_qhextri}
\end{equation}
where the operators act from right to left. The usual `horizontal' 
$\mathsf{H}$ and `vertical' $\mathsf{V}$ operators can be decomposed 
as products of sparse matrices 
\begin{equation}
  \mathsf{H} \;=\; \left(\prod_{i=1}^{L-1} \mathsf{Q}_{i,i+1} \right) \cdot 
                   \mathsf{Q}_{L,1}\,, \qquad 
  \mathsf{V} \;=\; \prod_{i=1}^L \mathsf{P}_i  \,, 
\label{def_h_v} 
\end{equation}
which in turn, are written in terms of the join/detach operators:
\begin{equation}
  \mathsf{Q}_{ij} \;=\;   \mathds{1}  - \mathsf{J}_{ij} \,, \qquad 
  \mathsf{P}_i    \;=\; - \mathds{1}  + \mathsf{D}_i \,,  
\label{def_q_p} 
\end{equation}
where $ \mathds{1}$ is the identity operator. The operator 
$\mathsf{V}_0^\text{(even)}$ is just the product of detach operators 
on the sites labelled by an even number:
\begin{equation}
  \mathsf{V}_0^\text{(even)} \;=\; \prod_{i=1}^{L/2} \mathsf{D}_{2i} \,. 
\label{def_v0} 
\end{equation}
 
%
%

These operators act on the space of non-crossing connectivities for $L$
points on a circle. Its dimension is given by the Catalan number $C_L$
\cite[table~2]{transfer1}. We can reduce this dimension by taking into account
that we are dealing with the chromatic polynomial, so we only need 
non-crossing non-nearest-neighbour connectivities, whose number is
given by the Riordan numbers $R_L$ for $L\ge 2$ \cite[table~2]{transfer1}. 
We can reduce this dimension by taking also into account the symmetries of
the $\mathcal{Q}(\text{hextri})$ strip: i.e. rotations by two units (because
of the non-equivalence of all points in the strip) and reflections by a 
line going through the vertices labelled $1$ and $1+L/2$ [see 
figure~\ref{fig_qhextri_tm}(b)].  
Once all these symmetries are accounted for, the dimension of the
corresponding connectivity space is given by $1$, $3$, $7$, $24$, $87$,
$437$, $2379$, and $14843$ for $L=2,4,\ldots,16$. 

We have implemented this procedure by writing a {\sc perl} script, and we
have been able to compute the TM for even widths $2\le L \le 14$. 
We have checked the results by computing symbolically the 
chromatic polynomial for strip graphs of size $L \times N$ 
for $2\le N\le N_\text{max}$, and comparing them to the results 
provided by other well-tested symbolic programs \cite{Bedini,HPR}. 
The agreement is indeed perfect. For $L\le 10$, $N_\text{max}=20$; but for 
larger values of the width, the above-mentioned programs were unable to obtain 
results for some `large' values of $N$ in a sensible amount of CPU time. 
In particular, for $L=12$, $N_\text{max}=18$, and for $L=14$, $N_\text{max}=4$. 

Then, for each width $L$, we have computed the free energy in the interval
$q\in [2,4]$ in terms of the dominant eigenvalue $\lambda_*$ of the 
corresponding TM:
\begin{equation}
f_L(q,-1) \;=\; \frac{1}{L} \, \log |\lambda_*|  \,.
\label{def.f_TM}
\end{equation}
We can extract the estimate for the central charge $c_L(q,-1)$ of a 
critical model by fitting the corresponding finite-size free energy 
\eqref{def.f_TM} to the CFT {\em Ansatz} 
\begin{equation}
f_L(q,-1)  \;=\; f_\text{bulk}(q,-1) + \frac{ G\, \pi\, c_L(q,-1)}{6\, L^2}\,,  
\label{def.f_Ansatz}
\end{equation}
where $G=4/\sqrt{3}$ is the geometrical factor for the 
$\mathcal{Q}(\text{hextri})$ lattice when using the above construction. 

\bigskip

\noindent
{\bf Remark.} Our definition of the free energy density \eqref{def.f_TM}
does \emph{not} coincide with the usual definition of the free energy 
density per vertex by a factor of $2$. This implies that the geometric factor 
$G$ is different for the one obtained in this latter case. See 
appendix~\ref{sec.app} for the computation of $G$.

\bigskip 
%
%
\begin{figure}[htb]
\begin{center}
\begin{tabular}{cc}
\includegraphics[width=200pt]{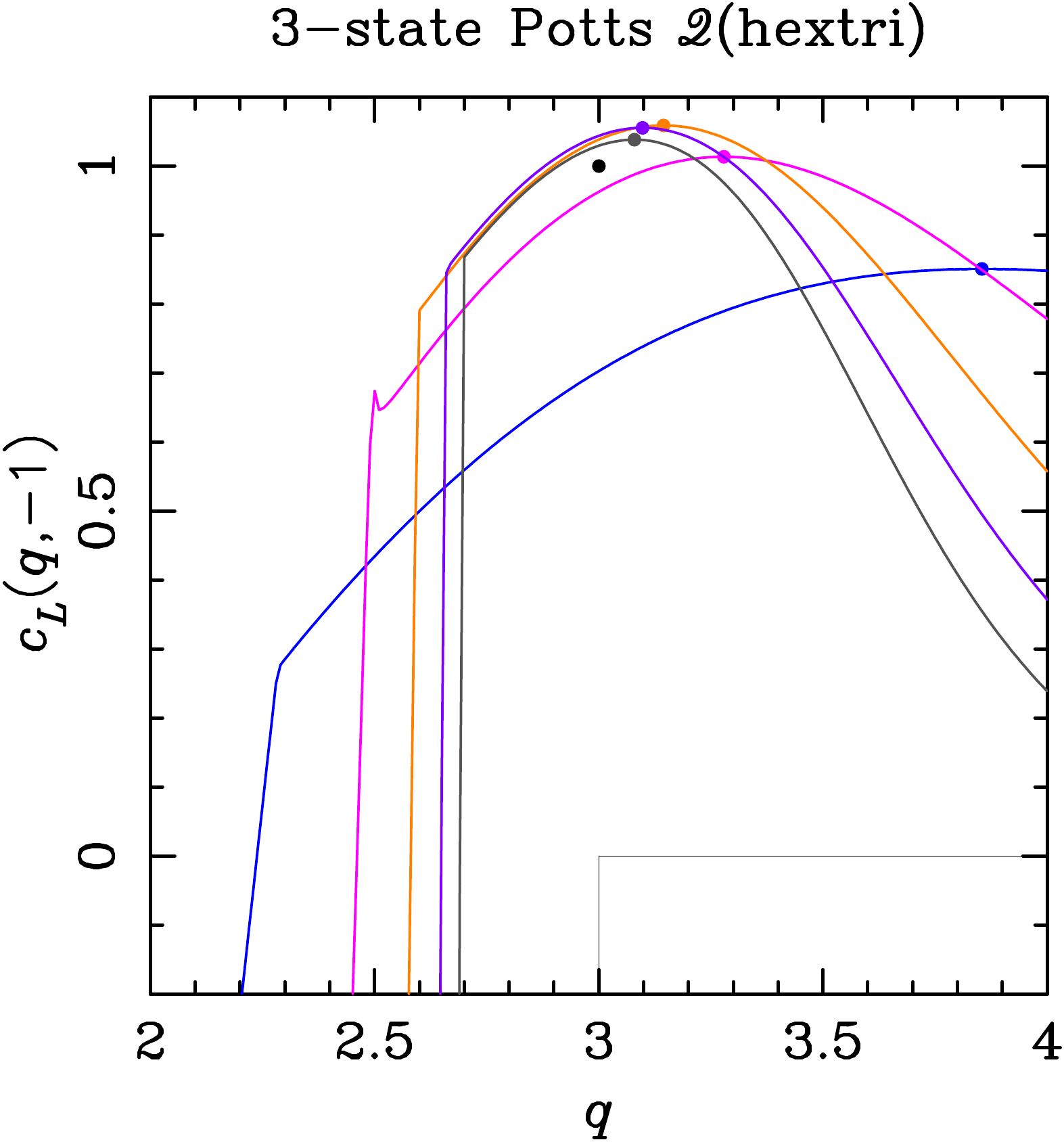} &   
\includegraphics[width=200pt]{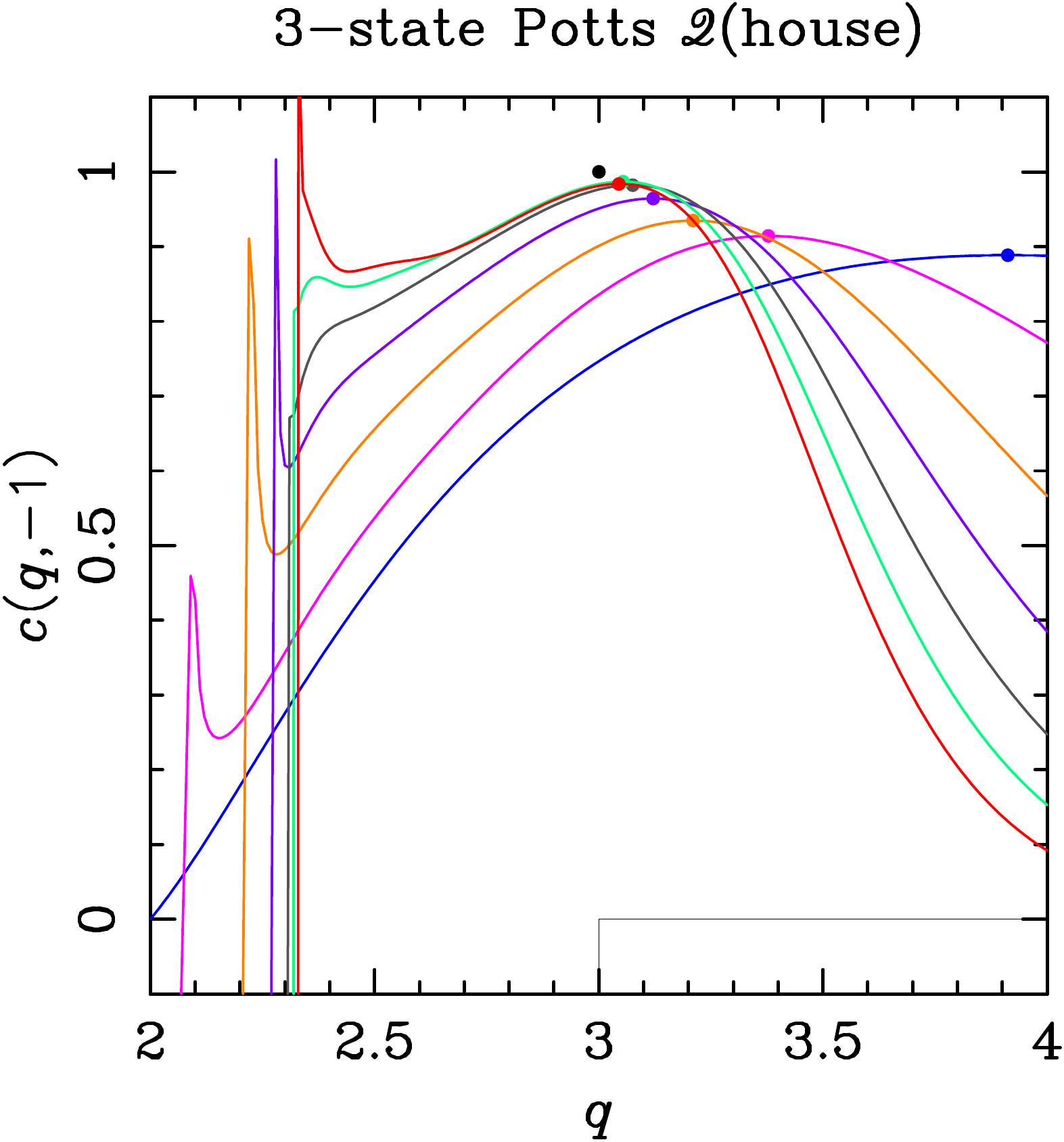} \\[2mm]
\qquad(a) & \qquad(b) \\
\end{tabular}
\end{center}
\vspace*{-0.5cm}
\caption{\label{fig_c_P}
   Estimates of the central charge $c_L(q,-1)$ for the 3-state Potts model 
   on the $\mathcal{Q}(\text{hextri})$ (a) and 
   $\mathcal{Q}(\text{house})$ (b) lattices with cylindrical boundary 
   conditions. We have used the {\em Ansatz} \eqref{def.f_Ansatz} for 
   $L=L_\text{min},L_\text{min}+4$ to avoid systematic parity effects. 
   We show the results for $L_\text{min}=2$ (blue), 
   $L_\text{min}=4$ (pink), $L_\text{min}=6$ (orange), 
   $L_\text{min}=8$ (violet), 
   $L_\text{min}=10$ (dark gray),
   $L_\text{min}=12$ (medium green), and  
   $L_\text{min}=14$ (red). 
   The dots at each curve mark the position 
   of the maximum value at $(q_\text{max}(L),c_\text{max}(L))$. 
   The black lines show the expected result 
   for $q>3$, and the black dot at $(q,c)=(3,1)$ shows the result
   predicted by conjecture~\ref{conj.main}(a). 
}
\end{figure}

Figure~\ref{fig_qhextri_tm}(b) shows clearly that the properties of the strip
under reflections depend on whether $L \equiv 0,2 \pmod{4}$. In the former 
(resp.\/ latter) case, the reflection axis goes trough two vertices
belonging to distinct (resp.\/ the same) sublattices. Therefore, we 
expect parity effects $\mathrm{mod}\, 4$ in the free energy $f_L(q,-1)$.
Again, as a precaution against subdominant FSS corrections, we have 
systematically varied the minimum value $L_\text{min}$ of the data 
included in the fit. Therefore, we have used in each fit to the \emph{Ansatz}
\eqref{def.f_Ansatz} the values $L= L_\text{min},L_\text{min}+4$. 
For each value of $L=L_\text{min}$, we obtained a curve $c_L(q,-1)$
that displays a maximum at the point $(q_\text{max}(L), c_\text{max}(L))$, 
where $c_\text{max}(L)=c(q_\text{max}(L),-1)$. These curves are depicted  
in figure~\ref{fig_c_P}(a), and the corresponding values for their
maxima are shown table~\ref{table_qhextri_P}. 
Notice that the existence of parity effects implies that the number of 
available data points is rather small: for $L \equiv 0 \pmod{4}$ it is two, and 
for $L \equiv 2 \pmod{4}$, three. Moreover, if we use the improved {\em Ansatz} 
\begin{equation}
f_L(q,-1)  \;=\; f_\text{bulk}(q,-1) + \frac{ G\, \pi\, c_L(q,-1)}{6\, L^2}
 + A\, L^{-4} \,,  
\label{def.f_AnsatzBis}
\end{equation}
for $L=L_\text{min}, L_\text{min}+4, L_\text{min}+8$, 
the number of data points $c_L(q,-1)$ is further reduced to one
and two, respectively (see table~\ref{table_qhextri_P}). 
In this table, the estimates $(q_\text{max}(L),c_\text{max}(L))$ seem to 
converge to the expected value $(3,1)$ as $L$ increases.

%
%
\begin{table}[htb]
\centering
\begin{tabular}{rlll}
\hline\hline \\[-4mm]
    & \multicolumn{2}{c}{{\em Ansatz} \eqref{def.f_Ansatz}} & \\
\hline    
$L$ & 
\multicolumn{1}{c}{$q_\text{max}(L)$} & 
\multicolumn{1}{c}{$c_\text{max}(L)$} & 
\multicolumn{1}{c}{$q_c(L)$} \\
\hline
2&  3.8544146155 & 0.8508786050 & \\
4&  3.2788982545 & 1.0133854086 & $1.8286335019 \pm 0.0559426374\,i$ \\
6&  3.1443621430 & 1.0588380075 & $2.2837808646$ \\
8&  3.0975518402 & 1.0554325766 & $2.4976597599$ \\
10& 3.0795627986 & 1.0383006482 & $2.5973897407$ \\ 
12&              &              & $2.6580810405 \pm 0.0007531318\, i$\\ 
14&              &              & $2.6959102812$ \\
\hline\\[-4mm] 
    & \multicolumn{2}{c}{{\em Ansatz} \eqref{def.f_AnsatzBis}} & \\
\hline
$L$ &
\multicolumn{1}{c}{$q_\text{max}(L)$} &
\multicolumn{1}{c}{$c_\text{max}(L)$} & \\
\hline 
2&  3.1224187094 & 1.1074883241 &  \\
4&  3.0645909602 & 1.0826966910 &  \\
6&  3.0547444360 & 1.0284408966 &  \\
\hline
\hline
\end{tabular}
\caption{
Finite-size results for the 
$\mathcal{Q}(\text{hextri})$ lattice with cylindrical boundary conditions
of widths $L$. The second and third columns show the position $q_\text{max}(L)$
of the maximum value of the central charge $c_L(q,-1)$, and the 
value $c_\text{max}(L)$ attained at this point. 
These estimates are obtained by fitting the free-energy $f_L(q,-1)$ 
\eqref{def.f_TM} using the {\em Ansatz} \eqref{def.f_Ansatz} on the top part
(resp.\/ \eqref{def.f_AnsatzBis} on the bottom part) at values of
the width $L,L+4$ (resp.\/ $L,L+4,L+8$) to avoid parity effects. 
The fourth column shows for each strip width $L$, the point(s) $q_c(L)$ of 
the limiting curve $\mathcal{B}_L$ closest to the real $q$-axis. 
The results in the first two columns for the \emph{Ansatz} \eqref{def.f_Ansatz} have already appeared in \cite[table~II]{Lv_17}. 
\label{table_qhextri_P}
}
\end{table}

We can also compute the largest real value of $q$ (if it exists at all!) 
where the limiting curve $\mathcal{B}_L$ crosses the real $q$-axis. 
This point $q_c(L)$ corresponds to two dominant eigenvalues of the TM 
becoming equimodular. These values are also displayed in 
table~\ref{table_qhextri_P}. For cylindrical boundary conditions, 
there is no such point in many cases \cite{transfer1,transfer2,JSS_tri}. 
This happens for strips of the $\mathcal{Q}(\text{hextri})$ lattice for 
$L=2$ (in this case the TM is one-dimensional), and $L=4, 12$. In these
latter cases, the limiting curve has a pair of complex conjugate endpoints 
rather close to the real $q$-axis. We will use the real part of such 
endpoints as our estimate for $q_c(L)$. As $L$ increases, the estimates 
for $\Re q_c(L)$ seem to grow towards $q_c=3$ (from below), but the FSS 
corrections seem to be larger than for the estimates $q_\text{max}(L)$.  
 
Our goal is to consider the limit $L\to\infty$ of these quantities. 
We first tried standard power-law fits:
\begin{subequations}
\label{power_laws}
\begin{align}
     \label{power_law_qmax}
     q_\text{max}(L) &=\; q_c(\mathcal{Q}(\text{hextri}))
                                    + A_q \, L^{-\omega_q} \,, \\ 
     \label{power_law_qc}
     \Re q_c(L)      &=\; q_c(\mathcal{Q}(\text{hextri}))
                                    + B_q \, L^{-\omega'_q} \,, \\ 
     \label{power_law_cmax}
     c_\text{max}(L) &=\; c(q_c,-1) + A_c \, L^{-\omega_c} \,, 
\end{align}
\end{subequations}
as we expect that 
$\displaystyle \lim_{L\to \infty} q_\text{max}(L) = \lim_{L\to \infty} q_c(L)=
q_c(\mathcal{Q}(\text{hextri}))$.

%
%
\begin{figure}[htb]
\begin{center}
\begin{tabular}{cc}
\includegraphics[width=200pt]{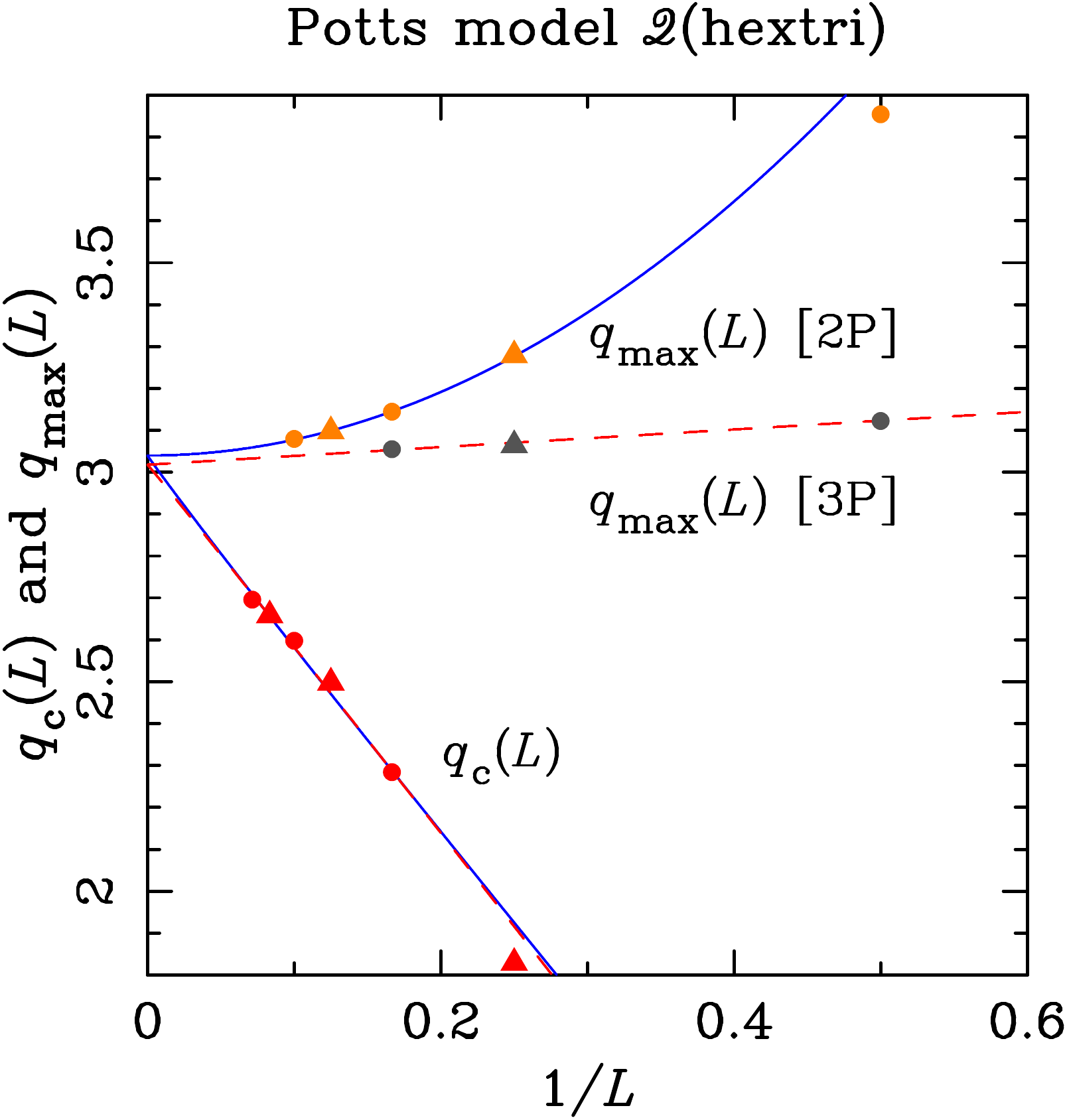} &  
\includegraphics[width=200pt]{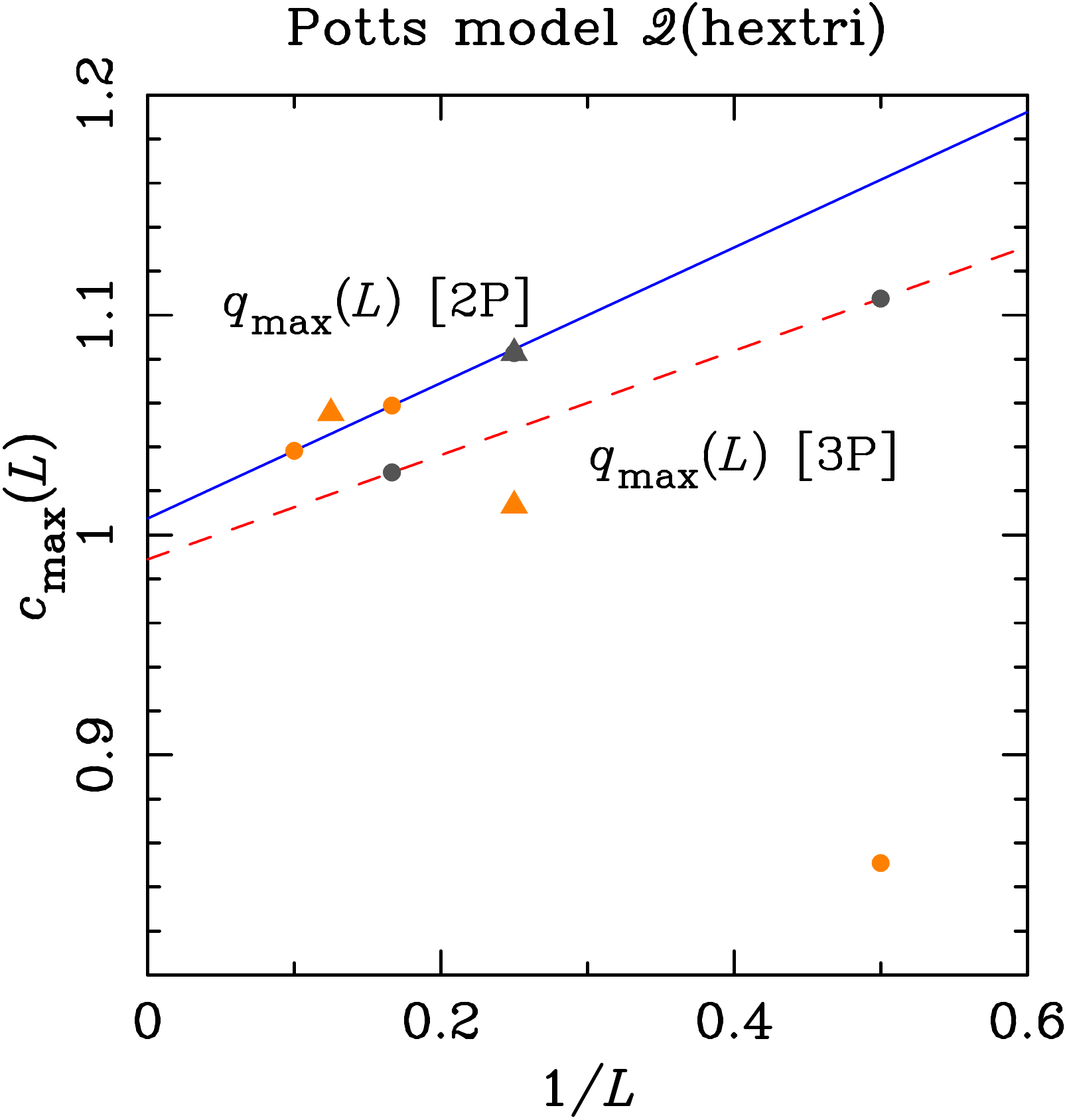} \\[2mm]
\qquad(a) & \qquad(b) \\
\end{tabular}
\end{center}
\vspace*{-0.5cm}
\caption{\label{fig_fits_c_qhextri_P}
   Fits for the critical parameters of the AF Potts model on the 
   $\mathcal{Q}(\text{hextri})$ lattice with cylindrical boundary conditions.
   (a) Value of $q_c(\mathcal{Q}(\text{hextri}))$. 
   The upper (orange) points points correspond to the estimates of 
   $q_\text{max}(L)$ using the {\em Ansatz} \eqref{def.f_Ansatz}, the middle
   (dark gray) points corresponds to the {\em Ansatz} \eqref{def.f_AnsatzBis}, 
   and the lower (red) points to those of $\Re q_c(L)$ (see 
   table~\ref{table_qhextri_P}). 
   The points corresponding to $L\equiv 2\pmod{4}$ (resp.\/ 
   $L\equiv 0\pmod{4}$) are depicted as solid circles $\bullet$ 
   (resp.\/ triangles $\blacktriangle$).
   The solid blue (resp.\/ dashed red) curves show the simultaneous fit of 
   the data sets $\Re q_c(L)$ and $q_\text{max}(L)$ obtained by 
   using the {\em Ansatz} \eqref{def.f_Ansatz} (resp.\/ the 
   {\em Ansatz} \eqref{def.f_AnsatzBis}). 
   (b) Value of $c(q_c,-1)$. The data points have the same colour/symbol 
   code is as in panel~(a) (see table~\ref{table_qhextri_P}). 
   The solid blue (resp.\/ dashed red) curve shows 
   the fit of the data set $c_\text{max}(L)$ obtained by using the 
   {\em Ansatz} \eqref{def.f_Ansatz} (resp.\/ the
   {\em Ansatz} \eqref{def.f_AnsatzBis}). 
}
\end{figure}

Let us start with the estimate for $q_c = q_c(\mathcal{Q}(\text{hextri}))$. 
First, we consider the data for $\Re q_c(L)$ (red points in 
figure~\ref{fig_fits_c_qhextri_P}(a)). Due to parity affects, we 
have fitted the data with $L\equiv 0 \pmod{4}$ and 
$L\equiv 2 \pmod{4}$ separately to the \emph{Ansatz} \eqref{power_law_qc}. 
In the former case we obtain $q_c \approx 2.84$ and $w'_q\approx 1.57$ 
for $L_\text{min}=4$, and in the latter, $q_c \approx 2.82$ and 
$\omega'_q \approx 1.71$ for $L_\text{min}=6$. In this latter fit, all data 
points have no imaginary part; so we tend to trust better this estimate 
than the other one. If we fix $\omega'_q=2$ in the last fit, we obtain
$q_c=2.80(6)$ for $L_\text{min}=10$.  
 
Now we consider the values of the estimates $q_\text{max}(L)$ obtained 
with the 2-parameter (2P) \emph{Ansatz} \eqref{def.f_Ansatz} (orange dots in
figure~\ref{fig_fits_c_qhextri_P}(a)). In this case, there are three points
with $L\equiv 2 \pmod{4}$. The power-law fit \eqref{power_law_qmax} gives
$q_c \approx 3.04$ and $\omega_q \approx 1.86$ for $L_\text{min}=2$. 
If we fix $\omega_q=2$ and repeat the fit, we obtain $q_c=3.04(4)$ for
$L_\text{min}=4$.

Finally, we have the estimates $q_\text{max}(L)$ obtained via the 
3-parameter (3P) \emph{Ansatz} \eqref{def.f_AnsatzBis} (dark gray points in 
figure~\ref{fig_fits_c_qhextri_P}(a)). There are only two data points
for $L\equiv 2 \pmod{4}$, so we fix $\omega_q=2$ and perform the fit to
the \emph{Ansatz} \eqref{power_law_qmax}; we obtain $q_c=3.05$ for
$L_\text{min}=2$. 

The above estimates look rather similar and close to the expected value
$q_c=3$. Moreover, the data set $\Re q_c(L)$ seems to increase monotonically 
as $L$ grows tending to $q_c = 2.80(6)$, while the data
set $q_\text{max}(L)$ (either coming from the 2P or 3P \emph{Ans\"atze} 
\eqref{def.f_Ansatz}/\eqref{def.f_AnsatzBis}, respectively) seems to 
decrease monotonically as $L$ grows tending to $q_c = 3.04(4)$. 
The expected value $q_c=3$ is just in between these two estimates (but many 
standard deviations away!).  If we fit these two estimates to a constant, 
we get $q_c=2.97(3)$ (with $\chi^2=11.08$, 1 DF, and $\text{CL}=0.087\%$). 

We might obtain an \emph{a priori} better estimate if we try to fit 
several data sets together with a single \emph{Ansatz}:  
\begin{equation}
q_c(L) \;=\; q_c(\mathcal{Q}(\text{hextri})) + \begin{cases} 
                   C_1 L^{-\omega_1} & \text{if data is $\Re q_c(L)$,} \\
                   C_2 L^{-\omega_2} & \text{if data is $q_\text{max}(L)$.}  
                   \end{cases}
\label{improved_qmax}
\end{equation}
We have used {\sc Mathematica}'s {\sc NonlinearModelFit} function to perform 
this more complex fit. If we consider the data sets $\Re q_c(L)$ and 
$q_\text{max}(L)$ coming from the 2P \emph{Ansatz} \eqref{def.f_Ansatz} 
for $L\equiv 2 \pmod{4}$, and fix $\omega_2=2$ (as we did above), we find 
$q_c= 3.04(3)$ and $\omega_1=0.97(10)$ for $L_\text{min}=6$. A better 
determination could be obtained by fixing $\omega_1=1$: the result is 
$q_c= 3.03(2)$ for $L_\text{min}=6$. 

If we repeat the same procedure with the data sets $\Re q_c(L)$ and 
$q_\text{max}(L)$ coming from the 3P \emph{Ansatz} \eqref{def.f_AnsatzBis}
for $L\equiv 2 \pmod{4}$, the slow variation of $q_\text{max}(L)$ versus 
$1/L$ suggests to fix $\omega_1=1$ in \emph{Ansatz} \eqref{improved_qmax}. 
Then we get $q_c=3.02(3)$ and $\omega_1=1.01(10)$ for $L_\text{min}=2$. If we
further fix $\omega_1=1$, we obtain $q_c=3.02(2)$.  

Notice that we refrain from merging both data sets for $q_\text{max}(L)$ 
(i.e., using \emph{Ans\"atze} \eqref{def.f_Ansatz}/\eqref{def.f_AnsatzBis}),
as they are not statistically independent. The curves obtained in the last 
two combined fits are depicted in figure~\ref{fig_fits_c_qhextri_P}(a). 
Moreover, the two distinct curves fitting the same data set $\Re q_c(L)$ are 
actually almost identical. In conclusion, our preferred estimated for
the combined \emph{Ansatz} \eqref{improved_qmax} is $q_c=3.03(2)$. 

If we merge the estimates $q_c=2.97(3)$ and $q_c=3.03(2)$, we obtain  
\begin{equation}
q_c( \mathcal{Q}(\text{hextri}) ) \;=\; 3.00(2) \,, 
\label{def_qc_Qhextri}
\end{equation} 
with $\chi^2=1.92$, 1 DF, and $\text{CL}=17\%$. This result is in full 
agreement with conjecture~\ref{conj.main}(a). Indeed, the error bar should be
taken with caution.  

Figure~\ref{fig_fits_c_qhextri_P}(b) shows the data points for the 
quantity $c_\text{max}(L)$. It is clear that both the parity effects and
the FSS corrections are rather strong in this case, while the number of 
data points is very small. Therefore, estimating the value of $c(3,-1)$ 
would be very difficult. 

We have first considered the points $c_\text{max}(L)$ (for 
$L\equiv 2 \pmod{4}$) obtained with the 2P \emph{Ansatz} \eqref{def.f_Ansatz} 
(orange dots in figure~\ref{fig_fits_c_qhextri_P}(b)). If we fix $\omega_c=1$ 
in the \emph{Ansatz} \eqref{power_law_cmax}, we get a sensible fit
$c(3,-1)\approx 1.01$ for $L_\text{min}=6$ (solid blue curve in 
figure~\ref{fig_fits_c_qhextri_P}(b)). If we add a second term $B L^{-2}$
to this \emph{Ansatz}, we get $c(3,-1)\approx 0.97$ for $L_\text{min}=2$. 
  
If we consider the estimates $c_\text{max}(L)$ obtained with the improved 3P 
\emph{Ansatz} \eqref{def.f_AnsatzBis} (dark gray points in 
figure~\ref{fig_fits_c_qhextri_P}(b)), we can fit them to the 
\emph{Ansatz} \eqref{power_law_cmax} by fixing $\omega_c=1$. We get 
$c(3,-1)\approx 0.99$ for $L_\text{min}=2$ (dashed red curve in 
figure~\ref{fig_fits_c_qhextri_P}(b)). 

If we consider all of the above, we conclude that our final estimate is    
\begin{equation}
c(3,-1) \;=\; 0.99(2) \,, 
\label{def_c_Qhextri}
\end{equation}
where the error bar should be taken with a grain of salt, as it takes into 
account the range of the few estimates discussed above. 
Again this results agrees well with conjecture~\ref{conj.main}(a). 

%
%
\subsubsection[Transfer matrix for Q(house)]%
  {Transfer matrix for $\bm{\mathcal{Q}(\text{house})}$} 
\label{sec.TM_Qhouse}
 
\begin{figure}[htb]
\begin{center}
\includegraphics[width=200pt]{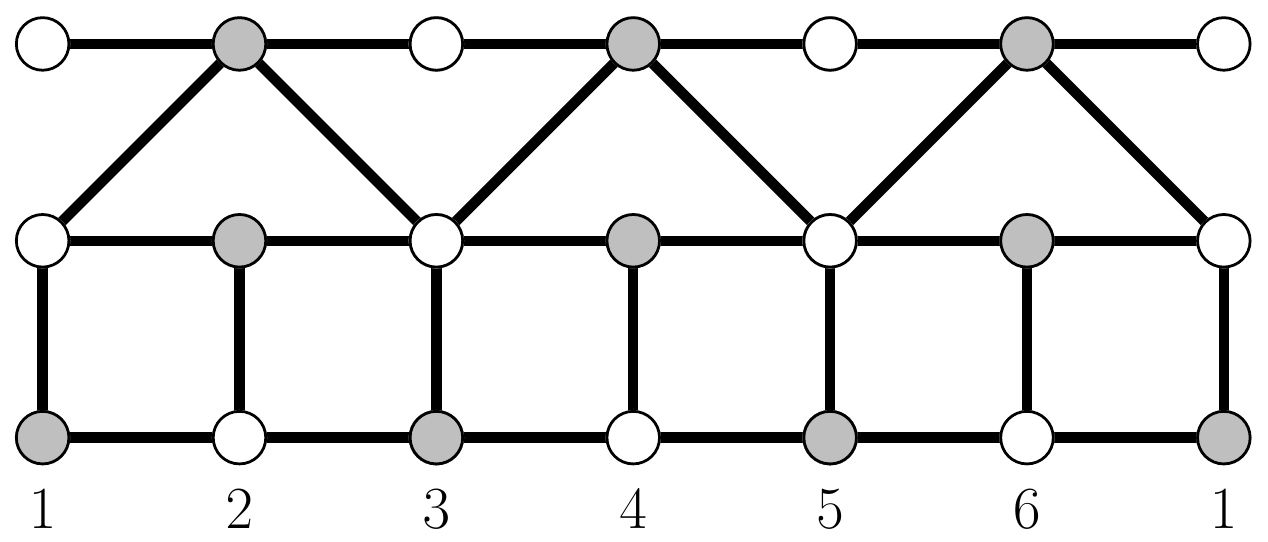}
\end{center}
\vspace*{-0.5cm}
\caption{\label{fig_qhouse_tm}
   TM construction for a layer of a strip graph of the 
   $\mathcal{Q}(\text{house})$ lattice of width $L=6$ and cylindrical  
   boundary conditions. There are two classes of 
   nonequivalent vertices, depicted respectively, as white and gray dots.
}
\end{figure}

We have repeated the previous computation for the $\mathcal{Q}(\text{house})$ 
lattice. In order to build this lattice, we have used the representation
depicted in figure~\ref{fig_qhouse_tm}. Again, the width of the strip should 
be an even integer $L$. If we label the vertices on the
bottom row of a strip width of size $L$ as $1,2,\ldots,L$, as in 
figure~\ref{fig_qhouse_tm}, then the TM for $\mathcal{Q}(\text{house})$
reads:
\begin{equation}
 \mathsf{T} \;=\; \mathsf{V}_0^\text{(odd)}  \cdot \mathsf{H} \cdot 
                  \mathsf{V}_0^\text{(even)} \cdot \mathsf{H} \cdot 
                  \mathsf{V}                 \cdot \mathsf{H} \,,
\label{def_tm_qhouse}
\end{equation}   
where the operators $\mathsf{H},\mathsf{V}$ and $\mathsf{V}_0^\text{(even)}$
are given by \eqref{def_h_v}/\eqref{def_v0}, and 
$\mathsf{V}_0^\text{(odd)}$ is expressed as
\begin{equation}
  \mathsf{V}_0^\text{(odd)} \;=\; \prod_{i=1}^{L/2} \mathsf{D}_{2i-1} \,. 
\label{def_v0_odd} 
\end{equation}

%
%

The space of connectivities is the same as for the $\mathcal{Q}(\text{hextri})$
lattice. We have also implemented the symbolic computation of the TM by a 
{\sc perl} script for widths $2\le L \le 16$. 
It is worth noticing that we found many rows that were exactly zero in all
TM for $L\ge 4$. In fact, these rows corresponded to connectivities for which
some of the odd-labelled sites were not singletons. Therefore, we could 
consider the smaller subspace of connectivities in which all odd-labelled 
sites are singletons. The dimensions of this `improved' TM are given by
by $1$, $2$, $3$, $6$, $10$, $24$, $49$, and $130$ for $L=2,4,\ldots,16$.
Indeed, we have checked the results by computing symbolically the
chromatic polynomial (using both TM) for strip graphs of size $L \times N$ 
with $2\le N\le 20$, and compare them to the results provided by other programs
\cite{Bedini,HPR}. The agreement is also perfect.

For each width $L$, we have computed the free energy $f_L(q,-1)$ in the 
interval $q\in [2,4]$ in terms of the dominant eigenvalue $\lambda_*$ of the 
corresponding TM (see \eqref{def.f_TM}). The central charge is then 
estimated by using the {\em Ansatz} \eqref{def.f_Ansatz}, where for this
lattice the geometrical factor is $G=2\sqrt{3}$ (see appendix~\ref{sec.app}). 
Following the same procedure as in the previous section, we obtain the 
results displayed in figure~\ref{fig_c_P}(b). 

%
%
\begin{table}[htb]
\centering
\begin{tabular}{rlll}
\hline\hline \\[-4mm]
    & \multicolumn{2}{c}{{\em Ansatz} \eqref{def.f_Ansatz}} & \\
\hline
$L$ &
\multicolumn{1}{c}{$q_\text{max}(L)$} &
\multicolumn{1}{c}{$c_\text{max}(L)$} & 
\multicolumn{1}{c}{$q_c(L)$} \\ 
\hline
 2& 3.7094424595 & 0.8808411388 & \\
 4& 3.3184514005 & 0.9155222019 & $1.7562125170 \pm 0.0575636262\, i$ \\
 6& 3.1770041039 & 0.9424270745 & $2.0770205951$\\
 8& 3.1034254588 & 0.9699972659 & $2.2109211636$\\
10& 3.0667634439 & 0.9837482525 & $2.2721580696$\\
12& 3.0500783475 & 0.9855564633 & $2.3029952390$\\
14&              &              & $2.3198098469$\\
16&              &              & $2.3297948774$\\
\hline\\[-4mm]
    & \multicolumn{2}{c}{{\em Ansatz} \eqref{def.f_AnsatzBis}} & \\
\hline
$L$ &
\multicolumn{1}{c}{$q_\text{max}(L)$} &
\multicolumn{1}{c}{$c_\text{max}(L)$} & \\
\hline
2&  3.1556222914 & 0.9623949139 &  \\
4&  3.0556456566 & 1.0064534361 &  \\
6&  3.0191585619 & 1.0209664138 &  \\
8&  3.0174550444 & 1.0040970911 &  \\
\hline
\hline
\end{tabular}
\caption{
Finite-size results for the 
$\mathcal{Q}(\text{house})$ lattice with cylindrical boundary conditions
of widths $L$. The second and third columns show the position $q_\text{max}(L)$
of the maximum value of the central charge $c_L(q,-1)$, and the
value $c_\text{max}(L)$ attained at this point.
These estimates are obtained by fitting the free-energy $f_L(q,-1)$
\eqref{def.f_TM} using the {\em Ansatz} \eqref{def.f_Ansatz} on the top part
(resp.\/ \eqref{def.f_AnsatzBis} on the bottom part) at values of
the width $L,L+4$ (resp.\/ $L,L+4,L+8$) to avoid parity effects.
The fourth column shows for each strip width $L$, the point(s) $q_c(L)$ of
the limiting curve $\mathcal{B}_L$ closest to the real $q$-axis (see text).
\label{table_qhouse_P}
}
\end{table}

It is clear from this figure that, for each value
$L=L_\text{min}$, the central charge presents a maximum 
$c_\text{max}(L)$ at $q=q_\text{max}(L)$. These values are shown in
table~\ref{table_qhouse_P}. Notice that the estimates for
$q_\text{max}(L)$ seem to converge rather quickly to the expected value $q_c=3$
from above, while the estimates for $c_\text{max}(L)$ seem also to increase as
$L$ increases. A FSS study is needed to tell whether these estimates
converge to the value $c=1$ predicted by conjecture~\ref{conj.main}(a).

We find that there is an endpoint of the limiting curve for $L=4$ very
close to the real $q$-axis. But contrary to what happens for the 
$\mathcal{Q}(\text{hextri})$ lattice,
for $6\le L\le 14$, we find that the limiting curve $\mathcal{B}_L$ 
contains an interval on the real $q$-axis. As $L$ increases, this interval 
moves to larger values of $q$ and its length decreases. It varies from 
$q \approx [2.056685,2.096902]$ for $L=6$ to 
$q \approx [2.329251,2.330336]$ for $L=16$. Therefore, we will not consider 
this estimate to try to compute $q_c$ for this lattice. The values displayed
in table~\ref{table_qhouse_P} correspond to the point in that interval where
the limiting curve emerges from the real $q$-axis.

%
%
\begin{figure}[htb]
\begin{center}
\begin{tabular}{cc}
\includegraphics[width=200pt]{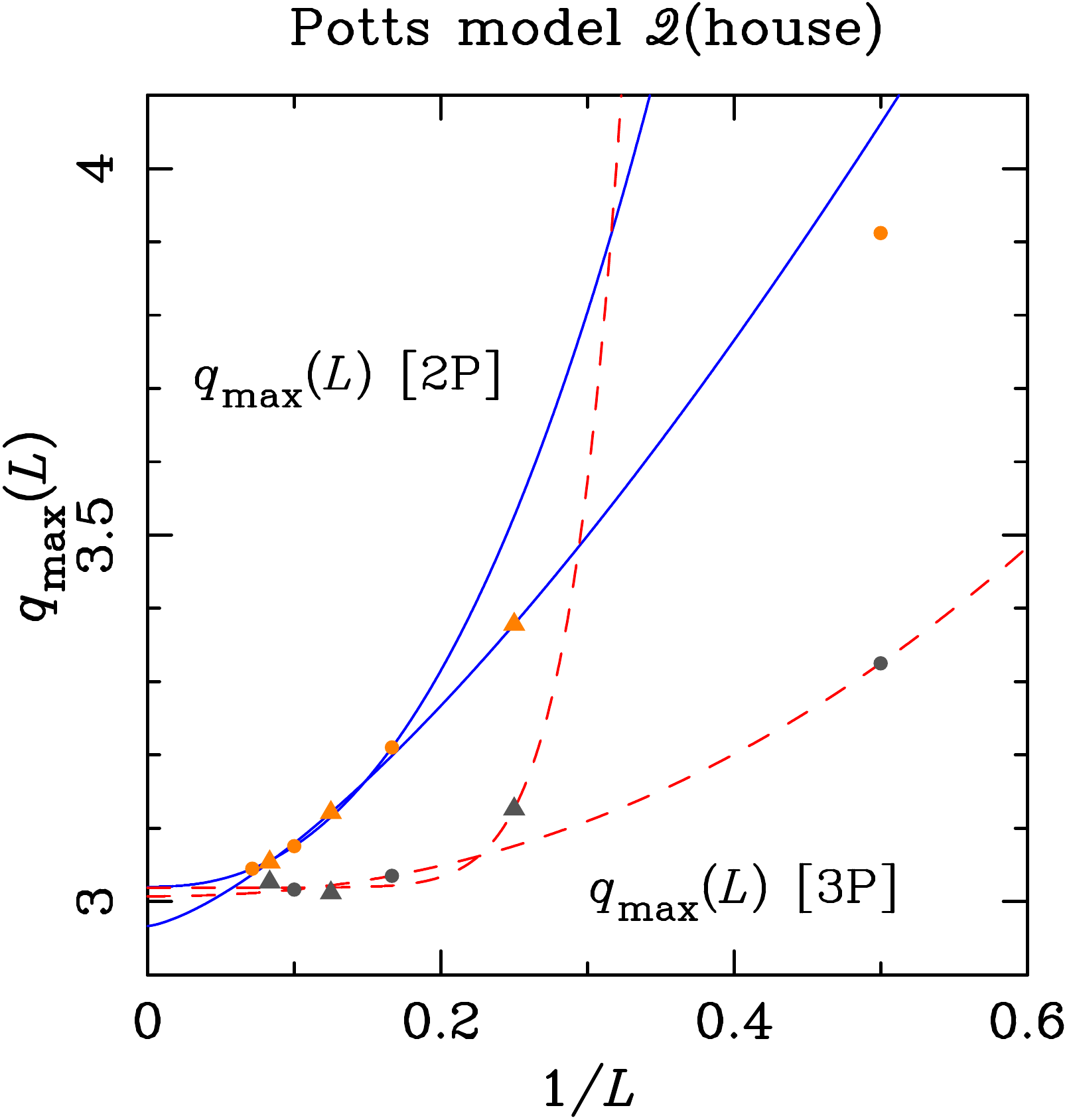} &  
\includegraphics[width=200pt]{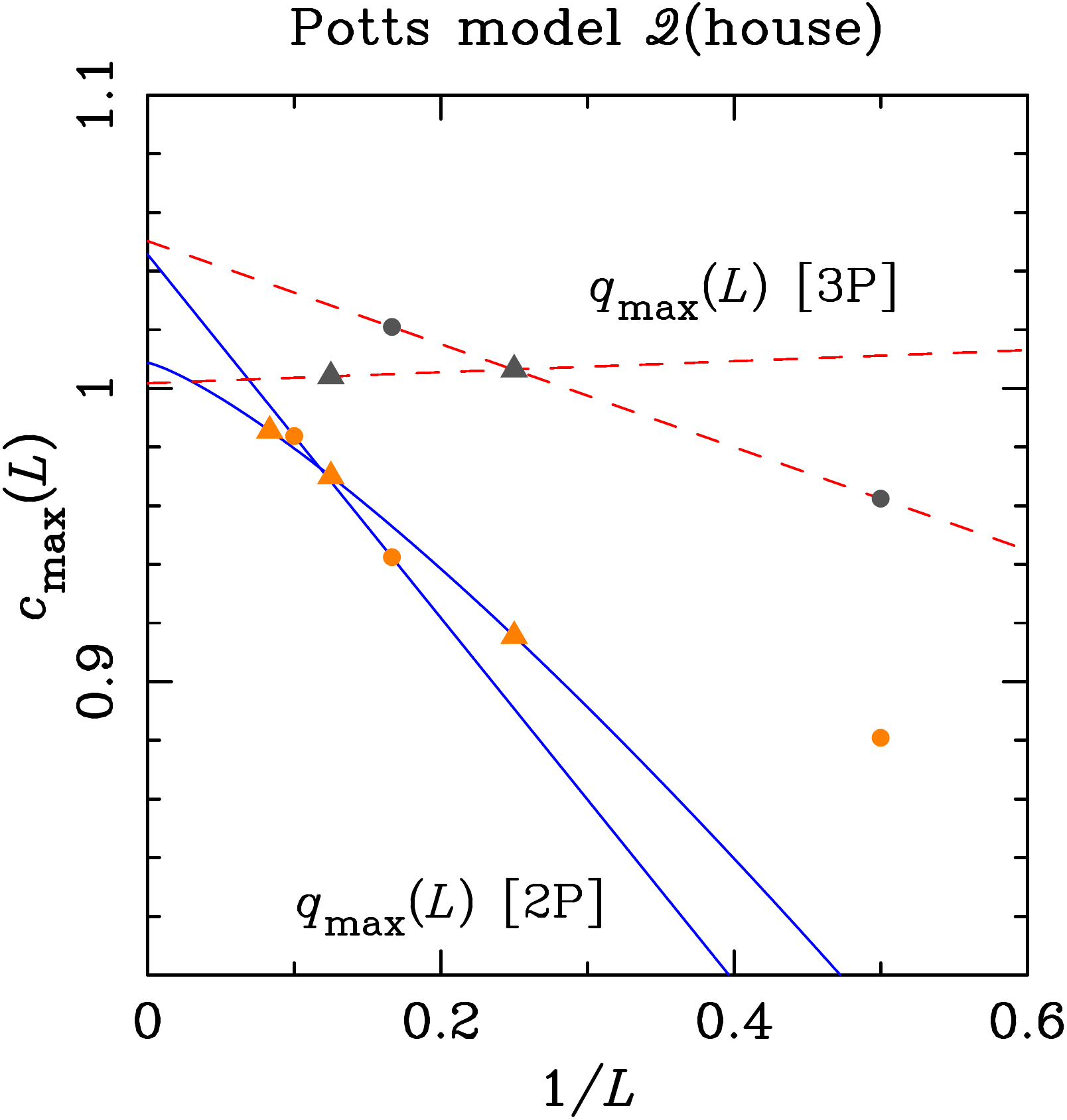} \\[2mm]
\qquad(a) & \qquad(b) \\
\end{tabular}
\end{center}
\vspace*{-0.5cm}
\caption{\label{fig_fits_c_house_P}
   Fits for the critical parameters of the AF Potts model on the 
   $\mathcal{Q}(\text{house})$ lattice with cylindrical boundary conditions.
   (a) Value of $q_c(\mathcal{Q}(\text{house}))$. 
   The upper (orange) points points correspond to the estimates of 
   $q_\text{max}(L)$ using the {\em Ansatz} \eqref{def.f_Ansatz}, and the lower 
   (dark gray) points corresponds to the {\em Ansatz} \eqref{def.f_AnsatzBis}. 
   (see table~\ref{table_qhouse_P}). 
   The points corresponding to $L\equiv 2\pmod{4}$ (resp.\/ 
   $L\equiv 0\pmod{4}$) are depicted as solid circles $\bullet$ 
   (resp.\/ triangles $\blacktriangle$).
   The solid blue (resp.\/ dashed red) curves show the fits of 
   the data set $q_\text{max}(L)$ obtained by 
   using the 2P {\em Ansatz} \eqref{def.f_Ansatz} (resp.\/ the 3P 
   {\em Ansatz} \eqref{def.f_AnsatzBis}). 
   (b) Value of $c(q_c,-1)$. The data points and curves have the same 
   colour/symbol code is as in panel~(a) (see table~\ref{table_qhouse_P}). 
}
\end{figure}

As for the $\mathcal{Q}(\text{hextri})$ lattice, we expect parity effects.
To obtain the estimate for $q_c = q_c(\mathcal{Q}(\text{house}))$,
we first consider the data $q_\text{max}(L)$ coming from the 2P \emph{Ansatz}
\eqref{def.f_Ansatz} (orange points in
figure~\ref{fig_fits_c_house_P}(a)). A power-law fit \eqref{power_law_qmax} 
gives $q_c \approx 2.97$ for $L_\text{min}=4$ for data with 
$L\equiv 0 \pmod{4}$, and the estimate $q_c \approx 3.02$ for $L_\text{min}=6$ 
for data with $L\equiv 2 \pmod{4}$.

If we perform similar fits for the data $q_\text{max}(L)$ coming from the 3P 
\emph{Ansatz} \eqref{def.f_Ansatz} (dark gray points in
figure~\ref{fig_fits_c_house_P}(a)), we obtain the estimates 
$q_c \approx 3.02$ for $L_\text{min}=4$, and 
$q_c \approx 3.01$ for $L_\text{min}=2$, respectively.

The lack of many data points discourage us to try to give an `objective'
error bar. By inspecting the above results, we conclude that our best 
estimate is 
\begin{equation}
q_c( \mathcal{Q}(\text{house}) ) \;=\; 2.99(3) \,, 
\label{def_qc_Qhouse}
\end{equation} 
in agreement with conjecture~\ref{conj.main}(a). Again, the error bar should
be taken with care.  
 
The estimate for $c(3,-1)$ can be obtained by considering the data 
$c_\text{max}(L)$ coming from the 2P \emph{Ansatz}
\eqref{def.f_Ansatz} (orange points in
figure~\ref{fig_fits_c_house_P}(b)). A power-law fit \eqref{power_law_qmax} 
gives $c(3,-1) \approx 1.01$ for $L_\text{min}=4$ and  
$c(3,-1) \approx 1.05$ for $L_\text{min}=6$ (after fixing $w_c=1$, because 
the power-law fit does not converge).  

If we use the data $q_\text{max}(L)$ coming from the 3P \emph{Ansatz} 
\eqref{def.f_Ansatz} (dark gray points in figure~\ref{fig_fits_c_house_P}(b)), 
we obtain the estimates $c(3,-1) \approx 1.05$ for $L_\text{min}=2$, and 
$c(3,-1) \approx 1.00$ for $L_\text{min}=4$. 

Finally, our `best' estimate is given by 
\begin{equation}
c(3,-1) \;=\; 1.03(3) \,, 
\label{def_cmax_Qhouse}
\end{equation} 
in agreement with conjecture~\ref{conj.main}(a). 
 
%
%
\subsection{Critical polynomials} \label{sec.CP}

The location of the phase transitions for the quadrangulations in 
figures~\ref{fig_Q_selfdual} and~\ref{fig_Q_non-selfdual} can also be studied 
by the method of critical polynomials 
\cite{JS_12a,JS_12b,JS_13,Jesper_14,JS_16}.
These polynomials $P_B(q,v)$ can in principle be computed for any lattice 
generated by the tessellation of the plane by some finite basis $B$.
However, we wish here to take advantage of the efficient TM algorithm of 
\cite{Jesper_14} which supposes
that $B$ admits a so-called four-terminal representation 
\cite[Section~3.4]{Jesper_14}.

%
%
\begin{figure}[htb]
\begin{center}
\begin{tikzpicture}[scale=1.0,>=stealth]
\foreach \xpos in {0,1,2,3}
\foreach \ypos in {0,1,2,3}
 \fill[black!20] (\xpos+0.5,\ypos) -- (\xpos+1,\ypos+0.5) -- %
                 (\xpos+0.5,\ypos+1) -- (\xpos,\ypos+0.5) -- cycle;

\foreach \xpos in {0,1,2,3}
\foreach \ypos in {0,1,2,3}
{
 \draw[blue,ultra thick] (\xpos+0.5,\ypos) -- (\xpos+0.75,\ypos+0.25) -- %
                         (\xpos+0.5,\ypos+1) -- (\xpos,\ypos+0.5) -- cycle;
 \draw[blue,ultra thick] (\xpos+0.75,\ypos+0.25) -- (\xpos+1,\ypos+0.5);
}

\foreach \ypos in {0,1,2,3}
{
 \draw[blue,ultra thick] (0,\ypos) -- (4,\ypos);
}

\foreach \xpos in {0,1,2,3}
\foreach \ypos in {0,1,2,3}
 \draw[black] (\xpos+0.5,\ypos) -- (\xpos+1,\ypos+0.5) -- %
              (\xpos+0.5,\ypos+1) -- (\xpos,\ypos+0.5) -- cycle;

\draw[very thick,->] (0,-0.5)--(4,-0.5);
\draw (4,-0.5) node[right] {$x$};
\foreach \xpos in {0,1,2,3}
{
 \draw[thick] (\xpos+0.5,-0.6)--(\xpos+0.5,-0.4);
 \draw (\xpos+0.5,-0.5) node[below] {$\xpos$};
}

\draw[very thick,->] (-0.5,0)--(-0.5,4);
\draw (-0.5,4) node[above] {$y$};
\foreach \ypos in {0,1,2,3}
{
 \draw[thick] (-0.6,\ypos+0.5)--(-0.4,\ypos+0.5);
 \draw (-0.5,\ypos+0.5) node[left] {$\ypos$};
}
 
\end{tikzpicture}
 \caption{Four-terminal representation of the quadrangulation 
          ${\cal Q}({\rm hextri})$.}
 \label{4term_Q_hextri}
\end{center}
\end{figure}
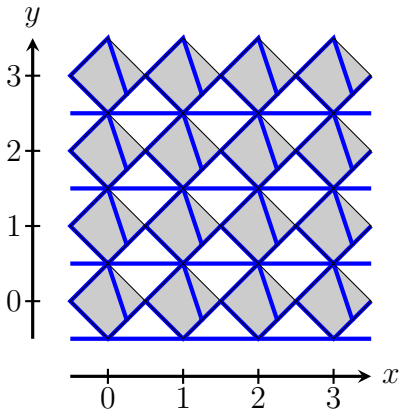

A four-terminal representation of $\mathcal{Q}(\text{hextri})$ is shown in 
figure~\ref{4term_Q_hextri}, for a basis of size $4 \times 4$ elementary cells.
In general, a four-terminal representation is a checkerboard alternation
of gray and white squares, such that the white squares are either empty or 
contain at most diagonally a horizontal edge from $B$, whereas all the 
remaining vertices and edges from $B$ reside within the gray squares. The most
efficient computations are obtained when the number of edges and vertices per 
gray square is as large as possible.
The representation in figure~\ref{4term_Q_hextri} contains 6 edges and 3 
vertices per gray square (this count includes by convention the diagonals in 
the white squares), which can be compared with \cite[table~3]{Jesper_14}.

Once a four-terminal representation for the lattice of interest has been found,
$P_B(q,v)$ can be computed by relatively minor modifications of the existing 
algorithm \cite{Jesper_14}.
It suffices to express the content within each gray square as an operator 
$\check{\sf R}_i$ \cite[section 3.4]{Jesper_14} which, in turn, is written as 
a product of elementary operators ${\sf H}_j$, ${\sf V}_j$ and ${\sf E}_j$. 
The first two of these, ${\sf H}_j$ (resp.\/  ${\sf V}_j$), correspond to the 
insertion of an edge of $B$ in the horizontal (resp.\/  vertical) time 
direction, with the usual convention that each gray square is traversed in 
the North-East direction under the time evolution that defines
the transfer matrix. The last operator, ${\sf E}_j$, is a
Temperley-Lieb generator that can be used in some cases (we do not need it 
here) to accomplish the basic join and detach operations.

For ${\cal Q}({\rm hextri})$ we can read off from figure~\ref{4term_Q_hextri} 
that
\be
 \check{\sf R}_i \;=\; {\sf V}_{i+2} \cdot {\sf H}_{i+1} \cdot {\sf V}_{i+2} 
                 \cdot {\sf V}_i \cdot {\sf H}_{i+1} \,,
\ee
which can be compared with the many examples given in 
\cite[Section~4]{Jesper_14}. In addition, there is a diagonal
edge on each white square.

In general, $\check{\sf R}_i$ is not necessarily given by the same expression 
for each gray square, and when this is not the case we will typically have an 
$n \times m$ pattern of different gray squares, with some periodicity in
the horizontal and vertical directions. As $B$ must be considered with 
implicit doubly periodic boundary conditions, for the
tiling of the whole plane to be valid this alternation
will impose parity constraints on $n$ and/or $m$, and in practice limit the 
number of sizes that can be studied.
For ${\cal Q}(\rm hextri)$ the expression for $\check{\sf R}_i$ is homogeneous,
and we can in particular consider bases of size $n \times n$ for any parity 
of $n$.

The critical polynomial for ${\cal Q}({\rm hextri})$ with a $1 \times 1$ 
basis is readily found as
\be
P_B(q,v) \;=\; -q^3 - 5 q^2 v - 8 q v^2 + 2 q v^3 + 12 v^4 + 6 v^5 + v^6 \,.
\ee
By modifying the algorithm of \cite{Jesper_14} we have similarly computed 
$P_B(q,v)$ for $n \times n$ bases with $n = 1,2,3,4$. They are polynomials of 
degree $3 n^2$ in $q$, and $6 n^2$ in $v$.
Setting $q=3$, we report in table~\ref{tab.PB_Q_hextri} the unique real zero 
of $P_B(3,v)$ in the ferromagnetic (FM) regime ($v > 0$), along
with any zero in the physical part of the AF regime ($-1 \le v < 0$). 
There are other real roots in the unphysical
AF regime ($v < -1$) that we do not report in table~\ref{tab.PB_Q_hextri}.

%
%
\begin{table}[htb]
\centering
\begin{tabular}{l|l|l}
\hline \hline \\[-4.5mm]
$n$ & \multicolumn{1}{|c|}{$v_{\rm c}^{\rm FM}(n)$} & 
      \multicolumn{1}{|c}{$v_{\rm c}^{\rm AF}(n)$} \\
\hline
1 & 1.53528140860819381729 & \\
2 & 1.53506577625594900753 & -0.91138737767231613799 \\
3 & 1.53500770075339075633 & \\
4 & 1.53499345908015323800 & -0.92997036994718707983 \\
\hline
$\infty$ 
   & 1.534987(1) & \\

\hline \hline
\end{tabular}
\caption{Real roots of $P_B(q,v)$, to 20 digit numerical precision, 
for the quadrangulation ${\cal Q}({\rm hextri})$.
We show the (unique) real roots, $v_{\rm c}^{\rm FM}(n)$ and 
$v_{\rm c}^{\rm AF}(n)$, in the intervals $v > 0$ and
$-1 \le v < 0$ respectively, for the $n \times n$ bases of 
figure~\ref{4term_Q_hextri}. The latter root only exists for even $n$.
The last row labeled `$\infty$' shows our extrapolation as $n\to \infty$.
\label{tab.PB_Q_hextri}
}
\end{table}

The FM zero $v_{\rm c}^{\rm FM}(n)$ converges rapidly to the critical 
coupling $v_{\rm c}$---a phenomenon which was already studied
extensively in \cite{Jesper_14} for a variety of lattices. A power-law fit 
\begin{equation}
v^\text{FM}_c(n) \;=\; v^\text{FM}_c + A\, n^{-\Delta} 
\label{def_vc_Ansatz}
\end{equation} 
for three consecutive data points (i.e., $n=n_\text{min},n_\text{min}+1,
n_\text{min}+2$) gives the estimate $v^\text{FM}_c=1.53498(6)$ for 
$n_\text{min}=2$. If we use Monroe's \cite{Monroe} implementation of the  
Bulirsh--Stoer \cite{BS} extrapolation scheme (MBS) with powers in the  
range $2.1$--$3.2$, we obtain a more precise result:
\begin{equation}
v_{\rm c}^{\rm FM} \;=\; 1.534987(1)\,.
\end{equation}
 
The physical AF zero $v_{\rm c}^{\rm AF}(n)$ only exists for even $n$
and appears to converge to the expected critical point, 
$v_{\rm c}^{\rm AF} = -1$, at a much slower rate. This contrasts with the
results of \cite{Jesper_14}, where $P_B(q,v)$ was seen to factorise in 
exactly solvable cases, thus giving one or more roots $v_{\rm c}$
with no $n$-dependence at all. In the present case, $P_B(3,v)$ does not 
factorise over the integers for any $n=1,2,3,4$. On the other
hand, our conviction that there is a critical point exactly at 
$v_{\rm c}^{\rm AF} = -1$ does not necessarily imply that this model would
be exactly solvable, in the sense that (for instance) the corresponding free 
energy could be computed exactly in the thermodynamic limit.

We have also found four-terminal representations for two of the other 
quadrangulations of self-dual type in figure~\ref{fig_Q_selfdual}: namely,  
$\mathcal{Q}(\text{house})$ and $\mathcal{Q}(\text{martini-B})$.
In both cases the operator $\check{\sf R}_i$ depends on the coordinates $(x,y)$
of the gray squares, see figure~\ref{4term_Q_hextri}.

For ${\cal Q}({\rm house})$ we have
\be
 \check{\sf R}_i \;=\; \begin{cases} 
 {\sf H}_{i+1} \cdot {\sf V}_{i+2} \cdot {\sf V}_i \cdot {\sf H}_{i+1} & 
  \text{for  $x \in 2 \mathbb{N}$,} \\
 {\sf H}_{i+1} \cdot {\sf E}_i \cdot {\sf H}_{i+1} \cdot {\sf V}_{i+2} & 
  \text{for $x \in 2 \mathbb{N}+1$,} 
 \end{cases}
\ee
and there are horizontal diagonals in the white squares with 
$x \in 2 \mathbb{N} + \tfrac12$. This representation is well-defined
for $n$ even, and it has 4 edges and 2 vertices per gray square. We have 
computed $P_B(q,v)$ for this lattice using $n \times n$
bases with $n=2,4$. The former case does not have any root in the physical 
AF regime. However, for $n=4$ we find a root at
$v_{\rm c}^{\rm AF}(4) = -0.94524856890048286949\cdots$, which is reasonably 
close, but not exactly equal, to $v_{\rm c}^{\rm AF} = -1$.

For ${\cal Q}(\text{martini-B})$ we have found
\be
 \check{\sf R}_i \;=\; \begin{cases} 
 {\sf H}_{i+1} \cdot {\sf V}_{i+2} \cdot {\sf V}_i \cdot {\sf H}_{i+1} & 
 \text{for  $x \in 4 \mathbb{N}$,} \\
 {\sf H}_{i+1} \cdot {\sf V}_{i+2} \cdot {\sf V}_i \cdot {\sf H}_{i+1} 
    \cdot {\sf V}_{i} & 
 \text{for  $x \in 4 \mathbb{N}+1$,} \\
 {\sf V}_i \cdot {\sf E}_{i+1} \cdot {\sf E}_{i+2} \cdot {\sf H}_{i+1} &  
 \text{for  $x \in 4 \mathbb{N}+2$,} \\
 {\sf H}_{i+1} \cdot {\sf E}_i \cdot {\sf H}_{i+1} \cdot {\sf V}_{i+2} & 
 \text{for  $x \in 4 \mathbb{N}+3$,} 
 \end{cases}
\ee
and there are again horizontal diagonals in the white squares with 
$x \in 2 \mathbb{N} + \tfrac12$.
This representation is well-defined only for $n \in 4 \mathbb{N}$ and has 
4 edges per gray square.

We now turn to the non-self-dual quadrangulations of 
figures~\ref{fig_Q_non-selfdual}--\ref{fig_Q_non-selfdual2}. The fact that 
most of these involve vertices of high degree has prevented us from finding
four-terminal representations.
There is of course no doubt that {\em some} TM formulation could be found for 
those lattices, but we refrain from studying this issue further here, 
since it would involve a more profound rewriting of the algorithm given 
in \cite{Jesper_14}.

The first of the non-self-dual quadrangulations, ${\cal Q}({\rm diced})$, 
has in fact already been treated in \cite[section 4.10]{Jesper_14} where a 
four-terminal representation and explicit polynomials $P_B(q,v)$
for several $n \times n$ bases were given. To find the corresponding AF 
critical point, we shall instead
resort to the eigenvalue method of \cite{Jesper_15}, which amounts to finding 
numerically the relevant root $v_{\rm c}^{\rm AF}(n)$ of $P_B(3,v)$ for a basis
of size $n \times m$, directly in the limit $m \to \infty$.
This TM method has the advantage over that of \cite{Jesper_14} that the size 
$n$ can be doubled with approximately the same computational effort.

%
%
\begin{table}[htb]
\centering
\begin{tabular}{r|l|l}
\hline \hline \\[-4.5mm]
$n$ & \multicolumn{1}{|c|}{$v_{\rm c}^{\rm FM}(n)$} &
      \multicolumn{1}{|c}{$v_{\rm c}^{\rm AF}(n)$} \\
\hline
  2 & 1.60090047247076387216 & $-0.93449469491145567949$ \\
  4 & 1.60092569962114202789 & $-0.93889690618313817225$ \\
  6 & 1.60092045368358355317 & $-0.93976678350525022210$ \\
  8 & 1.60091983209526681674 & $-0.94017098791714722205$ \\
 10 & 1.60091973715419384611 & $-0.94038789257375557598$ \\
 12 & 1.60091971679606195523 & $-0.94051494788357303489$ \\ 
\hline
$\infty$ 
    & 1.600919708(6)         & $-0.94080(1)$ \\
\hline \hline
\end{tabular}
\caption{Real roots of $P_B(q,v)$, to 20 digit numerical precision, for the 
quadrangulation ${\cal Q}({\rm diced})$.
We show the (unique) real roots, $v_{\rm c}^{\rm FM}(n)$ and 
$v_{\rm c}^{\rm AF}(n)$, in the intervals $v > 0$ and
$-1 \le v < 0$ respectively, for $n \times \infty$ bases. They are obtained 
from the eigenvalue method of \cite{Jesper_15}.
The last row labeled `$\infty$' shows our extrapolations as $n\to \infty$.
The results for the AF critical point $v_{\rm c}^{\rm AF}(n)$ have already 
appeared in \cite[table~IV]{Lv_17}.
\label{tab:PB_Q_diced}
}
\end{table}

The results are shown in table~\ref{tab:PB_Q_diced}. If we fit the 
data for $v_{\rm c}^{\rm FM}(n)$ to the power-law fit \eqref{def_vc_Ansatz}
using three consecutive data points (i.e., $n=n_\text{min},n_\text{min}+2,
n_\text{min}+4$), we obtain a rather accurate fit for $n_\text{min}=8$:
\begin{equation}
v_{\rm c}^{\rm FM} \;=\; 1.600919708(6) \,,
\end{equation}
where the error is twice the difference between the estimates with 
$n_\text{min}=6,8$. The convergence is rather fast, as the exponent $\Delta$ 
varies from $\Delta\approx 6.2$ for $n_\text{min}=6$ to
$\Delta\approx 6.5$ for  $n_\text{min}=8$. If we use the MBS 
extrapolation scheme with powers in the range $5.6$--$7.5$, we obtain a 
slightly less precise result:
\begin{equation}
v_{\rm c}^{\rm FM} \;=\; 1.60092049(1) \,.
\end{equation}

The convergence for the AF critical point is slower. If we do the same
procedure with the data in the third column in table~\ref{tab:PB_Q_diced},
we find an estimate 
\begin{equation}
v_{\rm c}^{\rm AF} \;=\; -0.9410 \pm 0.0002\,,  
\end{equation}
with the error bar estimated as twice the difference between the estimates for 
$n_\text{min}=6,8$. In this case the value of the exponent $\Delta$ stays
around $\Delta\approx 1.4$--$1.6$. 
We performed the MBS algorithm with powers in the range 
$1.2$--$1.5$, and obtained a more precise result:
\begin{equation}
v_{\rm c}^{\rm AF} \;=\; -0.94080(1) \,.
\end{equation}
This value is compatible with (and more precise than) the MC result 
shown in Table~\ref{table_results_non_self_dual}: 
$v_{\rm c}^{\rm AF}= -0.94075(12)$. 

Notice that for the ${\cal Q}({\rm diced})$ we have obtained estimates for
both the FM and AF regimes, and the convergence is a lot faster and the
result more precise when the corresponding critical theory is in the FM 
Potts universality class, rather than in the AF one.  

%
%
\section{Discussion} \label{sec.disc}

In this paper we have proposed a conjecture (conjecture~\ref{conj.main}) 
that \emph{predicts} the phase diagram of the 3--state Potts AF model
on any plane quadrangulation. This phase diagram depends on a single 
characteristic of these quadrangulations: namely, if they are of self-dual
or non-self-dual type. Notice that this prediction deals with an infinite
set of models. This conjecture follows from general arguments based on the 
height representation of these models at zero temperature, and on universality. 
Indeed, it generalises previous studies on two particular models: the square- 
\cite{Burton_Henley_97,Salas_98} and the diced-lattice 
\cite{Kotecky-Salas-Sokal} cases. 
 
We have tested the predictions of conjecture~\ref{conj.main} on four 
quadrangulations of self-dual type (see figure~\ref{fig_Q_selfdual}), and 
on seven quadrangulations of non-self-dual type (see 
figures~\ref{fig_Q_non-selfdual}--\ref{fig_Q_non-selfdual2}). 
We have made extensive MC simulations on all of these quadrangulations, as 
well as detailed TM computations on two of them 
(i.e., $\mathcal{Q}(\text{hextri})$ and $\mathcal{Q}(\text{house})$) and 
CP computations for $\mathcal{Q}(\text{hextri})$, $\mathcal{Q}(\text{house})$,
$\mathcal{Q}(\text{martini-B})$, and $\mathcal{Q}(\text{diced})$. All these
results are in good agreement with the predictions of 
conjecture~\ref{conj.main}. 
In table~\ref{table_G} we display all the quadrangulations $\Gamma$ 
considered in this work. We also include the two well-known cases studied 
in the literature: the square \cite{Burton_Henley_97,Salas_98} and the diced 
\cite{Kotecky-Salas-Sokal} lattices. 

%
%
\begin{table}[htb]
\centering
\small
\begin{tabular}{|c|cc|c|}
\hline
Quadrangulation  & Sublattice & Sublattice &  $v_c(3)$ \\
$\Gamma$         & $G_0$      & $G_1$      &           \\
\hline
\multicolumn{4}{c}{\quad} \\
\multicolumn{4}{c}{\bf Quadrangulations of self-dual type} \\
\hline
square                    & square    &square   &\multicolumn{1}{l|}{$-1$}\\
$\scrq(\text{hextri})$    & hextri    &hextri   &\multicolumn{1}{l|}{$-1$}\\ 
$\scrq(\text{house})$     & house     &house    &\multicolumn{1}{l|}{$-1$}\\ 
$\scrq(\text{martini-B})$ & martini-B &martini-B&\multicolumn{1}{l|}{$-1$}\\ 
$\scrq(\text{cmm-pmm})$   & cmm-pmm   &cmm-pmm  &\multicolumn{1}{l|}{$-1$}\\ 
\hline
\multicolumn{4}{c}{\quad} \\
\multicolumn{4}{c}%
{\bf Quadrangulations of non-self-dual type, $\bm{G_0}$ or $\bm{G_1}$ 
     triangulation \cite{Kotecky-Sokal-Swart}} \\
\hline
diced  & triangular  &   hexagonal   &   
       \multicolumn{1}{l|}{$-0.860599(4)$}  \\
$\mathcal{Q}(\text{martini})$  & dual martini   &  martini &
       \multicolumn{1}{l|}{$-0.77454(6)$}  \\
$\mathcal{Q}(\text{cross})$  & cross & bisected hexagonal & 
      \multicolumn{1}{l|}{$-0.80057(4)$}  \\
$\mathcal{Q}(\text{asanoha})$    & asanoha & 3--12 & 
      \multicolumn{1}{l|}{$-0.72033(3)$}  \\
$G''_2$ \cite{planar_AF_largeq}  & union-jack & four--eight  &  
       \multicolumn{1}{l|}{$-0.82278(4)$}  \\
\hline
\multicolumn{4}{c}{\quad} \\
\multicolumn{4}{c}%
{\bf Quadrangulations of non-self-dual type, $\bm{G_0}$ or $\bm{G_1}$ 
     quadrangulation} \\
\hline
$\mathcal{Q}(\text{diced})$ &  diced   &   kagome   & 
      \multicolumn{1}{l|}{$-0.94075(12)$}  \\
$\mathcal{Q}(\text{ruby})$ &  dual ruby   &   ruby  & 
      \multicolumn{1}{l|}{$-0.95588(9)$}  \\
$G''_3$ \cite{planar_AF_largeq} & $G'_2$ & decorated four-eight &
      \multicolumn{1}{l|}{$-0.72278(2)$} \\
\hline
\hline
\end{tabular}
\caption{
   Some plane quadrangulations $\Gamma$ arising from a dual pair $(G_0,G_1)$.
   For each of them, we display the critical temperature $v_c(3)$ 
   for the 3-state AF Potts model on $\Gamma$. The square 
   \cite{Burton_Henley_97,Salas_98} and diced \cite{Kotecky-Salas-Sokal} 
   lattices are added for completeness.  
}
\label{table_G}
\end{table}

Even though we have found an excellent empirical support for 
conjecture~\ref{conj.main}, there are a few issues (already raised in the
text) that might limit the generality of this conjecture:

\begin{itemize}

\item For the quadrangulations of self-dual type, we find that the 
      existence of a zero-temperature critical point depends on the fact that
      the stiffness of the effective Gaussian field theory satisfies 
      $K < K_c = \pi/2$. Even though
      we have seen that this is the case for the four quadrangulations of this 
      type considered in this work, we have found that the value of $K$ 
      grows smoothly with the maximum degree $\Delta$ of the quadrangulation. 
      Notice that the study of $\mathcal{Q}(\text{cmm-pmm})$ with $\Delta=8$ 
      was motivated to test whether increasing $\Delta$ would produce or not
      a large increment in $K$; but the answer was negative.  
      However, it is still possible that for quadrangulations of this type with 
      $\Delta$ large enough, $K > K_c$. This would imply that 
      conjecture~\ref{conj.main}(a) does not hold for such quadrangulations. 
      On the other hand, the solution~I found in 
      \cite{Delfino_17} seems to give additional theoretical support 
      to our conjecture.

\item For quadrangulations of non-self-dual type, we find that at $T=0$ 
      there are three coexisting (ordered) ideal states (= phases). This 
      implies that there should be some finite-temperature phase transition 
      separating the ordered phase from the disordered high-temperature one. 
      Even though, we have found that in all cases, this phase transition is of
      second order, we cannot rule out on general grounds the existence of a 
      first-order transition (like in the one undergone by the 3-state AF 
      Potts model on the triangular lattice \cite{Adler_95}).  

\item The argument leading to conjecture~\ref{conj.main}(b) assumes that 
      one sublattice has more vertices than the other one (i.e.
      $|V_0| < |V_1|$). This line of reasoning might be false when both
      sublattices have the same number of vertices. This observation led us 
      to study carefully two such cases, $\mathcal{Q}(\text{diced})$ and 
      $\mathcal{Q}(\text{ruby})$. We have seen that the fluctuations around
      the `naive' ideal states are enough to distinguish between the two
      sublattice: the one that contains the vertices with largest degree is
      the ordered one, while the other one is disordered. However, the 
      FM ordering of the former sublattice is only partial, contrary to what
      happens for the more generic case $|V_0| < |V_1|$.    

\end{itemize}

These questions motivate the study of more complicated lattices in order to 
find out whether there 
are any counter-examples to conjecture~\ref{conj.main}. 

As a side-effect of the MC computations, we have found that there is also
extensive empirical support for conjecture~\ref{conj.wsk}, which 
predicts the dynamic critical behavior of the WSK algorithm for the whole 
class of the 3-state AF Potts models on any quadrangulation. The most 
important part, in our view, is conjecture~\ref{conj.wsk}(a), which predicts the
total absence of CSD for quadrangulations of self-dual type. 
Quite often, the WSK algorithm (as well as its SW counterpart
for FM models) `only' reduces the characteristic dynamic critical exponent 
$z_\text{int}\approx 2$ for single-site algorithms to some smaller value; but
does not completely eliminate CSD. The only exception known in the 
literature was the square-lattice 3-state AF model 
\cite{Sokal_Ferreira,Salas_98}. 
Conjecture~\ref{conj.wsk}(a) implies that this algorithm eliminates CSD 
for an infinite class of models. On the other hand, for quadrangulations of 
non-self-dual type, conjecture~\ref{conj.wsk}(b) predicts that 
$z_\text{int}$ should be equal to that of the SW algorithm for the 3-state 
FM Potts model. Indeed, this is not surprising in the light of 
conjecture~\ref{conj.main}(b). 

Finally, it is worth mentioning that the convergence properties of the CP
method are far better for critical points in the FM regime, than for those
in the AF regime. Indeed, the estimates for the former critical points are 
several orders of magnitude more precise than the estimates for the latter.
In addition, to our knowledge this is the first time that the location of
the critical point for the 3-state FM Potts model on the 
$\mathcal{Q}(\text{hextri})$ and $\mathcal{Q}(\text{diced})$ have been 
computed to high precision.  

%
%
\appendix
\section{Computation of the geometric factor} \label{sec.app}

The computation of the geometric factor $G$ for the 
$\mathcal{Q}(\text{hextri})$ and $\mathcal{Q}(\text{house})$ is not
completely trivial (specially for the latter case).

\begin{figure}[htb]
\begin{center}
\includegraphics[width=200pt]{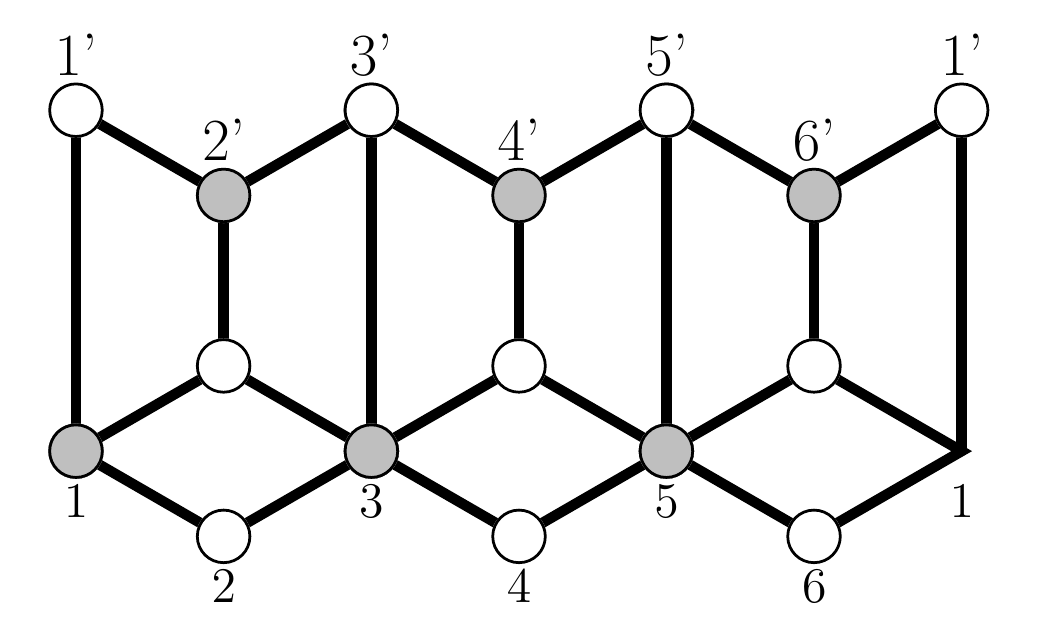}
\end{center}
\vspace*{-6mm}
\caption{\label{fig_G_hextri}
   Strip of width $L=6$ of the lattice $\mathcal{Q}(\text{hextri})$ with 
   cylindrical boundary conditions and depicted in an non-distorted way
   (compare to figure~\ref{fig_qhextri_tm}(a)). An application of
   the TM transforms a bottom-row $(1,2,\ldots,L)$ configuration into a 
   top-row $(1',2',\ldots,L')$ configuration. 
}
\end{figure}

Let us start with the simplest case: $\mathcal{Q}(\text{hextri})$ shown
in figure~\ref{fig_G_hextri}. First we should find an appropriate way of
representing this strip graph so that it is not distorted (like in 
figure~\ref{fig_qhextri_tm}(a)). In this case the choice is obvious:
take regular hexagons of lattice spacing $a$, so that all edges have the
same length, except for the vertical edges splitting each hexagon into two
equal halves (which have length $2a$). 
The area covered by one application of the TM is given by
$A_L = \smfrac{1}{2} d(1,3) \, d(1,1')\, L$, where the distances are 
$d(1,3)=2$, and $d(1,1')=2a$ (see figure~\ref{fig_G_hextri}). 
Because $d(1,3)=\sqrt{3}\, a=2$, we get that $a=2/\sqrt{3}$ and therefore, 
$A_L = (4/\sqrt{3})\, L$. This means that
\be
G(\mathcal{Q}(\text{hextri})) \;=\; \frac{4}{\sqrt{3}}\,. 
\ee
Notice that this is the right geometric factor for our definition of the
TM and free energy density \eqref{def.f_TM}. 

%
%

\begin{figure}[htb]
\begin{center}
\includegraphics[width=200pt]{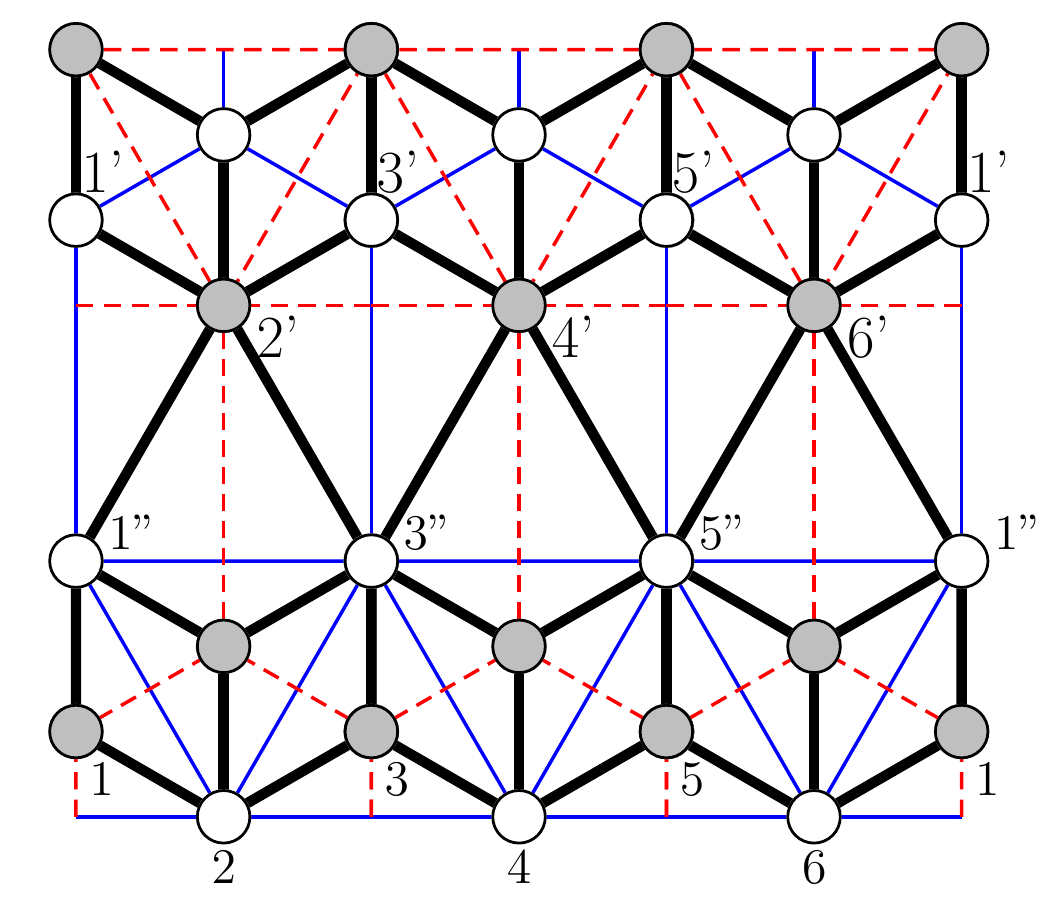}
\end{center}
\vspace*{-6mm}
\caption{\label{fig_G_house}
   Strip of width $L=6$ of the lattice $\mathcal{Q}(\text{house})$ with 
   cylindrical boundary conditions and depicted in an non-distorted way
   (compare to figure~\ref{fig_qhouse_tm}). An application of
   the TM transforms a bottom-row $(1,2,\ldots,L)$ configuration into a 
   top-row $(1',2',\ldots,L')$ configuration. We also show the two 
   underlying house lattices (white dots and thin solid blue lines, and 
   gray dots and thin dashed red lines, respectively). 
}
\end{figure}

The second case $\mathcal{Q}(\text{house})$ is more involved. Based on 
figure~2 of \cite{Okeeffe_92} (which represents the crystal structure of the
CrB binary compound), we have produced figure~\ref{fig_G_house}. 
One of the house lattices is depicted with gray dots
and thin solid blue lines, while its dual (which is also a house lattice) is 
displayed with white dots and thin dashed red lines. The quadrangulation 
$\mathcal{Q}(\text{house})$ corresponds to both sets of dots and thick solid
black lines. Notice that each house lattice contains triangular and pentagonal
faces. In figure~2 of \cite{Okeeffe_92} the triangles are equilateral, and 
the pentagons are not regular: i.e.,  they are symmetric with respect to an 
axis going through the `apex' vertex, and the two internal angles opposite to
the apex vertex are equal to $\pi/2$.  
The question is how to choose the free parameters in these
lattices so we can get the physical value of $G$. 
Figure~\ref{fig_G_house} shows a strip of width 6 and cylindrical boundary
conditions, so that an application of the TM transforms the 
bottom-row $(1,2,\ldots,L)$ configuration into a top-row $(1',2',\ldots,L')$
configuration. We have shown for clarity `half' of the next row.  

%
%

Again we should find an appropriate way of representing this strip graph so 
it is not distorted (like in  figure~\ref{fig_qhouse_tm}). 
We can first assume that the triangles in both house lattices are equilateral
with lattice spacing $d(1'',3'') = d(3'',2) = d(2,1'') = a$. 
The position of the dual vertex inside one
triangular face should be placed at its barycenter. This fixes the position of 
the apex vertex of each pentagon (say, vertex $3$ in figure~\ref{fig_G_house}). 
With this choice, the lattice spacing of the rhombi of the 
$\mathcal{Q}(\text{house})$ lattice is given by 
$d(1,1'') = d(1,2) = a/\sqrt{3}$. 
We can see two rows formed by these rhombi at the bottom and at the top of
the figure~\ref{fig_G_house}. 
However, the `height' $b$ of the pentagon (i.e., the distance $d(3',3'')$) 
remains unfixed. The only criterion we have found that makes sense is to
choose $b$ so that the triangles formed by vertices $1''$, $3''$ and $2'$ 
(or, equivalently, $2'$, $4'$ and $3''$) are also equilateral with lattice 
spacing equal to $a = d(1'',3'')=d(2',4')=d(2,4)$. This implies that 
$b=2a/\sqrt{3}$. Note that with this choice, $\mathcal{Q}(\text{house})$ is 
then nothing but an undistorted triangular lattice with extra vertices at
the barycenters of every second row of triangles.

The area covered by one application of the TM is given by
$A_L = \smfrac{1}{2} d(2,4) \, d(4,4')\, L$, where the distances are 
$d(2,4)=a=2$, and $d(4,4')=b+a/\sqrt{3}=a\sqrt{3}$ 
(see figure~\ref{fig_G_house}). 
Therefore, $A_L = (2\sqrt{3})\, L$. This means that
\be
G(\mathcal{Q}(\text{house})) \;=\; 2\sqrt{3}\,.
\label{GQhouseApp}
\ee

\bigskip

\noindent
{\bf Remark.} 
In general, the determination of the geometrical factor $G$ amounts to 
determining which `undistorted' representation of a given regular lattice
will lead to a continuum limit having \emph{continuous} rotational symmetry, 
which is clearly a minimum requirement for the critical model to
acquire conformal invariance. However, the precise mechanism by which a 
2, 3, 4 or 6-fold discrete rotational symmetry turns into a continuous one
has not, to our knowledge, been much studied in the literature. In particular, 
the principle what constitutes an undistorted representation is
not completely clear, except for the simplest Archimedean lattices.

The quadrangulation $\mathcal{Q}(\text{house})$ considered above is a good 
case in point: The naive 2-fold rotational symmetry is almost, but not quite, 
lifted to a 6-fold symmetry by the choice \eqref{GQhouseApp}. Indeed, in 
figure~\ref{fig_G_house}
every second horizontal row of triangles has an extra vertex inside each 
triangular face, while after a $\frac{\pi}{3}$ rotation, each horizontal 
row of triangles has an extra vertex inside every second triangular face.
It is not clear how to obtain such a `coarse grained' higher-order 
rotational symmetry for more general regular lattices.

%
%
\section*{Acknowledgments}

This paper is dedicated to the memory of Chris Henley. We thank Alan Sokal
for helpful comments and especially for sharing with us the initial ideas 
that led to this work and to reference \cite{Lv_17}. 
We are also indebted to Kun Chen and Yuan Huang who were involved in the 
first stages of this work. Last but not least, we thank Gesualdo Delfino 
for correspondence and for pointing out to us reference \cite{Delfino_17}. 

JS is grateful for the hospitality of the Laboratoire de Physique Th\'eorique 
at the \'Ecole Normale Sup\'erieure (from 2012 to 2018), and the 
Department of Mathematics at University College London (from 2012 to 2017), 
where part of this work was done.

This work has been supported in part by the National Natural Science
Foundation of China under grants No.~11774002 (JPL),
and No.~11625522 (YD), 
the Key Projects of Anhui Province University Outstanding Youth Talent
Support Program grant gxyqZD2017009 (JPL), 
the Ministry of Science and Technology of China grant No.~2016YFA0301600 (YD),
the Institut Universitaire de France, and 
the European Research Council through the Advanced Grant NuQFT (JLJ),
and the MINECO FIS2014-57387-C3-3-P and 
MINECO/AEI/FEDER, UE FIS2017-84440-C2-2-P grants (JLJ and JS). 

%
%

\end{document}